\documentclass[11pt,a4paper]{article}
\usepackage[T1]{fontenc}
\usepackage[utf8]{inputenc}
\usepackage{lmodern}
\usepackage[a4paper,left=2.35cm,right=2.35cm,top=2.25cm,bottom=2.25cm,headheight=15pt]{geometry}
\usepackage{amsmath,amssymb,mathtools,bm,amsthm}
\usepackage{graphicx,booktabs,tabularx,array,enumitem}
\usepackage{xcolor,tikz,tikz-cd}
\usetikzlibrary{arrows.meta,positioning,calc,decorations.markings,patterns}
\usepackage{float,needspace,microtype}
\usepackage[section]{placeins}
\usepackage{hyperref,bookmark,fancyhdr}
\definecolor{ink}{HTML}{193D50}
\definecolor{accent}{HTML}{16727A}
\definecolor{pale}{HTML}{F1F7F7}
\definecolor{linkblue}{RGB}{27,79,114}
\definecolor{muted}{RGB}{90,96,102}
\definecolor{speciesA}{RGB}{171,57,57}
\definecolor{speciesB}{RGB}{43,95,147}
\hypersetup{colorlinks=true,linkcolor=linkblue,citecolor=linkblue,urlcolor=linkblue,
pdftitle={A Duality Between Gauge Transformations and Time Evolutions},
pdfauthor={Hanyu Xue}}
\theoremstyle{plain}
\newtheorem{theorem}{Theorem}[section]
\newtheorem{proposition}[theorem]{Proposition}
\newtheorem{lemma}[theorem]{Lemma}
\newtheorem{corollary}[theorem]{Corollary}
\theoremstyle{definition}
\newtheorem{definition}[theorem]{Definition}
\newtheorem{remark}[theorem]{Remark}
\newcommand{\ZZ}{\mathbb Z}
\newcommand{\RR}{\mathbb R}
\newcommand{\RZ}{\RR/\ZZ}

\newcommand{\ii}{\mathrm i}

\newcommand{\dd}{\mathrm d}
\newcommand{\id}{\mathrm{id}}
\newcommand{\supp}{\operatorname{supp}}
\newcommand{\Hom}{\operatorname{Hom}}

\newcommand{\Sq}{\operatorname{Sq}}
\newcommand{\ket}[1]{\lvert #1\rangle}
\newcommand{\bra}[1]{\langle #1\rvert}

\newcommand{\OverlapWord}[2]{%
  \mathord{\vcenter{\hbox{%
    \begin{tikzpicture}[x=.8cm,y=.36cm]
      \draw[black,line width=.8pt,line cap=round] (0,0)--(3,0);
      \foreach \x in {0,2}
        \fill[speciesA] (\x,0) ellipse[x radius=.045,y radius=.12];
      \foreach \x in {1,3}
        \fill[speciesB] (\x,0) ellipse[x radius=.045,y radius=.12];
      \node[text=speciesA,inner sep=.5pt] at (0,.72) {$\scriptstyle -#1$};
      \node[text=speciesB,inner sep=.5pt] at (1,.72) {$\scriptstyle #2$};
      \node[text=speciesA,inner sep=.5pt] at (2,.72) {$\scriptstyle #1$};
      \node[text=speciesB,inner sep=.5pt] at (3,.72) {$\scriptstyle -#2$};
      \node[inner sep=.5pt] at (0,-.72) {$\scriptstyle x_1$};
      \node[inner sep=.5pt] at (1,-.72) {$\scriptstyle y_1$};
      \node[inner sep=.5pt] at (2,-.72) {$\scriptstyle x_2$};
      \node[inner sep=.5pt] at (3,-.72) {$\scriptstyle y_2$};
    \end{tikzpicture}%
  }}}%
}

\setlist{nosep,leftmargin=2.2em}

\numberwithin{equation}{section}
\title{A Duality Between Gauge Transformations and Time Evolutions}
\author{Hanyu Xue\\[0.35em]\normalsize Massachusetts Institute of Technology\\[0.2em]\normalsize\texttt{xhy2002@mit.edu}}
\begin{document}
\maketitle

\begin{abstract}
We propose that gauge transformations define an independent direction dual to time. Consider a matter field $\phi$ coupled to a background gauge field $A$. In the time direction, the path integral sums over matter-field histories at fixed $A$, with the covariant derivative $\dd\phi+A$ recording the excitations. In the gauge direction, $\dd\phi+A$ stays fixed while $A$ changes. This duality relates familiar concepts in field theory and condensed matter physics. In the time direction, $\dd\phi+A$ traces the world lines of domain walls. In the gauge direction, $A$ traces the world lines of symmetry defects. We introduce transport algebras to describe these motions and establish an antilinear isomorphism between them. This isomorphism unifies the statistics of particles and more general topological excitations with symmetry anomalies. As applications, we derive criteria for gauging $(d-1)$-form symmetries in $d$ spatial dimensions and ordinary symmetries in two spatial dimensions, and obtain a family of invariants that give simple measurements of anomalies and statistics.
\end{abstract}

\section{Introduction}
\label{sec:introduction}

Symmetry and gauge theory are closely related concepts in modern physics. In mathematics, a gauge transformation can be understood as a homotopy~\cite{KapustinThorngren2014,DelcampTiwari2019}. Concretely, a gauge transformation between two field configurations $\phi_0,\phi_1$ on a manifold $X$ is a field $\widetilde{\phi}$ on $X\times[0,1]$ whose restrictions to the two ends are $\phi_0$ and $\phi_1$, respectively. This suggests the following sequence of ideas:
\begin{equation}
  \text{symmetry}\to\text{gauge transformation}\to\text{homotopy}\to\text{fields in one higher dimension}.
\end{equation}
Here $X$ can itself be spacetime, so the interval $[0,1]$ supplies an extra dimension coming from the homotopy. We will give this dimension a physical interpretation and argue that it is dual to time.

Consider a $(q-1)$-form matter field $\phi$ coupled to a flat $q$-form gauge field $A$. With the gauge transformation written as $(\phi,A)\mapsto(\phi+s,A-\dd s)$, the covariant derivative is
\begin{equation}
	D_A\phi=\dd \phi+A.
\end{equation}
We work on a discrete lattice with a finite Abelian group $G$. The fields are cochains $\phi\in C^{q-1}(X,G)$ and cocycles $A\in Z^q(X,G)$, and $D_A\phi\in Z^q(X,G)$. For a simple picture, take $G=\ZZ_2$ and $q=1$ in one spatial dimension. The matter field $\phi$ describes the spins of a symmetry-breaking ferromagnet. The background $A$ marks the antiferromagnetic bonds in the Hamiltonian, which are symmetry defects. The covariant derivative $D_A\phi$ marks the domain walls that cost energy, which we call topological excitations. The two objects have similar geometry. Both are described by cocycles of the same degree, and their motion traces closed world lines with the same conservation law. Their roles differ, however. In a time evolution, we usually hold $A$ fixed and sum over the histories of the matter field $\phi$, including the world lines recorded by $D_A\phi$. In a gauge transformation, $D_A\phi$ stays fixed while the background field $A$ changes.

Topological excitations and symmetry defects can thus both move subject to a conservation law. The former move in time and the latter in the gauge direction. Table~\ref{tab:gauge-time-duality} pairs the concepts that arise in the two directions.

\begin{table}[H]
	\centering
  \caption{Duality between time evolution and gauge transformations. At fixed background, matter-field histories are summed with dynamical weights. The background path of a gauge transformation is specified externally.}
	\label{tab:gauge-time-duality}
	\begin{tabular}{@{}l@{\qquad\qquad}l@{}}
		\toprule
    \textbf{Time direction (dynamical)} & \textbf{Gauge direction (no independent dynamics)}\\
		\midrule
    covariant derivative $D_A\phi$ & background field $A$\\
		topological excitation & symmetry defect\\
		hopping & gauge transformation \\
		statistics & anomaly\\
		condensation & gauging\\
		\bottomrule
	\end{tabular}
\end{table}

Both columns contain familiar concepts in field theory and topological phases of matter. Their resemblance can make it difficult to tell when one can substitute for the other. Treating gauge transformations as a direction alongside time makes their relationship precise. We study the operator algebra of hoppings that move topological excitations and the operator algebra of gauge transformations that move symmetry defects. We find three properties. First, the two algebras are mutual commutants: the motions in the time and gauge directions can be performed independently. Second, the two algebras are antilinearly isomorphic. In this sense the two directions have equal standing, and an operation in one has a counterpart in the other. Third, a single algebraic structure, which we call a \emph{transport algebra}, describes both motions. Its invariant obstructs condensation when applied to topological excitations and gauging when applied to symmetry defects. In this sense it unifies statistics and anomaly. As an application, we show that a $(d-1)$-form symmetry in $d$ spatial dimensions can be gauged when its class in $H^{d+2}(K(G,d),\RZ)$ is trivial. For an ordinary symmetry in two spatial dimensions, one must also consider its restriction to a disk. Applying $a$ and then $b$ in the disk differs from applying $a+b$ only along the boundary. These fusion errors define a known obstruction to gauging, $[\nu]\in H^2(G,\mathbb Q_{>0}^{\times})$~\cite{KawagoeShirley2025,ShirleyZhangJiLevin2026}. We prove that when $[\nu]$ is trivial, compatible restrictions can be chosen and assembled into gauge transformations. The symmetry can then be gauged precisely when the class $[\omega]\in H^4(G,\RZ)$ measured by those transformations is trivial (Theorem~\ref{thm:gauging-2d-patches}).

Our proposal to put the gauge direction on the same footing as time grew out of a difficulty in reconciling two notions of robustness. To see the difficulty, consider the braiding statistics of Abelian anyons. The two strings in Fig.~\ref{fig:boundary-avoidance}(a) cross once. The hopping operator $U_1(L_1)$ moves the first anyon to the right along the red string, and $U_2(L_2)$ moves the second upward along the blue string. Consider the commutator
\begin{equation}
[U_2(L_2),U_1(L_1)]
 =U_2(L_2)^{-1}U_1(L_1)^{-1}U_2(L_2)U_1(L_1).
 \label{eq:intro-commutator}
\end{equation}
Following the relative motion of the two particles shows that one winds once around the other. This gives the geometric interpretation of braiding. What requires explanation is the robustness of the braiding phase. Neither the precise positions of the strings $L_1,L_2$ nor the precise formula for $U(L)$ is fixed, yet every admissible choice of hopping operators must give the same commutator for braiding statistics to be well defined. We can test this by perturbing $U_1(L_1)$ and asking whether $[U_2(L_2),U_1(L_1)]$ changes.

\begin{figure}[htbp]
\centering
\begin{tikzpicture}[>=Latex,x=1cm,y=1cm,font=\small,line cap=round]
  \node at (0,1.48) {(a) Braiding};
  \node at (6.65,1.48) {(b) Type-II perturbation at $x$};
  \draw[speciesA,line width=1pt] (-1.43,0)--(1.43,0);
  \draw[speciesB,line width=1pt] (0,-.94)--(0,.94);
  \draw[speciesA,-{Latex[length=4pt]},line width=1pt] (.60,0)--(1.02,0);
  \draw[speciesB,-{Latex[length=4pt]},line width=1pt] (0,.43)--(0,.73);
  \fill[speciesA] (-1.43,0) circle (3pt);
  \fill[speciesB] (0,-.94) circle (3pt);
  \draw[speciesA,fill=white,line width=.8pt] (1.43,0) circle (2.8pt);
  \draw[speciesB,fill=white,line width=.8pt] (0,.94) circle (2.8pt);
  \node[speciesA,above] at (1.0,.04) {$L_1$};
  \node[speciesB,left] at (-.02,.83) {$L_2$};
  \node at (0,-1.40) {$U_1,\ U_2,\ U_1^{-1},\ U_2^{-1}$};

  \foreach \offset/\bluey in {4.70/-.94,8.60/.94} {
    \begin{scope}[shift={(\offset,0)}]
      \filldraw[fill=gray!18,draw=gray!65,thin] (0,0) circle (.41);
      \draw[speciesA!43,thin] (-1.12,0)--(1.12,0);
      \draw[speciesB!43,thin] (0,-.94)--(0,.94);
      \node[font=\footnotesize] at (.16,-.16) {$x$};
      \fill[speciesA] (1.12,0) circle (3pt);
      \fill[speciesB] (0,\bluey) circle (3pt);
    \end{scope}
  }
  \draw[-{Latex[length=4.5pt]},muted] (6.14,0)--node[above] {$U_2$}(7.16,0);
  \node at (4.70,-1.40) {$O_x$ contributes $e^{\ii\alpha}$};
  \node at (8.60,-1.40) {$O_x^{-1}$ contributes $e^{-\ii\alpha}$};
\end{tikzpicture}
\caption{Filled dots mark the initial anyon positions. (a) The red and blue anyons move in turn along the arrows to the open dots, then return in turn. The four hopping operators are listed in order of application. (b) Make the type-II perturbation $U_1\mapsto O_xU_1$, where $O_x$ multiplies a state by $e^{\ii\alpha}$ if the gray disk is empty and by $1$ if it contains an anyon. The configuration near $x$ is read after $U_1$ (left) and again after $U_2$ (right). The blue anyon moves from below the disk to above it. The disk is empty at both readings, so the two phases cancel.}
\label{fig:boundary-avoidance}
\end{figure}

Choose a point $x$ on the red string in Fig.~\ref{fig:boundary-avoidance}(b) and replace $U_1$ by $O_xU_1$, where $O_x$ is a local unitary near $x$. Then
\[
 [U_2,O_xU_1]=U_2^{-1}U_1^{-1}\bigl(O_x^{-1}U_2O_x\bigr)U_1 .
\]

If $x$ lies off the blue string, as it does at either endpoint of the red string, $O_x$ commutes with $U_2$ and $[U_2,U_1]$ is unchanged. At the crossing, a generic local operator $O_x$ can change the commutator. A hopping operator, however, transports an excitation between the endpoints without leaving a trace in the interior. In this Abelian anyon example, once the anyon configuration near $x$ is fixed, the topological states are locally indistinguishable there. An $O_x$ that preserves this subspace therefore acts on it by a phase determined by the local configuration. Since $U_2$ changes the anyon configuration only at the endpoints of $L_2$, the phases contributed by $O_x$ and $O_x^{-1}$ cancel at the crossing (Fig.~\ref{fig:boundary-avoidance}(b)).

This argument involves two kinds of perturbations. The first allows arbitrary local operators near the endpoints. The second allows the hopping operator to acquire a phase anywhere along the string, provided the phase depends only on the local anyon configuration. Robustness under both perturbations follows from the same geometric condition, which we call \emph{boundary avoidance}:
\begin{equation}
	\partial L_1\cap L_2=L_1\cap\partial L_2=\varnothing.
	\label{eq:mutual-avoidance}
\end{equation}
The interiors of the two operations can still intersect.

The distinction between the two perturbations matters in other settings. For topological excitations of codimension $q$ and fusion group $G$ in $d$ spatial dimensions, Ref.~\cite{Xue2026Statistics} constructed operator sequences called \emph{statistical processes}. They generalize braiding and the loop self-statistics of Fidkowski, Haah, and Hastings~\cite{FidkowskiHaahHastings2022}. Evaluating them on hopping operators $U$ yields invariants classified by $H^{d+2}(K(G,q),\RZ)$. We call these invariants statistics: they generalize the $H^4(K(G,2),\RZ)$ classification of anyon statistics, which includes both braiding and the exchange of identical particles. These generalized statistics are robust under the second kind of perturbation but need not be robust under the first. Example III.2 of Ref.~\cite{Xue2026Statistics} illustrates the distinction in a one-dimensional $\ZZ_2$ ferromagnet with all spins up or all spins down. Its domain walls have fusion statistics (an $F$-symbol) classified by $H^3(\ZZ_2,\RZ)\simeq\ZZ_2$. For adjacent intervals $I=[k,l]$ and $J=[l+1,m]$, the sign of $[U(J),U(I)^2]=\pm1$ measures these statistics. The simplest hopping operator is a product of Pauli operators,
\begin{equation}
	U([\ell,r])=\prod_{j=\ell}^r X_j.
\end{equation}
Its statistics are trivial. Adding controlled-$Z$ gates at the two endpoints gives

\begin{equation}\label{eq:Z2 nontrivial hopping}
	U'([\ell,r])=\mathrm{CZ}_{\ell-1,\ell}
	\Bigl(\prod_{j=\ell}^r X_j\Bigr)\mathrm{CZ}_{r,r+1}.
\end{equation}

A direct calculation gives nontrivial statistics for this domain-wall hopping operator.

This differs from the anyon example. A perturbation of the first kind acts locally near an anyon. It leaves the anyon type intact, so one would expect it to preserve the statistics as well. The $H^{d+2}(K(G,q),\RZ)$ classification challenges this intuition. The two perturbations distinguish two physical operations that can have the same string geometry. For example, consider the simplest $\ZZ_2$ symmetry on a chain,
\begin{equation}
	S=\prod X_j
\end{equation}
The operator $U([\ell,r])$ plays two roles. It implements the symmetry in the interior of $[\ell,r]$ and is therefore a symmetry patch. It is also a symmetric hopping operator that creates domain walls at the endpoints. The two requirements can be separated. The endpoint perturbation in Eq.~\eqref{eq:Z2 nontrivial hopping} gives an equally valid symmetry patch $U'([\ell,r])$ that fails to commute with the symmetry. Alternatively, choose a short interval $[k,k+1]$ inside $[\ell,r]$ and define
\begin{equation}
	U''([\ell,r])=\left(\prod_{j=\ell}^r X_j\right) Z_kZ_{k+1}
\end{equation}
Relative to $U([\ell,r])$, this operator assigns a phase according to whether a domain wall is present at $k$, a perturbation of the second kind. The resulting operator is still symmetric and is an equally valid domain-wall hopping operator. More generally, a global symmetry transformation can be restricted to an interval. The choice of restriction is a perturbation of the first kind. Any quantity extracted from a symmetry patch that characterizes the global symmetry must be invariant under this freedom. The Else--Nayak construction of the anomaly from one-dimensional symmetry patches has precisely this property~\cite{ElseNayak2014}. Likewise, the choice of symmetric domain-wall hopping operators admits perturbations of the second kind, and a quantity that characterizes the symmetry must be invariant under those choices. A statistical process is invariant under perturbations of the second kind, but its value can change under the first. Evaluating the same process directly on symmetry patches can therefore give a result that depends on how the symmetry was restricted. The two kinds of perturbations correspond to two approaches to symmetry:
\begin{itemize}
  \item First kind: symmetry transformation $\to$ symmetry patch $\to$ anomaly.
  \item Second kind: symmetric operator $\to$ domain-wall hopping $\to$ statistics.
\end{itemize}

The cohomological 't~Hooft anomalies of a finite Abelian $(q-1)$-form symmetry in $d$ spatial dimensions are also classified by $H^{d+2}(K(G,q),\RZ)$~\cite{GaiottoKapustinSeibergWillett2015}, the same group that classifies the statistics of codimension-$q$ excitations. In many examples, including the one-dimensional ferromagnet above, the statistics of symmetric hoppings agree with the anomaly of the global symmetry. Why should two constructions involving different operators and different freedoms give consistent results? We return to the definition of an anomaly as an obstruction to promoting a global symmetry to a gauge symmetry. Coupling to a gauge field turns the commutant relation between symmetry transformations and the symmetric hopping algebra into a commutant relation between gauge transformations and the gauge-invariant hopping algebra. It also reveals an additional structure: the gauge-invariant hoppings that move excitations and the gauge transformations that move symmetry defects generate isomorphic transport algebras. A statistical process invariant under perturbations of the second kind can thus be evaluated on hoppings to measure statistics or on gauge transformations to measure the anomaly. The isomorphism guarantees their consistency.

It is useful to compare this view with that of Chatterjee and Wen, who write~\cite[p.~2]{ChatterjeeWen2023Shadow}:
\begin{quote}
\textquotedblleft a symmetry is not about transformations. What a symmetry really does is to select a set of local symmetric operators which form an algebra.\textquotedblright
\end{quote}
In our description, symmetry transformations and the algebra of symmetric operators have equal standing. The former act in the gauge direction and the latter in time. Coupling to a background gauge field makes their algebraic relationship explicit.

One question remains. Why is two-dimensional braiding robust under both kinds of perturbations? Invariance under the second kind allows us to evaluate a statistical process on hoppings to obtain statistics and on gauge transformations to obtain an anomaly. Invariance under the first kind also lets us use symmetry patches to measure that anomaly. To see how these statements fit together, consider a patch $P(u)$ labeled by a cochain $u$. Acting on the zero-background Hilbert space $\mathcal H_0$, it creates a defect with background $A=-\dd u$ at its boundary. Representing the initial and final backgrounds by separate copies of the Hilbert space gives a map
\begin{equation}
	P(u):\mathcal{H}_0\to \mathcal{H}_{-\dd u}.
\end{equation}
The two copies have the same lattice matter degrees of freedom, so the map can still be represented by a matrix on the matter space.
Moving a defect already present in a background $A$ requires a map $T_A(u):\mathcal H_A\to\mathcal H_{A-\dd u}$, which generally differs from $P(u)$. One must either construct the general gauge transformation from patches or account for the effects of identifying the source and target spaces. In fact, a perturbation of $P(u)$ of the first kind at its boundary $\dd u$ comes from an automorphism of $\mathcal H_{-\dd u}$. When gauge transformations are composed with matching backgrounds around a closed process, the automorphisms at the intermediate backgrounds cancel. The local choice of phase of a gauge transformation gives the second kind of perturbation.

The geometry of anyon braiding lets us use the zero-background patch at each step: the background within the region read by that step is the same as in the zero-background calculation. Gauge transformations and symmetry patches therefore give the same result. This holds for every commutator satisfying boundary avoidance.

Section~\ref{sec:spacetime} extracts hopping operators from symmetric time evolution and introduces the transport algebra to characterize excitation statistics. Section~\ref{sec:statistics_anomaly} couples the same system to a flat background field, proves that the hopping and gauge transformation algebras are mutual commutants, and relates statistics to anomalies by their antilinear isomorphism. Section~\ref{sec:patch-compatibility} studies the gauging of a general symmetry. Section~\ref{sec:geometry} extends anyon braiding to commutators satisfying boundary avoidance and classifies the statistics and anomalies that they can detect. Section~\ref{sec:bockstein_word} constructs a process $W_N$ that measures Bockstein statistics even when the four-step commutator loses its protection, and compares its hopping and gauge transformation realizations on a ring. Section~\ref{sec:typeiii-zero} uses a type-III model on a circle to compare the original symmetry with the symmetry defined by closed hoppings, distinguishing their anomalies through twisted boundary conditions. Appendix~\ref{app:wn-homology} gives an integral homology interpretation of $W_N$. Appendix~\ref{app:patch-transport} proves the results of Section~\ref{sec:patch-compatibility} on the existence of compatible restrictions and the uniqueness of their statistical class.

\Needspace{4\baselineskip}
\phantomsection\label{note:independence}
\textbf{Relation to the literature and statement of independence}

The present work grew out of a collaboration with the authors of Ref.~\cite{HsinChen2026Bockstein}. Some of the results on Bockstein statistics overlap with those of that reference. The author of the present paper obtained all of the overlapping results independently.\footnote{Our access to Slack and Overleaf was revoked before we could preserve the records there. We retain partial records from WeChat, which are available from the author on request. We requested the Overleaf records from Yu-An Chen by email and received no reply.} At the end of its introduction, Ref.~\cite{HsinChen2026Bockstein} states:
\textit{Closing
an excitation operator turns it into a generalized-symmetry generator. The same junction-linking phase
therefore gives a direct microscopic diagnosis of the mixed anomaly $A\cup\beta_N B$.} 
This assertion identifies a closed excitation operator as a symmetry generator, then interprets $W_N$ for open hoppings as the mixed anomaly of that symmetry. However, $W_N$ is not robust under perturbations of the first kind and therefore admits counterexamples similar to Eq.~\ref{eq:Z2 nontrivial hopping}. A related issue arises in Ref.~\cite{KobayashiEtAl2026Generalized}. We therefore do not support publication of Ref.~\cite{HsinChen2026Bockstein}. In Ref.~\cite{Xue2026Comparison}, we proved that, for a type-III anomaly, the available freedom of choice cannot resolve this inconsistency. Here we use the duality between the gauge and time directions to explain the relationship between statistics and anomalies.

Among the overlapping results on Bockstein statistics, Yu-An Chen was the first to obtain a statistical process for the mutual statistics of loops in three dimensions. The author independently obtained the process $W_N$ used here.

\section{Symmetric hopping and the transport algebra}
\label{sec:spacetime}

Let $X$ be a finitely triangulated closed oriented space and $G$ a finite Abelian group. The matter field is $\phi\in C^{q-1}(X,G)$, and the states $\ket{\phi}$ form an orthonormal basis.

On a $(d+1)$-dimensional spacetime bordism $M:X_0\to X_1$, let $\tilde\phi$ denote a matter-field history connecting the boundary configurations $\phi_0,\phi_1$. The path integral for a system with $(q-1)$-form symmetry $\phi\mapsto \phi+z$, where $z$ is a cocycle of degree $q-1$, takes the form

\begin{equation}
 \bra{\phi_1}Z(M)\ket{\phi_0}
 =\sum_{\tilde\phi\,:\,\tilde\phi|_{X_i}=\phi_i}
 \exp\!\left[\ii\int_M\mathcal L(\dd\tilde\phi)\right]
 \exp\!\left[-2\pi\ii\int_M\Theta(\tilde\phi)\right].
 \label{eq:matter-evolution}
\end{equation}

In an Ising chain, $\tilde{\phi}$ describes the spin orientations throughout spacetime, and $\dd\tilde\phi$ describes the domain-wall worldlines (the conserved current). The action has two parts. The dynamical part comes from a local Lagrangian density $\mathcal L(\dd\tilde\phi)$, which is manifestly invariant under $\tilde\phi\mapsto\tilde\phi+z$. The topological term $\Theta(\tilde\phi)$ allows a symmetry transformation to change the Lagrangian density by a total derivative. Accordingly, $\Theta$, viewed as a natural transformation from $C^{q-1}(-,G)$ to $C^{d+1}(-,\RZ)$, must satisfy
\begin{equation}
 \dd\Theta(\phi)=\omega(\dd\phi),\qquad \dd\omega=0 .
 \label{eq:first-descent}
\end{equation}
The local topological term $\Theta(\phi)$ and the cocycle $\omega$ each encode the full statistics class $[\omega]\in H^{d+2}(K(G,q),\RZ)$~\cite{Xue2026Statistics,XueWen2026Holographic}.

Fix the phase $\bra{z}S(z)\ket{0}$ of the global shift acting on $\ket{0}$. In Fig.~\ref{fig:symmetry-state-square}, the history $\tilde\phi$ from $0$ to $\phi$ and its translate $\tilde\phi+z$ from $z$ to $\phi+z$ have the same dynamical weight. By Eq.~\eqref{eq:matter-evolution}, the ratio of their topological amplitudes determines the phase of the global transformation on the final state,
\begin{equation}
 \boxed{S(z)\ket{\phi}
 =\bra{z}S(z)\ket{0}\exp\!\left[-2\pi\ii\int_{I_{\rm t}\times X}
       \bigl(\Theta(\tilde\phi+z)-\Theta(\tilde\phi)\bigr)\right]\ket{\phi+z}.}
 \label{eq:symmetry-from-hopping}
\end{equation}
If, for a given closed shift $z$, there is a local $d$-cochain $\lambda_z$ satisfying
$\Theta(\phi)-\Theta(\phi+z)=\dd \lambda_z(\phi)$, Stokes' theorem reduces the history integral in Eq.~\eqref{eq:symmetry-from-hopping} to a difference of endpoint terms,
\begin{equation}
 S(z)\ket{\phi}=\bra{z}S(z)\ket{0}\exp\!\left[2\pi\ii\int_X
      \bigl(\lambda_z(\phi)-\lambda_z(0)\bigr)\right]\ket{\phi+z}.
 \label{eq:symmetry-descent-primitive}
\end{equation}

\begin{figure}[htbp]
\centering
\begin{tikzpicture}[>=Stealth,line cap=round,every node/.style={font=\small,text=ink},
  trans/.style={->,accent,thick},hist/.style={->,ink,thick}]
\node (L) at (0,0) {$\ket{0}$};
\node (R) at (3.4,0) {$S(z)\ket{0}$};
\node (T) at (0,1.6) {$\ket{\phi}$};
\node (F) at (3.4,1.6) {$\ket{\phi+z}$};
\draw[trans] (L)--node[midway,below] {$S(z)$} (R);
\draw[trans,dashed] (T)--node[midway,above] {$S(z)$} (F);
\draw[hist] (L)--node[midway,left] {$\tilde\phi$} (T);
\draw[hist] (R)--node[midway,right] {$\tilde\phi+z$} (F);
\end{tikzpicture}
\caption{The vertical edges represent two histories in physical time. At the lower right, $S(z)\ket{0}$ differs from $\ket{z}$ by the prescribed initial phase. The ratio of the amplitudes of the two histories determines the phase of $S(z)\ket{\phi}$.}
\label{fig:symmetry-state-square}
\end{figure}

On a finite lattice, $\phi$ takes values in the finite group $G$. We interpolate between two discrete configurations on the simplicial cylinder $I_{\rm t}\times X$. Let $\eta_{\rm t}$ be the $0$-cochain on the interval with values $0,1$ at the lower and upper endpoints, respectively. Pull the spatial cochains back to the cylinder and set
\[
 \tilde\phi=\phi_0+\eta_{\rm t}\smile(\phi_1-\phi_0),
 \qquad \tilde\phi|_{0\times X}=\phi_0,\quad
 \tilde\phi|_{1\times X}=\phi_1 .
\]
The cup product also assigns field values to simplices spanning the two layers. Using this interpolation for the elementary evolution $\phi\to\phi+s$ and setting its dynamical weight to one gives the standard hopping operator
\begin{equation}
 \boxed{
 U(s)\ket{\phi}
 =\exp\!\left[-2\pi\ii\int_{I_{\rm t}\times X}
 \Theta(\phi+\eta_{\rm t}\smile s)\right]\ket{\phi+s}.
 }
 \label{eq:raw-hopping}
\end{equation}
Append this step to a preparation history $0\to\phi$ and shift the entire history by $z$. Comparing the amplitude ratios in Eq.~\eqref{eq:symmetry-from-hopping} before and after appending the step gives $S(z)U(s)=U(s)S(z)$. Shifting all the histories in Eq.~\eqref{eq:matter-evolution} and summing their amplitudes then gives the commutation relation between symmetric time evolution and $S(z)$ shown in Fig.~\ref{fig:zero-background-square}.

\begin{figure}[htbp]
\centering
\begin{tikzpicture}[
  x=1cm,y=.85cm,>=Stealth,line cap=round,line join=round,
  every node/.style={text=ink,font=\small,inner sep=1.5pt},
  horizontal/.style={->,accent,line width=.85pt,shorten <=2pt,shorten >=2pt},
  vertical/.style={->,ink,line width=.85pt,shorten <=2pt,shorten >=2pt},
  axis/.style={-{Stealth[length=6pt,width=5pt]},muted!78,line width=.6pt},
  axis label/.style={font=\footnotesize\sffamily,text=muted},
  axis tip/.style={font=\small\sffamily,text=ink}
]
\coordinate (O) at (0,0);       \coordinate (G) at (3.35,0);
\coordinate (T) at (0,3.35);    \coordinate (C) at (3.35,3.35);
\draw[horizontal] (O) -- node[below=3pt] {$S(z)$} (G);
\draw[horizontal] (T) -- node[above=3pt,align=center,font=\small\sffamily,text=accent] {symmetry transformation\\[-1pt]$S(z)$} (C);
\draw[vertical] (O) -- node[left=4pt] {$Z(I_{\rm t}\times X)$} (T);
\draw[vertical] (G) -- node[right=4pt,align=left,font=\small\sffamily] {symmetric evolution\\[-1pt]$Z(I_{\rm t}\times X)$} (C);
\foreach \p in {O,G,T,C} \fill[ink] (\p) circle[radius=1.25pt];
\node[font=\Large] at (1.675,1.675) {$=$};
\draw[axis] (-2.35,-.72) -- (-2.35,3.35)
  node[midway,above=3pt,sloped,axis label] {time direction};
\draw[axis] (-2.35,-.72) -- (3.35,-.72)
  node[midway,below=3pt,axis label] {transformation direction};
\end{tikzpicture}
\caption{At zero background $A=0$, the symmetric time evolution $Z(I_{\rm t}\times X)$ commutes with the global transformation $S(z)$.}
\label{fig:zero-background-square}
\end{figure}

Fix $\phi$. As $s$ ranges over $C^{q-1}(X,G)$, the states $U(s)\ket{\phi}$ form an orthonormal basis, so
\begin{equation}
 \sum_{s\in C^{q-1}(X,G)} U(s)\ket{\phi}\bra{\phi}U(s)^{\dagger}
 =\sum_s\ket{\phi+s}\bra{\phi+s}=\id .
 \label{eq:hopping-resolution-identity}
\end{equation}
Write $Z=Z(I_{\rm t}\times X)$. Applying this resolution of the identity to $Z\ket{\phi}$ groups the matter histories in Eq.~\eqref{eq:matter-evolution} by the net shift $s$ between their initial and final states. Each group has total amplitude $\bra{\phi+s}Z\ket{\phi}$. Factoring out the standard hopping phase for this shift leaves the coefficient
\[
 \frac{\bra{\phi+s}Z\ket{\phi}}{\bra{\phi+s}U(s)\ket{\phi}}
 =\bra{\phi}U(s)^{-1} Z\ket{\phi}.
\]
Since both $Z$ and $U(s)$ commute with $S(z)$, the ratio is invariant under a simultaneous closed shift $z$ of the initial and final states. It therefore depends only on $s$ and the initial domain-wall configuration $a=\dd\phi$. Denote it by $Z_s(a)$. Summing the groups gives
\begin{equation}
 Z=\sum_s U(s)\,Z_s(\hat a).
 \label{eq:symmetric-expansion-main}
\end{equation}
Here $Z_s(\hat a)\ket{\phi}=Z_s(\dd\phi)\ket{\phi}$. The hat marks the argument read from the state, whereas $s$ specifies the move. For the single-layer history chosen in Eq.~\eqref{eq:raw-hopping}, $U(s)$ already contains the topological phase, so the coefficient is just the dynamical weight. In general, $Z_s(a)$ collects the amplitudes of all histories from $\phi$ to $\phi+s$ relative to this standard hopping operator.

This expansion used only the fact that $Z$ commutes with every $S(z)$, so it applies to any symmetric operator. Conversely, for arbitrary configuration functions $Z_s$, each term of Eq.~\eqref{eq:symmetric-expansion-main} commutes with $S(z)$ because a closed shift preserves $a$. For each $\ket{\phi}$, distinct $s$ give orthogonal final states, so the coefficients $Z_s$ are unique.

The same move $s$ admits other interpolations. Choose any history $0\to s$ and add $\phi$ to the entire history when starting from $\phi$. A closed shift of the initial state then shifts the chosen history along with it, so the resulting $U'(s)$ is also symmetric. It sends each $\phi$ to $\phi+s$, leaving only the term for $s$ in Eq.~\eqref{eq:symmetric-expansion-main}. Unitarity requires its coefficient to be a phase, so
\begin{equation}
 U'(s)=U(s)\,e^{2\pi\ii f(s;\hat a)},
 \qquad f(s;a)\in\RZ .
 \label{eq:zero-hopping-rephasing}
\end{equation}
Any such configuration-dependent phase commutes with closed shifts. If the family $f(s;a)$ is a normalized local function in the sense of Definition~\ref{def:local-function}, we call this a local basis change.

The relative phase between doing $s$ followed by $s'$ and doing $s+s'$ directly can be computed on a triangle in parameter space. Assign the matter fields $\phi,\phi+s,\phi+s+s'$ to its three vertices, in that order. Take two degree-zero cochains with vertex values $\eta_1=(0,1,1)$ and $\eta_2=(0,0,1)$. The field $\tilde\phi=\phi+\eta_1\smile s+\eta_2\smile s'$ restricts on each edge to the interpolation in Eq.~\eqref{eq:raw-hopping}. By Eq.~\eqref{eq:first-descent}, integrating $\Theta$ around the boundary gives the integral of $\omega$ over the interior. Hence
\begin{equation}
 U(s')U(s)\ket{\phi}
 =\exp\!\left[-2\pi\ii\int_{\Delta^2\times X}
             \omega(\dd\tilde\phi)\right]U(s+s')\ket{\phi}.
 \label{eq:hopping-triangle}
\end{equation}
Combining adjacent moves successively gives the phase of an entire hopping sequence. If the net shift vanishes, the direct path gives $U(0)=\id$.

The operators $S(z')S(z)$ and $S(z+z')$ implement the same shift, so their ratio is diagonal in the matter-field basis. It commutes with the hopping operators that connect all basis states and must therefore be a constant phase. Thus the $S(z)$ span a unital $*$-algebra.

\begin{theorem}
\label{thm:elementary-hopping-algebra}
Let $X$ be a finitely triangulated closed oriented space and $G$ a finite Abelian group. On the matter-field space with basis labeled by $\phi\in C^{q-1}(X,G)$, an operator $Z$ commutes with all $S(z)$ if and only if it admits the expansion~\eqref{eq:symmetric-expansion-main}, with uniquely determined functions $Z_s$ on $B^{q}(X,G)$. Denote the algebra of these operators by $\mathcal A_{\rm hop}$. By the finite-dimensional double commutant theorem,
\[
 \mathcal A_{\rm hop}'=\operatorname{span}\{S(z):z\in Z^{q-1}(X,G)\}.
\]
\end{theorem}

Given the initial phase $\bra{z}S(z)\ket{0}$, we can also recover the specified $S(z)$ from the hopping operators. Apply $U(s)$ to $\ket{0}$ and $\ket{z}$ to transport the matrix element on the bottom edge of Fig.~\ref{fig:symmetry-state-square} to the top edge. Summing over $s$ gives
\[
 S(z)=\bra{z}S(z)\ket{0}\sum_s U(s)\,\ket{z}\bra{0}\,U(s)^{-1} .
\]
The two inputs differ by the closed parameter $z$, so the phases depending on $\dd\phi$ introduced by a local basis change cancel between the ket and the bra.

\subsection{Two kinds of domain-wall hopping in a ferromagnetic chain}
\label{subsec:z2-hop-symmetry}

Take a $\ZZ_2$-valued vertex field $\phi$ on a periodic chain $X=S^1$, with bonds oriented around the circle as $j\to j+1$. For the nontrivial $\ZZ_2$ class, choose
\begin{equation}
 \Theta(\phi)=\tfrac12\phi\smile\dd\phi\smile\dd\phi,\qquad
 \omega(b)=\tfrac12 b\smile b\smile b .
 \label{eq:z2-matter-cocycle}
\end{equation}
The global flip $\phi\mapsto\phi+\mathbf{1}$ changes $\Theta$ by $\tfrac12\dd(\phi\smile\dd\phi)$, so Eq.~\eqref{eq:symmetry-descent-primitive} gives
\begin{equation}
 S(\mathbf{1})\ket{\phi}
 =(-1)^{\int_X\phi\smile\dd\phi}\ket{\phi+\mathbf{1}} .
 \label{eq:z2-matter-symmetry}
\end{equation}
The exponent counts bonds directed from $1$ to $0$, so each pair of domain walls contributes a minus sign to the global flip. This is a familiar anomalous symmetry. For hopping, substituting the shift $s$ into Eq.~\eqref{eq:raw-hopping} and summing over the two triangles swept out by each bond gives
\begin{equation}
 U(s)\ket{\phi}
 =(-1)^{\int_X[\phi\smile s\smile\dd(\phi+s)+\phi\smile\dd\phi\smile s]}
   \ket{\phi+s}.
 \label{eq:z2-matter-hop}
\end{equation}
For the move $e_j$ that is $1$ only at site $j$, we get $U_j=U(e_j)=\mathrm{CZ}_{j-1,j}\,X_j\,\mathrm{CZ}_{j,j+1}$. It commutes with $S(\mathbf{1})$. For the trivial topological term $\omega=\Theta=0$, the same formulas give the ordinary Ising symmetry $\prod_j X_j$ and the single-site hopping operators $X_j$.

Both symmetries preserve the ferromagnetic Hamiltonian
\[
 H_0=-\sum_j Z_jZ_{j+1}
\]
This energy depends only on the domain walls $\dd \phi$ and contains no hopping terms. Both symmetries exchange the two ground states $\ket{0}$ and $\ket{\mathbf{1}}$ in exactly the same way.

Now add a symmetric hopping term to each model, with amplitude $\lambda\geq 0$. For the ordinary symmetry this gives the transverse-field Ising model
\[
 H_\lambda=H_0-\lambda\sum_j X_j .
\]
As $\lambda$ increases, the chain becomes critical at $\lambda=1$ and then enters the paramagnetic phase~\cite{Pfeuty1970TransverseIsing}. In the strong-field limit its ground state is $\ket{+}^{\otimes L}$, where $L$ is the number of sites. Each $X_j$ preserves this state, so flipping any finite interval, which inserts domain walls at its ends, also preserves it. This is the fixed point of domain-wall condensation.

For the anomalous symmetry, adding $U_j+U_j^{\dagger}$ gives
\begin{equation}
 H'_\lambda=H_0-\lambda\sum_j X_j(Z_{j-1}+Z_{j+1}).
 \label{eq:z2-wall-hamiltonians}
\end{equation}
This chain is exactly solvable. After applying one layer of $\mathrm{CZ}$ gates on alternating bonds of an even periodic chain, $H'_\lambda/\lambda$ is precisely the model of Ref.~\cite[Eqs.~(5) and (6)]{KoppChakravarty2005Criticality} at $\lambda_1=1/\lambda$ and $\lambda_2=1$. Along this path the bulk gap stays open for every finite $\lambda>0$, so in the thermodynamic limit the system remains in the ferromagnetic phase connected to $\lambda=0$.

From the anomaly viewpoint, a generalized Lieb--Schultz--Mattis theorem rules out a unique gapped ground state preserving the anomalous $\ZZ_2$ symmetry~\cite{KapustinSopenko2025LSM}. From the hopping viewpoint, nontrivial domain-wall statistics obstruct condensation. Take adjacent finite intervals $I=[\ell,r]$ and $J=[r+1,m]$, and write $U(I)=U_\ell\cdots U_r$. The $\mathrm{CZ}$ gates in the interior cancel in pairs, leaving only the two endpoint gates in Eq.~\eqref{eq:Z2 nontrivial hopping}. The nontrivial domain-wall statistics follow from
\[
 [U(J),U(I)^2]=-\id .
\]
The domain-wall hopping operators $U(I),U(J)$ therefore have no common eigenstate. We interpret this as an obstruction to domain-wall condensation.

Closing the hopping operator over the entire space fails to capture this dynamical property. Taking $s=\mathbf{1}$ in Eq.~\eqref{eq:z2-matter-hop}, the two terms in the exponent cancel and give $U(\mathbf{1})=\prod_j X_j$. One can regard this as a $\ZZ_2$ symmetry, but it does not commute with $H'_\lambda$ for $\lambda>0$ and hence has no direct bearing on the properties of the system under study.

\begin{remark}
The ground states alone do not determine the domain-wall statistics in this example. The two symmetries act identically on the entire two-dimensional ground space of $H_0$. Distinguishing their anomalies, their allowed Hamiltonian terms, and their domain-wall statistics requires the action of the symmetry on the full Hilbert space. For intrinsic topological order in two dimensions, the anyon theory is usually understood to be fully determined by the ground state. The entanglement bootstrap extracts the anyon fusion rules and the topological $S$ matrix from a single ground state satisfying suitable entanglement conditions~\cite{ShiKatoKim2020Fusion,Shi2020Verlinde}. We do not yet fully understand the origin of this difference.
	
\end{remark}

\subsection{Transport algebra and statistical processes}
\label{subsec:hopping-statistics}
In Eqs.~\eqref{eq:symmetric-expansion-main} and
~\eqref{eq:hopping-triangle}, we started with a cocycle $\omega$ and constructed the algebra generated by the corresponding domain-wall hopping operators. We now describe the structure of this algebra axiomatically, without specifying $\omega$ in advance. We require that the structure itself yield a statistical invariant and that, in the example of Eq.~\eqref{eq:hopping-triangle}, the invariant recover $[\omega]\in H^{d+2}(K(G,q),\RZ)$.
 
The physical picture is a set of domain-wall configurations labeled by $a\in B^q(X,G)$, with a Hilbert space for each configuration,
\begin{equation}
	\mathcal H_a=\operatorname{span}\{\ket{\phi}:\dd\phi=a\},
\end{equation}
and the full Hilbert space is their orthogonal direct sum,
\begin{equation}
	 \mathcal H=\bigoplus_{a\in B^{q}(X,G)}\mathcal H_a.
\end{equation}

The algebra of symmetric operators has two kinds of generators. The configuration-preserving functions
$f(\hat a)$ act on $\mathcal H_a$ by multiplication by $f(a)$, and the hopping operators $U(s)$ map $\mathcal H_a$ unitarily to $\mathcal H_{a+\dd s}$. As in Eq.~\eqref{eq:hopping-triangle}, composing hopping operators can give a phase. This phase must encode the locality of hopping.

\Needspace{7\baselineskip}
\begin{definition}\label{def:local-function}
Let $f(s_1,\ldots,s_m;a)$ be an $\RZ$-valued function whose arguments are cochains, cocycles, or coboundaries on $X$. We call $f$ a \textbf{local function} \textbf{normalized} in $s_1,\ldots,s_m$ if $f$ decomposes over top-dimensional simplices $\Delta\in X_d$ as
\begin{equation}
 f(s_1,\ldots,s_m;a)=
 \sum_{\Delta\in X_d} f_\Delta(s_1|_\Delta,\ldots,s_m|_\Delta;a|_\Delta),
 \label{eq:transport-local-multiplier}
\end{equation}
with $f_\Delta$ vanishing whenever some $s_i|_\Delta=0$.
\end{definition}

Here $s_1,\cdots,s_m$ label operations and $a$ describes the configuration on which they act. The definition expresses the fact that an operation can be performed locally. On any region $\Delta$ where some $s_i|_\Delta=0$, that operation is trivial and contributes nothing, regardless of whether $a|_\Delta$ vanishes.

The phase in Eq.~\eqref{eq:hopping-triangle} comes from integrating a normalized cochain operation over $X$, so it satisfies the definition of a local function. This locality is central to the following definition.
\begin{definition}\label{def:hopping-algebra}
Let $\mathcal{D}$ be the commutative $C^*$-algebra of all complex-valued functions on $B^{q}(X,G)$. We write a function $f:B^q(X,G)\to \mathbb C$ as $f(\hat{a})$, with $\hat{a}$ denoting its coboundary argument.

Let $\mathcal{A}$ consist of all formal sums
\begin{equation}
	\sum_{s\in C^{q-1}(X,G)} U(s)f_s(\hat a).
\end{equation}
Equivalently, it is the free right $\mathcal{D}$-module with basis given by all $U(s)$.

To define multiplication on $\mathcal{A}$, set
\begin{equation}
	\boxed{\begin{aligned}
			f(\hat a)\,U(s)&=U(s)\,f(\hat a+\dd s),\\
			U(t)\,U(s)&=U(s+t)\,e^{2\pi\ii c(t,s;\hat a)}.
	\end{aligned}}
	\label{eq:hopping-algebra-product}
\end{equation}
where $c(t,s;a)$ is a normalized local function. Its first two arguments follow the order of the factors in $U(t)U(s)$, and $a$ is the configuration before $U(s)$ acts. Associativity requires $c$ to satisfy
\[
 c(t,s;a)+c(r,s+t;a)=c(r,t;a+\dd s)+c(t+r,s;a).
\]

Each $U(s)$ is invertible, and $U(0)=\id$. The $*$-structure is defined by
\[
f(\hat a)^*=\overline f(\hat a),\qquad
U(s)^*=U(s)^{-1}=U(-s)\,e^{-2\pi\ii c(s,-s;\hat a)},
\]
Then $\mathcal{A}$ is a finite-dimensional $C^*$-algebra. We call it the \textbf{transport algebra} generated by the basis $U(s)$.

For several excitation species with codimensions $q_i$ and groups $G_i$, take
\[
 s\in\prod_i C^{q_i-1}(X,G_i),\qquad a\in\prod_i B^{q_i}(X,G_i),
\]
with addition, restriction, and $\dd$ acting componentwise. In Eq.~\eqref{eq:transport-local-multiplier}, a move vanishes on $\Delta$ when all its components vanish on $\Delta$.
\end{definition}

\begin{definition}\label{def:local-basis-change}
A \textbf{local basis change} replaces each $U(s)$ by $U(s)e^{2\pi\ii\lambda(s;\hat a)}$, where $\lambda(s;a)$ is a normalized local function. The multiplication coefficients then change according to
\begin{equation}
 c'(t,s;a)-c(t,s;a)=\lambda(s;a)+\lambda(t;a+\dd s)-\lambda(s+t;a).
 \label{eq:hopping-local-choice}
\end{equation}

\end{definition}
In Fig.~\ref{fig:boundary-avoidance}(b), $O_x$ appears to the left of $U_1$. If $O_x$ multiplies configuration $b$ by $e^{2\pi\ii g(b)}$, then for the step $U_1=U(s)$ the phase exponent in terms of the input configuration $a$ is $\lambda(s;a)=g(a+\dd s)$.

\begin{definition}\label{def:statistical-process}
A sequence of hopping operators $W=U(s_n)^{\epsilon_n}\cdots U(s_1)^{\epsilon_1}$ with $\epsilon_j=\pm1$ and $\sum_j\epsilon_j s_j=0$ is a \textbf{closed process}. If the product of the same sequence is invariant under every local basis change of $U$, we call it a \textbf{statistical process}.
\end{definition}
On each top-dimensional simplex, it suffices to check that every nontrivial local move at each local configuration appears equally often in the forward and inverse directions. The two phases then cancel under any local basis change. The configuration for an inverse step is read at the starting point of the corresponding forward step.

\begin{theorem}[Initial-state independence of statistical processes]\label{thm:process-initial-state}
The product of every statistical process is a phase times $\id$.
\end{theorem}
\begin{proof}
By the multiplication law~\eqref{eq:hopping-algebra-product}, the product of a closed process is a unitary configuration function $w(\hat a)$. Fix $r\in C^{q-1}(X,G)$. Comparing the two ways of associating the factors in each of $U(t)U(s)U(r)$, $U(t)U(r)U(s)$, and $U(r)U(t)U(s)$ gives
\[
 \begin{aligned}
 \lambda_r(s;a)&=c(s,r;a)-c(r,s;a),\\
 c(t,s;a+\dd r)-c(t,s;a)&=\lambda_r(s;a)+\lambda_r(t;a+\dd s)-\lambda_r(s+t;a).
 \end{aligned}
\]
For fixed $r$, $\lambda_r$ is a local basis change. The product is therefore invariant under replacing $a$ with $a+\dd r$, so $w(a+\dd r)=w(a)$. Starting from $a=0$, these shifts reach every configuration.
\end{proof}

Our definitions of a transport algebra and a statistical process reformulate Ref.~\cite{Xue2026Statistics} in algebraic terms. That work effectively assumes that the Hilbert space $\mathcal{H}_a$ for each configuration $a\in B^q(X,G)$ is one-dimensional, so the hopping operator $U(s): \mathcal{H}_a\to\mathcal{H}_{a+\dd s}$ is specified by a single phase $\theta(s,a)$. Here we need to allow arbitrary dimensions for $\mathcal{H}_a$, and encode the phase information in the multiplication coefficients $c(t,s;a)$. One distinction remains. When every $\mathcal{H}_a$ is one-dimensional, the hopping operators for any $z,w\in Z^{q-1}(X,G)$ necessarily commute, $[U(w),U(z)]=1$. Under the transport algebra definition, $[U(w),U(z)]$ is a statistical process whose phase can be nontrivial. Here we impose $[U(w),U(z)]=1$ as an additional condition on the transport algebra,
\begin{equation}
 \mathrm{(ZZ)}\qquad c(w,z;a)=c(z,w;a)
 \quad\text{for all }z,w\in Z^{q-1}(X,G).
 \label{eq:closed-commutation-conditions}
\end{equation}

With these definitions we obtain the following theorem.

\begin{theorem}[Statistics as equivalence classes of transport algebras~{\cite{XueStatisticalProcessLean}}]\label{thm:hopping-statistics}
Let $X$ be a simplicial triangulation of the sphere $S^d$, take $1\le q\le d$, and impose condition \textup{(ZZ)} on the vacuum orbit. The equivalence classes of multiplication laws in Definition~\ref{def:hopping-algebra} under local basis changes~\eqref{eq:hopping-local-choice} are classified by
\[
 H^{d+2}\bigl(K(G,q),\RZ\bigr)
\]
The hopping multiplication in Eq.~\eqref{eq:hopping-triangle} realizes the class $[\omega]$. There is a finite set of statistical processes whose phases distinguish all equivalence classes.
A single local basis change makes $U(t)\,U(s)=U(s+t)$ hold for all $s,t$ if and only if $[\omega]=0$.
\end{theorem}

With the help of GPT 5.6-Sol, we proved this theorem and formally verified it in Lean. Two observations seem to us more significant than the proof itself. First, the classification is independent of the triangulation $X$ of the sphere, which suggests that the transport algebra definition captures physical information. Second, for fixed $X$, one can solve for $c$ using the local decomposition~\eqref{eq:transport-local-multiplier}, associativity in Definition~\ref{def:hopping-algebra}, and condition~\eqref{eq:closed-commutation-conditions}, and then take the quotient by Eq.~\eqref{eq:hopping-local-choice}. This is a finite system of linear equations over $\RZ$ with integer coefficients, and its predicted classification $ H^{d+2}\bigl(K(G,q),\RZ\bigr)$ can be checked directly by computation.

\begin{remark}
In some settings it is physically meaningful to relax \textup{(ZZ)} and allow a nontrivial commutator $[U(w),U(z)]$ when $z,w$ represent nontrivial cohomology classes, as in Sec.~\ref{sec:typeiii-zero}. In general one should still require the commutator of two coboundary hopping operators to be trivial. Otherwise, with the same definitions of transport algebra and local basis change, the classification acquires an extra term that depends on the triangulation of $X$ and grows rapidly under subdivision.
\end{remark}

\begin{remark}
There is another way to express Definition~\ref{def:hopping-algebra}. Expand each configuration function $f(\hat{a})$ in indicator functions (delta functions),
	\begin{equation}
		f(\hat{a})=\sum_{a\in B^q(X,G)}f(a)\Pi_a,
	\end{equation}
so that $\mathcal A$ can be expressed as a complex vector space. Setting $\tau_{s,a}=U(s)\Pi_a$, we can write each element of $\mathcal{A}$ as
	\[
	O=\sum_s U(s)f_s(\hat a)=\sum_{s,a}f_s(a)\,\tau_{s,a}.
	\]
For computing products, the relation with the original notation is
$U(s)=\sum_a\tau_{s,a}$ and $f(\hat a)=\sum_a f(a)\tau_{0,a}$. Equation~\eqref{eq:hopping-algebra-product} gives
	\[
	\tau_{t,b}\tau_{s,a}
	=\delta_{b,a+\dd s}\,e^{2\pi\ii c(t,s;a)}\,\tau_{s+t,a},
	\qquad
	\tau_{s,a}^*=e^{-2\pi\ii c(-s,s;a)}\,\tau_{-s,a+\dd s}.
	\]
Physically, $\tau_{s,a}$ describes hopping from $\mathcal{H}_a$ to $\mathcal{H}_{a+\dd s}$. Its restriction to every $\mathcal{H}_{a'}$ with $a'\ne a$ is zero. This projection onto a prescribed configuration makes $\tau_{s,a}$ nonlocal.
\end{remark}

\subsection{\texorpdfstring{Holographic duality between $\ZZ_4$ and $\ZZ_2\times\ZZ_2$ with a mixed anomaly}{Holographic duality between Z4 and Z2 x Z2 with a mixed anomaly}}
\label{subsec:z4-holographic-duality}

Symmetries admit a notion of equivalence weaker than an isomorphism of group actions. One example relates an anomaly-free $\ZZ_4$ symmetry in one dimension to a $\ZZ_2\times\ZZ_2$ symmetry with a mixed anomaly. This equivalence has several interpretations. The fusion categories of symmetry defects are weakly Morita equivalent~\cite{Uribe2017Pointed}. Partial gauging of the $\ZZ_2$ subgroup of $\ZZ_{4}$ relates the two symmetries~\cite{BhardwajTachikawa2018Gauging}. In one higher dimension, untwisted $\ZZ_4$ gauge theory and $\ZZ_2\times\ZZ_2$ gauge theory with a mixed twist have the same topological order~\cite{HuWan2020ElectricMagnetic,KongEtAl2020Algebraic}. At the lattice level, Chatterjee and Wen~\cite{ChatterjeeWen2023Shadow} computed the algebras of symmetric operators in models with $\ZZ_4$ and anomalous $\ZZ_2\times\ZZ_2$ symmetry, respectively, and exhibited an isomorphism between them. In their words, ``Two global symmetries are holoequivalent if their algebras of local symmetric operators are isomorphic.''~\cite{ChatterjeeWen2023Transitions}

We think this is the right direction. An isomorphism preserving only addition and multiplication, however, would identify physically inequivalent symmetries. For example, take $G=\ZZ_4$ or $\ZZ_2\times\ZZ_2$ and let each act through its ordinary onsite regular action on $n$ sites. Global shifts partition the configurations into $4^{n-1}$ orbits. Each orbit gives a copy of the regular representation, so every character has multiplicity $4^{n-1}$. The two symmetric operator algebras are thus both isomorphic to $\bigoplus_{j=1}^4 M_{4^{n-1}}(\mathbb C)$. The excitations of the corresponding untwisted two-dimensional gauge theories are labeled by group elements and characters, with fusion group $G\times\widehat G$. The first fusion group, $\ZZ_4^2$, contains elements of order four, whereas every element of the second, $\ZZ_2^4$, squares to the identity. The two symmetries are not holographically equivalent.

The algebra of local symmetric operators should carry additional structure, and an isomorphism should preserve it. The transport algebra takes a step in this direction by encoding locality and conservation laws in the algebra itself. Statistical processes can then extract the domain-wall statistics intrinsically. Coupling to a background field in Sec.~\ref{sec:statistics_anomaly} will relate these statistics to the anomaly of the global symmetry. Our definition does not yet give a general criterion for holographic equivalence or a proof of its agreement with Morita equivalence. To gain some intuition, we work out the example of $\ZZ_4$ and anomalous $\ZZ_2\times\ZZ_2$ in the language of transport algebras. In constructing a transport algebra $\mathcal{A}$, we start with a commutative $C^*$-algebra $\mathcal{D}$ of configuration functions $B^q(X,G)\to \mathbb C$ and introduce operators $U(s)$ that act on it by shifts. A local basis change holds $\mathcal{D}$ fixed and changes the choice of $U(s)$. Holographic equivalence calls for choosing a different commutative $C^*$-subalgebra as $\mathcal{D}$.

Take an oriented periodic chain $S^1$, with vertex field $\phi\in C^0(S^1,\ZZ_4)$ and domain-wall configuration $a=\dd\phi\in B^1(S^1,\ZZ_4)$. By Eq.~\eqref{eq:symmetric-expansion-main}, the symmetric operator algebra is generated by the domain-wall functions $f(\hat a)$ and the vertex shifts $X(s)\ket{\phi}=\ket{\phi+s}$. Orient each bond as $[j,j+1]$. On this bond $t\smile a$ takes the value $t(j)a[j,j+1]$, whereas $a\smile t$ takes the value $a[j,j+1]t(j+1)$. Write $\oint$ for integration over $S^1$. For $s,t\in C^0(S^1,\ZZ_4)$, set
\begin{equation}
 V(s,t)=X(s)\,\ii^{\oint\hat a\smile t},\qquad
 V(s',t')V(s,t)=\ii^{\oint\dd s\smile t'}\,V(s+s',t+t').
 \label{eq:z4-weyl-product}
\end{equation}
The second identity follows directly from $f(\hat a)X(s)=X(s)f(\hat a+\dd s)$.

When both labels are even, the phase in Eq.~\eqref{eq:z4-weyl-product} is $1$. Set
\[
 \mathcal D=\operatorname{span}\{V(2r,2w)\},\qquad
 U(u,v)=V(\widetilde u,\widetilde v),
 \qquad r,w,u,v\in C^0(S^1,\ZZ_2),
\]
where the tilde denotes the pointwise lift to $0,1$. The algebra $\mathcal D$ is a commutative $C^*$-algebra. Multiplying two $U$'s gives carries $2uu'$ and $2vv'$ in the two labels, with $uu'$ and $vv'$ denoting pointwise products. Placing even shifts on the left and configuration functions with even labels on the right gives
\begin{equation}
 U(u',v')U(u,v)=\ii^{\oint\dd\widetilde u\smile\widetilde v'}\,
 V(2uu',0)\,U(u+u',v+v')\,V(0,2vv').
 \label{eq:z4-even-residual}
\end{equation}

We now interpret the points in the spectrum of $\mathcal D$ as new configurations $(p,q)$. The operator $V(0,2w)$ reads $p=a\bmod 2=\dd(\phi\bmod 2)$. At each spectral point, the value of $X(2r)$ is a character of $r$. Since each vertex has exactly one outgoing bond, this character can be written uniquely as $(-1)^{\oint r\smile q}$ with $q\in C^1(S^1,\ZZ_2)$. Taking $r=\mathbf1$ shows that $q$ is a coboundary precisely when $X(2\mathbf1)$ has value $+1$. We now impose the central relation $X(2\mathbf1)=\id$ in the $\ZZ_4$ algebra. The spectral points of $\mathcal D$ then satisfy
\[
 \eta_{p,q}\bigl(V(2r,2w)\bigr)=(-1)^{\oint(p\smile w+r\smile q)},
 \qquad p,q\in B^1(S^1,\ZZ_2).
\]

By Eq.~\eqref{eq:z4-weyl-product}, commuting $V(2r,2w)$ past $U(u,v)$ gives the sign $(-1)^{\oint(\dd u\smile w+\dd r\smile v)}$. Stokes' theorem gives $\oint\dd r\smile v=\oint r\smile\dd v$ modulo two. Thus $U(u,v)$ transports the spectral point $(p,q)$ to $(p+\dd u,q+\dd v)$. Starting from any spectral point, these shifts reach all of $B^1(S^1,\ZZ_2)^2$. The image of $\mathcal D$ in the quotient is therefore the full function algebra on this configuration space. For any $F$,
\begin{equation}
 F(\hat p,\hat q)\,U(u,v)=U(u,v)\,F(\hat p+\dd u,\hat q+\dd v).
 \label{eq:z4-spectrum-transport}
\end{equation}

Let $e_j$ be the indicator cochain of a single vertex $j$. Then $U(e_j,0)$ is the original vertex shift, while $U(0,e_j)$ reads $\ii^{a[j-1,j]}$ on the incoming bond. Their squares belong to $\mathcal D$ and have the following values at the same spectral point,
\[
 \eta_{p,q}\bigl(U(e_j,0)^2\bigr)=\eta_{p,q}\bigl(X(2e_j)\bigr)=(-1)^{q[j,j+1]},
 \qquad
 \eta_{p,q}\bigl(U(0,e_j)^2\bigr)=\eta_{p,q}\bigl(V(0,2e_j)\bigr)=(-1)^{p[j-1,j]} .
\]
Thus the first reads $q$ on the outgoing bond and the second reads $p$ on the incoming bond. Equation~\eqref{eq:z4-spectrum-transport} shows that they now move the first and second species of domain wall, respectively.

In Eq.~\eqref{eq:z4-even-residual}, the right factor $V(0,2vv')$ reads the input configuration $p$, and the left factor $V(2uu',0)$ reads the configuration $q+\dd(v+v')$ after the move. Thus the multiplication exponent in Definition~\ref{def:hopping-algebra} is
\[
\begin{aligned}
 c\bigl((u',v'),(u,v);(p,q)\bigr)
 ={}&\tfrac14\oint\dd\widetilde u\smile\widetilde v'\\
 &+\tfrac12\oint\bigl[p\smile(vv')+
       (uu')\smile\bigl(q+\dd(v+v')\bigr)\bigr]\pmod 1 .
\end{aligned}
\]
Each term is a sum over bonds. If either move vanishes at both ends of a bond, that bond contributes zero. Hence $c$ is a local function normalized in each move.

Let $\mathcal A_4$ and $\mathcal A_{2\times2}$ be the transport algebras of Definition~\ref{def:hopping-algebra} on the coboundary configurations of the two sides, with the multiplication for the latter specified by $c$ above. Since $a=\dd\phi$, we have $V(0,\mathbf1)=\ii^{\oint\hat a}=\id$. Thus $U(u,v)\mapsto V(\widetilde u,\widetilde v)$ induces the isomorphism\footnote{Write $s=\widetilde u+2r$ and $t=\widetilde v+2w$. Then each $V(s,t)$ is a product of $V(\widetilde u,\widetilde v)$ with an element of $\mathcal D$, so the map is surjective. By the multiplication laws, the central involution on each side pairs the move labels of the free module. For a ring of $n$ vertices, the two dimensions thus both decrease from $4^n\,4^{n-1}$ by a factor of two.}
\[
 \frac{\mathcal A_{2\times2}}{\langle U(0,\mathbf1)-\id\rangle}
 \;\cong\;
 \frac{\mathcal A_4}{\langle X(2\mathbf1)-\id\rangle}.
\]
All configuration functions survive in the quotient, so it retains the statistical phases.

Section~\ref{sec:bockstein_word} gives a statistical process $W_2$ that detects the mixed statistics of $\ZZ_2\times\ZZ_2$, see Eq.~\eqref{eq:bockstein-word}. To evaluate $W_2$, we can substitute the $\ZZ_4$ Pauli expressions for the two hopping operators directly, without expanding the dependence of $c$ on $p,q$. Take two intervals $I=[L,R)$ and $J$ with interlaced endpoints, separated by more than the local readout range. With $\chi_I,\chi_J$ denoting their indicator $0$-cochains,
\[
 \mathsf X=U(\chi_I,0)=X(\chi_I),\qquad
 \mathsf Y=U(0,\chi_J)=\ii^{\oint\hat a\smile\chi_J}.
\]
By Eq.~\eqref{eq:z4-weyl-product}, $\mathsf Y\mathsf X=\ii^{\oint\dd\chi_I\smile\chi_J}\mathsf X\mathsf Y$, with the integer lifts of the indicator functions understood in the exponent. Since $J$ contains exactly one of $L,R$, we have $\oint\dd\chi_I\smile\chi_J=\chi_J(L)-\chi_J(R)=\pm1$. Thus
\[
 W_2=(\mathsf X\mathsf Y)^{-2}(\mathsf Y\mathsf X)^2
     =(-1)^{\oint\dd\chi_I\smile\chi_J}\id=-\id .
\]
This gives the nontrivial mixed statistics of the two species of $\ZZ_2$ domain wall on coboundary configurations.

\Needspace{12\baselineskip}

\section{Duality between gauge-covariant hopping and gauge transformations}
\label{sec:statistics_anomaly}
\label{sec:comparison}

In Sec.~\ref{sec:spacetime}, we constructed the global symmetry $S(z)$ from the topological term $\Theta$ and obtained the multiplication law for symmetric hopping operators from the cocycle $\omega$. To relate their statistical class to the anomaly of $S(z)$, we couple the same system to a background gauge field and ask whether local gauge transformations can act as redundancies.

\subsection{Background coupling and gauge transformations}

On a closed spacetime $M$, replace $\dd\phi$ in the dynamical term by $a=D_A\phi$ and write the topological term as $\Theta^1(\phi;A)$, to be determined. The partition function is
\[
 Z(M,A)=\sum_\phi
 \exp\!\left[\ii\int_M\mathcal L(D_A\phi)\right]
 \exp\!\left[-2\pi\ii\int_M\Theta^1(\phi;A)\right].
\]
Once $A$ is specified throughout spacetime, summing over matter histories $\phi$ also sums over excitation histories $a=A+\dd\phi$. The background history that moves a symmetry defect is specified externally. For two gauge-equivalent backgrounds, the map $(\phi,A)\mapsto(\phi+u,A-\dd u)$ pairs the matter histories. Paired histories have the same covariant derivative and hence the same dynamical weight.

The cochain interpolation of Sec.~\ref{sec:spacetime} expresses this gauge transformation as a homotopy. On $I_{\rm g}\times M$, interpolate from $(\phi,A)$ to $(\phi+u,A-\dd u)$ while keeping the covariant derivative $a$ fixed. Let $\eta_{\rm g}$ be a $0$-cochain taking the values $0,1$ at the two ends of the interval, and write $\eta_{\rm g}u$ for $\eta_{\rm g}\smile u$. The interpolating fields satisfy
\begin{equation}
 \bigl(\phi+\eta_{\rm g}u,\ A-\dd(\eta_{\rm g}u)\bigr),\qquad
 D_{A-\dd(\eta_{\rm g}u)}(\phi+\eta_{\rm g}u)=a .
\label{eq:gauge-prism-fields}
\end{equation}
Even when $\dd u=0$, integrating the background along $I_{\rm g}$ gives $-u$. With $u=-\phi$, the gauge transformation sends $(\phi,A)$ to $(0,a)$, as shown in Fig.~\ref{fig:background-to-covariant}.

\begin{figure}[H]
\centering
\begin{tabular}{@{}>{\sffamily\small\color{ink}}l@{\hspace{1.8em}}l@{}}
Gauge transformation &
  $\displaystyle (\phi,A)\mathrel{\color{accent}\xrightarrow{\;u=-\phi\;}}(0,D_A\phi)$ \\[.7em]
Gauge-field history &
  $\displaystyle \widetilde A=A+\dd(\eta_{\rm g}\phi)\quad\text{on }I_{\rm g}\times M$
\end{tabular}
\caption{The gauge-field history $\widetilde A$ restricts to $A$ and $D_A\phi=A+\dd\phi$ on the two end faces.}
\label{fig:background-to-covariant}
\end{figure}

We construct the background coupling from the cocycle $\omega$ of Sec.~\ref{sec:spacetime}. Integrating along the gauge-field history from $A$ to $a$ gives a $(d+1)$-cochain $\Theta^1$ on $M$. Its value on each $(d+1)$-simplex $\Delta$ is the integral of $\omega$ over $I_{\rm g}\times\Delta$. Summing over spacetime gives the topological action,
\begin{equation}
 \boxed{\begin{aligned}
 \Theta^1(\phi;A)&=\int_{I_{\rm g}}\omega\bigl(A+\dd(\eta_{\rm g}\phi)\bigr),\\
 \int_M\Theta^1(\phi;A)&=\int_{I_{\rm g}\times M}\omega\bigl(A+\dd(\eta_{\rm g}\phi)\bigr).
 \end{aligned}}
\label{eq:background-coupling-construction}
\end{equation}
When $\phi=0$, the background along the homotopy is pulled back from $M$, so $\Theta^1(0;A)=0$. Since $\dd\omega=0$, Stokes' theorem on the cylinder gives
\begin{equation}
 \boxed{\dd\Theta^1(\phi;A)=\omega(a)-\omega(A).}
 \label{eq:omega-descent-explicit}
\end{equation}
At $A=0$, this reduces to $\dd\Theta^1(\phi;0)=\omega(\dd\phi)$, as in Eq.~\eqref{eq:first-descent}. We can therefore choose the same representative, $\Theta^1(\phi;0)=\Theta(\phi)$.

To compute the gauge variation, observe that $a$ on the cylinder~\eqref{eq:gauge-prism-fields} is pulled back from $M$, so $\omega(a)=0$. For closed $M$, applying Stokes' theorem to Eq.~\eqref{eq:omega-descent-explicit} on the cylinder gives $-\int_{I_{\rm g}\times M}\omega\bigl(A-\dd(\eta_{\rm g}u)\bigr)$ for the difference between the integrals of $\Theta^1$ at the two ends. This difference is independent of the matter history, so its phase factors out of the path integral and gives
\begin{equation}
\boxed{Z(M,A-\dd u)=
\exp\!\left[2\pi\ii\int_{I_{\rm g}\times M}
\omega\bigl(A-\dd(\eta_{\rm g}u)\bigr)\right]Z(M,A).}
\label{eq:closed-prism-comparison}
\end{equation}
A local counterterm $\Theta^1(\phi;A)\mapsto\Theta^1(\phi;A)+\nu(A)$ replaces $\omega$ in the background response by $\omega-\dd\nu$. The response thus defines the cohomology class $[\omega]$.

Now allow $M$ to have a boundary and regard it as a bordism $M:X_0\to X_1$ with $\partial M=X_1-X_0$. Write $A_i,u_i$ for the restrictions of $A,u$ to $X_i$. Since a gauge transformation changes the spatial background, we include the background in the basis label $\ket{\phi,A}$. On each spatial slice $X$, take the direct sum over all flat backgrounds,
\[
 \mathcal H(X)=\bigoplus_{\dd A=0}\mathcal H(X,A),\qquad
 \ket{\phi,A}\in\mathcal H(X,A).
\]
The spacetime background $A$ is still specified externally. The path integral $Z(M,A):\mathcal H(X_0,A_0)\to\mathcal H(X_1,A_1)$ sums only over matter histories.

For a bordism, the derivation of Eq.~\eqref{eq:closed-prism-comparison} acquires two boundary terms, one from $I_{\rm g}\times X_0$ and one from $I_{\rm g}\times X_1$. On each side face $I_{\rm g}\times X$, evaluate $\Theta^1$ on the fields in~\eqref{eq:gauge-prism-fields}. The resulting topological phase defines the gauge transformation on the direct sum,
\begin{equation}
\boxed{T(u)\ket{\phi,A}=
\exp\!\left[-2\pi\ii\int_{I_{\rm g}\times X}
   \Theta^1\bigl(\phi+\eta_{\rm g}u;A-\dd(\eta_{\rm g}u)\bigr)\right]\ket{\phi+u,A-\dd u}.}
\label{eq:R-prism-definition}
\end{equation}
Write $T_A(u):\mathcal H(X,A)\to\mathcal H(X,A-\dd u)$ for its restriction to the initial background $A$. For a closed cochain $u=z$, the initial and final backgrounds agree, and we denote this global action by $S_A(z)=T_A(z)$. Starting with background $A_0$ on the incoming boundary and placing the two boundary actions on opposite sides of the equation gives
\begin{equation}
\boxed{Z(M,A-\dd u)\,T(u_0)
=\exp\!\left[2\pi\ii\int_{I_{\rm g}\times M}\omega\bigl(A-\dd(\eta_{\rm g}u)\bigr)\right]
T(u_1)\,Z(M,A).}
\label{eq:bordism-prism-comparison}
\end{equation}
Figure~\ref{fig:bordism-prism-comparison} shows the two compositions.

\begin{figure}[htbp]
\centering
\begin{tikzpicture}[>=Stealth,line width=.65pt,line cap=round,line join=round,
  every node/.style={font=\small,text=ink}]
\begin{scope}[draw=ink]
  \draw (-1.06,.28) .. controls (-1.06,1.42) and (-.53,1.62) .. (-.53,2.52);
  \draw ( 1.06,.28) .. controls ( 1.06,1.42) and ( .53,1.62) .. ( .53,2.52);
  \draw (-.28,.28) .. controls (-.28,.75) and (-.16,1.16) .. (0,1.16)
                    .. controls ( .16,1.16) and ( .28,.75) .. (.28,.28);
  \draw (0,2.52) ellipse [x radius=.53,y radius=.15];
  \foreach \x in {-.67,.67}{
    \begin{scope}[shift={(\x,.28)}]
      \draw (-.39,0) arc[start angle=180,end angle=360,x radius=.39,y radius=.12];
      \draw[dashed,thin] (-.39,0) arc[start angle=180,end angle=0,x radius=.39,y radius=.12];
    \end{scope}
  }
  \node at (0,1.79) {$M$};
  \node[above] at (0,2.68) {$X_1$};
  \node[below] at (0,.12) {$X_0$};
\end{scope}
\node (inA) at (4.15,.28) {$\mathcal H(X_0,A_0)$};
\node (outA) at (4.15,2.52) {$\mathcal H(X_1,A_1)$};
\node (inB) at (9.85,.28) {$\mathcal H(X_0,A_0-\dd u_0)$};
\node (outB) at (9.85,2.52) {$\mathcal H(X_1,A_1-\dd u_1)$};
\draw[->,accent] (inA)--node[left=3pt] {$Z(M,A)$} (outA);
\draw[->,accent] (inB)--node[right=3pt] {$Z(M,A-\dd u)$} (outB);
\draw[->,accent] (inA)--node[below=2pt] {$T_{A_0}(u_0)$} (inB);
\draw[->,accent] (outA)--node[above=2pt] {$T_{A_1}(u_1)$} (outB);
\node[font=\footnotesize] at (7,1.40)
  {$\displaystyle\exp\!\left[2\pi\ii\int_{I_{\rm g}\times M}
      \omega\bigl(A-\dd(\eta_{\rm g}u)\bigr)\right]$};
\end{tikzpicture}
\caption{The bordism $M:X_0\to X_1$ on the left merges two circles into one. On the right, the vertical arrows are the path integrals in the two backgrounds,
and the horizontal arrows are the gauge transformations on the incoming and outgoing boundaries. Going right and then up gives the phase inside the square relative to going up and then right.}
\label{fig:bordism-prism-comparison}
\end{figure}

\subsection{Gauge-covariant time evolution and the two transport algebras}
\label{subsec:two-operator-algebras}

In Fig.~\ref{fig:zero-background-square}, the global symmetry $S(z)$ commutes with the time evolution $Z(I_{\rm t}\times X)$. This allowed us to expand symmetric evolution according to the net displacement of the matter field. Now couple the same system to a background gauge field. For a closed spatial manifold $X$, write $Z_A=Z(I_{\rm t}\times X,A)$ for time evolution in a fixed spatial background $A$. When both $A$ and the gauge parameter $u$ are pulled back from $X$, the background phase in Eq.~\eqref{eq:bordism-prism-comparison} is unity, giving
\begin{equation}
 \boxed{T_A(u)\,Z_A=Z_{A-\dd u}\,T_A(u).}
 \label{eq:static-evolution-covariance}
\end{equation}
Equation~\eqref{eq:static-evolution-covariance} expresses gauge covariance of the evolution operators $Z_A$ across different backgrounds. On the direct sum $\mathcal H(X)$ over all flat backgrounds, they define a single gauge-invariant evolution operator $Z$,
\[
 Z=\bigoplus_{A\in Z^q(X,G)}Z_A,\qquad T(u)Z=ZT(u).
\]
Since $Z$ preserves each background sector, it also commutes with every function $g(\hat A)$ of the background.

Take a closed parameter $u=z$ at $A=0$. Since $\Theta^1(\phi;0)=\Theta(\phi)$, Eqs.~\eqref{eq:bordism-prism-comparison} and~\eqref{eq:symmetry-from-hopping} give the same ratio of endpoint phases for each matter history. Thus $T_0(z)$ and $S(z)$ differ only by a constant phase. Choosing the same reference matrix element, $\bra{z}T_0(z)\ket{0}=\bra{z}S(z)\ket{0}$, gives $T_0(z)=S(z)$, and Eq.~\eqref{eq:static-evolution-covariance} reduces at zero background to Fig.~\ref{fig:zero-background-square}.

As in Sec.~\ref{sec:spacetime}, the one-step history $\phi\to\phi+s$ with unit dynamical weight defines the standard hopping operator. Its action on the direct sum over all flat backgrounds is
\begin{equation}
 U(s)\ket{\phi,A}
 =\exp\!\left[-2\pi\ii\int_{I_{\rm t}\times X}
       \Theta^1(\phi+\eta_{\rm t}s;A)\right]\ket{\phi+s,A}.
 \label{eq:raw-U}
\end{equation}
Denote its restriction to the initial background $A$ by $U_A(s)$. Since Eq.~\eqref{eq:bordism-prism-comparison} holds for each matter history separately, applying it to this one-step history gives the square in Fig.~\ref{fig:two-directions-covariance},
\begin{equation}
 T_A(u)\,U_A(s)=U_{A-\dd u}(s)\,T_A(u),\qquad T(u)U(s)=U(s)T(u).
 \label{eq:raw-U-background-covariance}
\end{equation}
The second equality collects the relations for all backgrounds on the direct sum.

\begin{figure}[htbp]
\centering
\begin{tikzpicture}[
  x=1cm,y=1cm,>=Stealth,line cap=round,line join=round,
  every node/.style={text=ink,font=\small,inner sep=1.5pt},
  horizontal/.style={->,accent,line width=.85pt,shorten <=2pt,shorten >=2pt},
  vertical/.style={->,ink,line width=.85pt,shorten <=2pt,shorten >=2pt},
  axis/.style={->,muted!78,line width=.55pt}
]
\coordinate (O) at (0,0);   \coordinate (G) at (3.45,0);
\coordinate (T) at (0,3.45); \coordinate (C) at (3.45,3.45);
\draw[axis] (-1.15,-.8) -- (-1.15,3.45) node[above] {$a$};
\draw[axis] (-1.15,-.8) -- (3.45,-.8) node[right] {$A$};
\draw[horizontal] (O) -- node[above=3pt] {$T_A(u)$} (G);
\draw[horizontal] (T) -- node[below=3pt] {$T_A(u)$} (C);
\draw[vertical] (O) -- node[left=3pt] {$U_A(s)$} (T);
\draw[vertical] (G) -- node[right=3pt] {$U_{A-\dd u}(s)$} (C);
\foreach \p in {O,G,T,C} \fill[ink] (\p) circle[radius=1.25pt];
\node[below=4pt] at (O) {$(\phi,A)$};
\node[below=4pt] at (G) {$(\phi+u,A-\dd u)$};
\node[above=4pt] at (T) {$(\phi+s,A)$};
\node[above=4pt] at (C) {$(\phi+s+u,A-\dd u)$};
\node[font=\Large] at (1.725,1.725) {$=$};
\end{tikzpicture}
\caption{Gauge covariance of time evolution for a single hopping step. The two time edges preserve their respective backgrounds, $A$ and $A-\dd u$. Each gauge edge preserves the corresponding covariant derivative.}
\label{fig:two-directions-covariance}
\end{figure}

Apply the resolution of the identity in Eq.~\eqref{eq:hopping-resolution-identity} in each background. For fixed initial matter field $\phi$, divide the evolution amplitude from $\phi$ to $\phi+s$ by the phase of the standard hopping operator. Denote the resulting coefficient by
\[
 F_s(\phi,A)=
 \frac{\bra{\phi+s,A}Z_A\ket{\phi,A}}
      {\bra{\phi+s,A}U_A(s)\ket{\phi,A}} .
\]
By Eqs.~\eqref{eq:static-evolution-covariance} and~\eqref{eq:raw-U-background-covariance}, a gauge transformation of both the initial and final states gives the same endpoint phases in the numerator and denominator. Hence $F_s(\phi+u,A-\dd u)=F_s(\phi,A)$. Taking $u=-\phi$ gives $F_s(\phi,A)=F_s(0,A+\dd\phi)$. Writing $Z_s(x)=F_s(0,x)$ and summing over the groups of histories gives
\begin{equation}
 Z=\sum_{s\in C^{q-1}(X,G)} U(s)\,Z_s(\hat a),\qquad a=A+\dd\phi .
 \label{eq:gauge-covariant-evolution-expansion}
\end{equation}
At zero background this is the worldline expansion of Eq.~\eqref{eq:symmetric-expansion-main}. With a background field, the coefficient of each group of histories depends on the covariant derivative $a$ of the initial state.

Define two unital $*$-algebras by
\[
 \mathcal A_T=\operatorname{Alg}\{T(u),g(\hat A)\},\qquad
 \mathcal A_U=\operatorname{Alg}\{U(s),f(\hat a)\},
\]
where $f,g$ range over the complex-valued functions on $Z^q(X,G)$. An operator commutes with all $g(\hat A)$ precisely when it preserves each background sector. Commutation with all $T(u)$ imposes the covariance relation~\eqref{eq:static-evolution-covariance} on its components. These are exactly the two properties used in the expansion above, so every operator in $\mathcal A_T'$ belongs to $\mathcal A_U$. Conversely, $U(s)$ preserves $A$, $T(u)$ preserves $a$, and the two moves satisfy Eq.~\eqref{eq:raw-U-background-covariance}. Thus every generator of $\mathcal A_U$ belongs to $\mathcal A_T'$. We obtain $\mathcal A_T'=\mathcal A_U$, and the finite-dimensional double commutant theorem gives $\mathcal A_U'=\mathcal A_T$.

The background coupling also relates the generators of the two algebras. The hopping operator $U(s)$ moves the excitation configuration $a$ to $a+\dd s$, while $T(-s)$ moves the background $A$ to $A+\dd s$. Writing $x$ for the initial configuration on either side, both operations induce the same substitution $x\mapsto x+\dd s$ on configuration functions. To compare the products of two successive moves in Eq.~\eqref{eq:hopping-algebra-product}, use the triangle of Eq.~\eqref{eq:hopping-triangle} with $\eta_1=(0,1,1)$ and $\eta_2=(0,0,1)$. On the time triangle, $A$ is fixed, so only $\omega(a)$ survives in the descent equation~\eqref{eq:omega-descent-explicit}. On the gauge triangle, $a$ is fixed, so only $-\omega(A)$ survives. With the product phases evaluated on the input, we obtain
\begin{align}
 c_U(s_2,s_1;a)
 &=-\int_{\Delta^2\times X}\omega\bigl(a+\dd(\eta_1s_1+\eta_2s_2)\bigr),
 \label{eq:dual-triangle-products}\\
 c_T(u',u;A)
 &=+\int_{\Delta^2\times X}\omega\bigl(A-\dd(\eta_1u+\eta_2u')\bigr).
 \label{eq:normalized-R-multiplication}
\end{align}
\begin{figure}[H]
\centering
\begin{tikzpicture}[
  x=1cm,y=1cm,>=Stealth,line cap=round,line join=round,
  every node/.style={font=\small,text=ink,inner sep=1.4pt},
  path edge/.style={->,line width=.85pt,shorten <=2pt,shorten >=2pt},
  fused/.style={path edge,dashed}
]
\foreach \panel/\offset/\edgecolor in {1/0/ink,2/7.0/accent}{
  \begin{scope}[xshift=\offset cm]
    \coordinate (b0) at (0,0);
    \coordinate (b1) at (2.3,1.85);
    \coordinate (b2) at (4.6,0);
    \fill[\edgecolor!5] (b0)--(b1)--(b2)--cycle;
    \ifnum\panel=1
      \node at (2.3,2.76) {(a) Fixed background $A$};
      \draw[path edge,ink] (b0)--node[above,sloped] {$U(s_1)$} (b1);
      \draw[path edge,ink] (b1)--node[above,sloped] {$U(s_2)$} (b2);
      \draw[fused,ink] (b0)--node[below=2pt] {$U(s_1+s_2)$} (b2);
      \node at (2.3,.53) {$\exp\!\bigl(-2\pi\ii\!\int\!\omega\bigr)$};
    \else
      \node at (2.3,2.76) {(b) Fixed excitation configuration $a$};
      \draw[path edge,accent] (b0)--node[above,sloped] {$T(-s_1)$} (b1);
      \draw[path edge,accent] (b1)--node[above,sloped] {$T(-s_2)$} (b2);
      \draw[fused,accent] (b0)--node[below=2pt] {$T(-s_1-s_2)$} (b2);
      \node at (2.3,.53) {$\exp\!\bigl(+2\pi\ii\!\int\!\omega\bigr)$};
    \fi
    \foreach \point in {b0,b1,b2}
      \fill[\edgecolor] (\point) circle[radius=1.4pt];
    \node[below left=1.5pt] at (b0) {$b$};
    \node[above=3pt] at (b1) {$b+\dd s_1$};
    \node[below=3pt] at (b2) {$b+\dd(s_1+s_2)$};
  \end{scope}
}
\end{tikzpicture}
\caption{On the left, $b$ is the initial excitation configuration for hopping. On the right, $b$ is the initial background for the gauge transformation.
The solid arrows give two successive steps, and the dashed arrow gives a single step. Both triangles have the same three vertices, and $\omega$ is integrated over the same $\Delta^2\times X$.}
\label{fig:duality-product-triangles}
\end{figure}
As shown in Fig.~\ref{fig:duality-product-triangles}, the moves $s$ and $r$ starting at $x$ correspond, in that order, to gauge parameters $-s$ and $-r$. The cochains in the two integrals are identical, so
\[
 c_T(-r,-s;x)=-c_U(r,s;x)\pmod 1 .
\]
Both exponents are integrals of normalized cochains over the triangles in the figure. Summing over the top-dimensional simplices of $X$ gives the local decomposition in Definition~\ref{def:local-function}. The two algebras therefore satisfy the following relations.

\begin{theorem}\label{thm:dual-operator-algebras}
For a model on a finite lattice with finite Abelian group $G$, take the direct sum of the state spaces over all flat backgrounds. The two operator algebras defined above have the following properties.
\begin{enumerate}[label=\textup{(\arabic*)}]
\item They are mutual commutants,
\begin{equation}
 \mathcal A_U'=\mathcal A_T,\qquad \mathcal A_T'=\mathcal A_U .
 \label{eq:dual-commutants}
\end{equation}
\item The following correspondence on generators extends uniquely to an antilinear $*$-isomorphism $\alpha:\mathcal A_U\to\mathcal A_T$ that preserves the order of multiplication,
\begin{equation}
 \boxed{\alpha\bigl(U(s)\bigr)=T(-s),\qquad
        \alpha\bigl(f(\hat a)\bigr)=\overline f(\hat A).}
 \label{eq:dual-operator-expansion}
\end{equation}
\end{enumerate}
\end{theorem}

\begin{proof}
The expansion in Eq.~\eqref{eq:gauge-covariant-evolution-expansion} proves the first assertion. For the second, define a map on expansions by
\[
 \sum_s U(s) f_s(\hat a)\ \longmapsto\
 \sum_s T(-s)\,\overline{f_s}(\hat A).
\]
The shift and product relations on each side express every element as a sum of the indicated form. For any flat configuration $x$, applying either sum to $\ket{0,x}$ gives a distinct final state for each $s$. The coefficient functions on each side are therefore unique. For $f_s=Z_s$, the right-hand side is the image of the full evolution operator $Z$ in the gauge transformation algebra. The map is an antilinear bijection. It preserves the order of multiplication because the shifts agree and the product phases are complex conjugates. It preserves $*$ on the function algebra and maps each unitary generator to a unitary, so it preserves $*$ on the entire algebra.
\end{proof}

Theorem~\ref{thm:dual-operator-algebras} gives a gauge transformation measurement for every statistical process of Sec.~\ref{subsec:hopping-statistics}. Given a hopping process $W_U$, replace each $U(s)$ by $T(-s)$, keeping the order of the steps and their inverses, to obtain $W_T$. Since $a=A+\dd\phi$, the configurations on the two sides lie in the same cohomology class. Restricting the isomorphism to their common vacuum orbit $[a]=[A]=0$, antilinearity gives
\[
 W_U=w\,\id\quad\Longrightarrow\quad W_T=\alpha(W_U)=\overline w\,\id .
\]
After a step $T(-s)$ in the gauge direction, the next step starts from $A+\dd s$. If every local gauge transformation is a redundancy, each step must fix the physical states satisfying the Gauss constraints. Under $\alpha$, the constraints $U(e)-\id$ for simultaneous condensation of the excitations correspond to $T(-e)-\id$. If a closed process has a protected phase $\overline w\ne 1$, the Gauss constraints have no common nonzero solution on this vacuum orbit. This is the 't~Hooft anomaly detected by the process.

Once the matter-field shift at each step is fixed, the unitary transport operators on each side admit phase redefinitions,
\begin{equation}
 U'(s)=U(s)e^{2\pi\ii f(s;\hat a)},\qquad
 T'(u)=T(u)e^{2\pi\ii g(u;\hat A)} .
 \label{eq:background-hopping-rephasing}
\end{equation}
If the same $\alpha$ pairs the two sets of representatives, then $g(-s;x)=-f(s;x)\bmod 1$. When evaluating a statistical process, the local basis changes of Eq.~\eqref{eq:hopping-local-choice} can be made independently on the two sides. The paired sequences read the same configuration at each step, so the same local cancellation condition holds on both sides.

Starting from $\ket{0,0}$, with no excitations or defects, consider the physical operations paired by the isomorphism on two slices of the state space. Hopping in the time direction keeps the background at $A=0$. The paired moves in the gauge direction keep the covariant derivative at $a=0$. The two sets of states span
\begin{equation}
 \mathcal H_{\rm exci}=\operatorname{span}\{\ket{\phi,0}\}_\phi,\qquad
 \mathcal H_{\rm defect}=\operatorname{span}\{\ket{-\phi,\dd\phi}\}_\phi .
 \label{eq:dual-slice-bases}
\end{equation}
On the left, $U(s)$ shifts the matter label $\phi$ to $\phi+s$, and $\dd\phi$ labels the excitations. On the right, $T(-s)$ takes $\ket{-\phi,\dd\phi}$ to $\ket{-\phi-s,\dd(\phi+s)}$ up to a phase. Here $\dd\phi$ labels an externally imposed twist, which the matter field $-\phi$ exactly compensates.

Consider the ferromagnetic Ising ring in Fig.~\ref{fig:ising-hopping-defect-slices}. In the spin basis $Z_j\ket{\phi}=(-1)^{\phi_j}\ket{\phi}$, take
\[
 H[A]=-K\sum_{b=(j,j+1)}(-1)^{A_b}Z_jZ_{j+1},\qquad K>0 .
\]
Starting from the uniform vacuum, flip the spins in an interval $I$. In $H[0]$, a domain wall appears on each of the two end bonds, raising the energy by $4K$ in total. Reversing the couplings on those two bonds at the same time imposes $A=\dd\chi_I$, and the same spin configuration is now a ground state of $H[\dd\chi_I]$. Extend $I$ by one site. In the top row, one wall moves at fixed Hamiltonian. In the bottom row, flipping the added spin and reversing the couplings on its two neighboring bonds takes the original ground state to a ground state of the newly twisted Hamiltonian.

\begin{figure}[htbp]
\centering
\begin{tikzpicture}[x=.94cm,y=1cm,every node/.style={font=\small}]
  \foreach \yy/\slice in {1.0/{\mathcal H_{\mathrm{exci}}\ (A=0)},0/{\mathcal H_{\mathrm{defect}}\ (a=0)}} {
    \node[anchor=east] at (-.35,\yy) {$\slice$};
    \draw[muted,line width=.7pt] (-.2,\yy)--(5.2,\yy);
    \foreach \xx/\spin in {0/0,1/0,2/1,3/1,4/0,5/0} {
      \node[draw=ink,fill=white,circle,inner sep=1.6pt] at (\xx,\yy) {$\spin$};
    }
  }
  \foreach \xx in {1.5,3.5} {
    \draw[speciesA,line width=1.3pt] (\xx,0.87)--(\xx,1.13);
    \draw[speciesB,line width=1.2pt,fill=white] (\xx,0) circle (2.5pt);
  }
  \node[anchor=west,text=speciesA] at (5.45,1) {unsatisfied bond $(a=1)$};
  \node[anchor=west,text=speciesB] at (5.45,0) {twisted bond $(A=1)$};
\end{tikzpicture}
\caption{The same spin configuration for two choices of couplings. The vertical marks in the top row are domain walls on ferromagnetic bonds. The open circles in the bottom row are externally twisted bonds. Every bond in this row is in its lowest-energy configuration.}
\label{fig:ising-hopping-defect-slices}
\end{figure}

\begin{remark}[Closed hopping on the two slices]
On the $A=0$ slice, the prescribed global symmetry is $S(z)=T_0(z)$. The closed hopping operator $U_0(z)$ preserves the excitation configuration but can change the Hamiltonian, as illustrated by $H'_\lambda$ for $\lambda>0$ in Sec.~\ref{subsec:z2-hop-symmetry}. On the $a=0$ slice, $U(-z)$ preserves the slice and commutes with every defect transport operator $T(-s)$. Take the nontrivial $\ZZ_2$ topological term of that section and substitute the constant shift $\mathbf 1$ into Eq.~\eqref{eq:raw-U}. The term $A\smile\phi\smile a/2$ in $\Theta^1$ gives
\begin{equation}
 U_A(\mathbf 1)\ket{\phi}
      =(-1)^{\int_{S^1}A\smile\phi}\ket{\phi+\mathbf 1}.
 \label{eq:z2-closed-hop-two-slices}
\end{equation}
On the $A=0$ slice this is $\prod_j X_j$. On the $a=0$ slice, $A=\dd\phi$ and $\int\dd\phi\smile\phi=\int\phi\smile\dd\phi$, so the resulting matrix is precisely the anomalous symmetry in Eq.~\eqref{eq:z2-matter-symmetry}. On the same slice, $T(\mathbf 1)$ itself gives the global spin flip $\prod_j X_j$.
\end{remark}

\begin{remark}[Realization on the state space]
For the prism representatives above, take the reference states $\ket{0,A}$ in each background and adopt the return convention $T(-s)U(s)\ket{0,A}=\ket{0,A+\dd s}$. On the basis of prepared states, define
\[
 J\,U(\phi)\ket{0,A}=T(-\phi)\ket{0,A},
\]
Since each side forms an orthonormal basis, antilinear extension defines an antiunitary operator $J$. The matter and background labels on the right are $(-\phi,A+\dd\phi)$, so $J$ exchanges $A$ and $a=A+\dd\phi$. The triangle multipliers on the two sides are complex conjugates, giving $J O J^{-1}=\alpha(O)$ for $O\in\mathcal A_U$. The return convention gives $J^2=\id$.
\end{remark}

\section{Gauging a generic global symmetry}
\label{sec:patch-compatibility}

In the preceding two sections, we started from a cocycle $\omega$ and integrated the topological term $\Theta^1(\tilde{\phi},\tilde{A})$ in the time and gauge directions. This gave the transport algebras of symmetric hopping and gauge transformations, respectively, with complex-conjugate phases for statistics and anomaly (Theorem~\ref{thm:dual-operator-algebras}). We now consider a question that arises more often in a microscopic description. Given a symmetry $S$ implemented by a finite-depth local unitary circuit (FDQC), how can we determine its anomaly without knowing $\omega$ in advance, and gauge it when the anomaly is trivial? In our framework this requires constructing the transport algebra of gauge transformations from the global symmetry. This section examines what that construction entails.

\subsection{From a symmetry patch to a gauge transformation}

Given an FDQC for a global symmetry, a natural starting point is to truncate the circuit. In one dimension, for instance, applying a group element $g$ only on an interval $I$ gives a symmetry patch $P(u)$ with parameter $u=g\chi_I$. A pair of symmetry defects appears at the endpoints. Gauge symmetry acts locally, suggesting a close relation between a symmetry patch and a gauge transformation. To keep track of the background when composing operators, we label copies $\mathcal H_A$ of the matter Hilbert space by $A\in Z^q(X,G)$. The global symmetry $S$ preserves the zero background, whereas applying a symmetry patch at zero background gives a map
\begin{equation}
	P(u):\mathcal H_0\longrightarrow\mathcal H_{-\dd u}.
\end{equation}

Let $\Omega$ be the set of parameters for which patches are supplied. From $\{P(r)\}_{r\in\Omega}$ we want to assemble a family of gauge transformations satisfying
\begin{equation}
 T_A(u):\mathcal H_A\longrightarrow\mathcal H_{A-\dd u},
 \qquad T_0(u)\propto P(u):\mathcal H_0\longrightarrow\mathcal H_{-\dd u}.
 \label{eq:patch-background-copies}
\end{equation}
The second relation identifies the step from zero background whenever $P(u)$ is given. Here $\propto$ denotes equality up to a phase independent of the matter state. All copies have the same lattice matter degrees of freedom, so we multiply their matter matrices directly and use the subscript to track the initial background at each step. We can choose a unitary automorphism $V_A$ independently on each copy $\mathcal H_A$. With the convention $V_0=\id$, the preparation and transport matrices then transform as
\[
P(r)\ \longmapsto\ V_{-\dd r}P(r),\qquad
T_A(e)\ \longmapsto\ V_{A-\dd e}\,T_A(e)\,V_A^{-1} .
\]
For these matrices on the fixed lattice Hilbert space to represent local truncations of the same symmetry, the new $P(r)$ must retain its original action away from the truncation boundary. Likewise, the new $T_A(e)$ must act and read the background only within a fixed neighborhood of the elementary move $e$. These locality-preserving changes of basis on the background copies give rise to perturbations of the first kind.

The gauge transformations should satisfy the multiplication law of a transport algebra,
\begin{equation}
 T_{A-\dd s}(t)\,T_A(s)=e^{2\pi\ii c(t,s;A)}\,T_A(s+t).
 \label{eq:patch-required-product}
\end{equation}
where $c(t,s;A)$ is a local function normalized in $s$ and $t$ separately. The multiplication law already suggests how to construct a general gauge transformation from patch operators.

In Eq.~\eqref{eq:patch-required-product}, take $A=0$, $s=r$, and $t=u$. If the two preparations from zero background are represented by $P(r)$ and $P(r+u)$, substituting $T_0\propto P$ gives
\begin{equation}
 \boxed{T_{-\dd r}(u)\ \propto\ P(r+u)P(r)^{-1} .}
 \label{eq:patch-reconstruct}
\end{equation}

For example, taking $r=u$ with $2u=0$ gives $T_{-\dd u}(u)\propto P(u)^{-1}$, which generally differs from $T_0(u)=P(u)$. For the anomalous $\ZZ_2$ symmetry in Eq.~\eqref{eq:z2-matter-symmetry}, the truncation to $I=[L,R]$ is $P_I\ket{\phi}=(-1)^{\int_I\phi\smile\dd\phi}\ket{\phi+\chi_I}$. Hence
\begin{equation}
 T_{-\dd\chi_I}(\chi_I)\ \propto\ P_I^{-1}= Z_LZ_R P_I .
 \label{eq:patch-interval-return}
\end{equation}
The endpoint factor $Z_LZ_R$ relating this operator to $P_I$ plays a central role in the Else--Nayak calculation of the anomaly index~\cite{ElseNayak2014}.

For another example, consider moving a symmetry defect from point $1$ to point $2$ in Fig.~\ref{fig:patch-defect-move}. Equation~\eqref{eq:patch-reconstruct} gives the matrix $P_{[0,2]}P_{[0,1]}^{-1}$. Both patches extend to the common endpoint $0$. For the product to describe a move localized near $1$ and $2$, their actions at the distant endpoint must cancel.
\begin{figure}[htbp]
\centering
\begin{tikzpicture}[
  x=1cm,y=1cm,>=Stealth,line cap=round,line join=round,
  every node/.style={font=\small,text=ink,inner sep=2pt},
  lattice/.style={draw=muted!70,line width=.6pt},
  empty site/.style={draw=ink!70,fill=white,line width=.7pt},
  occupied site/.style={draw=speciesA!85!black,fill=speciesA,line width=.7pt},
  transition/.style={->,draw=accent,line width=.9pt}
]
  \foreach \offset/\state in {0/{Initial state},4.8/{Intermediate state},9.6/{Final state}}{
    \begin{scope}[xshift=\offset cm]
      \draw[lattice] (-.28,0)--(2.28,0);
      \foreach \site in {0,1,2}{
        \draw[empty site] (\site,0) circle[radius=2.7pt];
        \node[below=5pt] at (\site,0) {$\site$};
      }
      \node[below=2pt,font=\footnotesize,text=muted] at (1,-.48) {\state};
    \end{scope}
  }
  \foreach \site in {0,1}{\draw[occupied site] (\site,0) circle[radius=3.8pt];}
  \begin{scope}[xshift=9.6cm]
    \foreach \site in {0,2}{\draw[occupied site] (\site,0) circle[radius=3.8pt];}
  \end{scope}
  \draw[transition] (2.52,0)--node[above=3pt] {$P_{[0,1]}^{-1}$} (4.28,0);
  \draw[transition] (7.32,0)--node[above=3pt] {$P_{[0,2]}$} (9.08,0);
  \draw[transition,rounded corners=3pt]
    (1,.24)--(1,1.02)--node[above=2pt] {$T_A(e)$} (11.6,1.02)--(11.6,.24);
\end{tikzpicture}
\caption{A filled dot marks a $\ZZ_2$ defect and an open dot an empty site. Along the upper arrow, $T_A(e)$ moves the defect at $1$ to $2$ while leaving the defect at $0$ in place. Along the lower path, $P_{[0,1]}^{-1}$ annihilates the pair at $0,1$, and $P_{[0,2]}$ creates a pair at $0,2$. The two paths satisfy $T_A(e)\propto P_{[0,2]}P_{[0,1]}^{-1}$.}
\label{fig:patch-defect-move}
\end{figure}

\subsection{From compatible symmetry patches to a transport algebra}
\label{subsec:local-patch-transport}

An elementary cochain $e$ is supported on a single $(q-1)$-simplex. Fix an integer $D\geq1$. In any simplicial complex, let $N_e$ be the subcomplex spanned by the vertices reachable from some vertex of that simplex along at most $D$ edges. All elementary moves on the same simplex share this neighborhood. We take $D$ large enough to contain the support of one move and the background data it reads.

It suffices to construct the elementary moves $T_A(e)$. The projective multiplication law then assembles the general $T_A(s)$, including the prescribed action $T_0(z)\propto S(z)$ for closed parameters. By Eq.~\eqref{eq:patch-reconstruct}, to choose $T_A(e)$ using only $A|_{N_e}$, we require
\begin{equation}
 T_A(e)\ \propto\ P(r+e)P(r)^{-1},
 \qquad r,r+e\in\Omega,\quad (-\dd r)|_{N_e}=A|_{N_e}.
 \label{eq:compatible-elementary-patches}
\end{equation}
For each $e$ and $A|_{N_e}$, at least one such pair of patches must exist, and all available pairs must give the same product up to a scalar phase. For each input we choose one matter matrix as $T_A(e)$ and take the identity for the zero move. For the gauge transformations assembled from these moves to obey the projective multiplication law~\eqref{eq:patch-required-product}, it suffices to check fusion on the same cell and interchange on distinct cells. For $e,f$ supported on $\sigma,\rho$, respectively, these relations read
\begin{equation}
 T_{A-\dd e}(f)T_A(e)=\exp(2\pi\ii\gamma(f,e;A))
 \begin{cases}
 T_A(e+f),&\sigma=\rho,\\
 T_{A-\dd f}(e)T_A(f),&\sigma\ne\rho .
 \end{cases}
 \label{eq:elementary-gauge-relations}
\end{equation}
Indeed, in the product of any two sequences of elementary moves we may first reorder the moves and then fuse them cell by cell to obtain the sum of their parameters. Conversely, restricting Eq.~\eqref{eq:patch-required-product} to elementary moves gives Eq.~\eqref{eq:elementary-gauge-relations}. Since $D\geq1$, $N_e$ also contains every cell on which $\dd e$ is nonzero. We consider disjoint and overlapping neighborhoods in turn.

\emph{Disjoint neighborhoods.} Suppose $N_e\cap N_f=\varnothing$. The operators $T_A(e)$ and $T_A(f)$ are supported in $N_e$ and $N_f$, respectively, and hence commute. Their local dependence on $A$ gives
\[
 T_{A-\dd e}(f)T_A(e)=T_A(f)T_A(e)=T_A(e)T_A(f)=T_{A-\dd f}(e)T_A(f).
\]
Thus $\gamma(f,e;A)=0$ in Eq.~\eqref{eq:elementary-gauge-relations}, and the preparations for these moves can be chosen independently.

\emph{Overlapping neighborhoods.} Suppose $N_e\cap N_f\ne\varnothing$. To interchange moves on distinct cells, choose $r$ such that $-\dd r$ agrees with $A$ on $N_e\cup N_f$ and $r,r+e,r+f,r+e+f\in\Omega$. Equation~\eqref{eq:compatible-elementary-patches} gives
\[
 \begin{aligned}
 T_{A-\dd e}(f)T_A(e)
 &\propto \bigl[P(r+e+f)P(r+e)^{-1}\bigr]\bigl[P(r+e)P(r)^{-1}\bigr]
 =P(r+e+f)P(r)^{-1},\\
 T_{A-\dd f}(e)T_A(f)
 &\propto \bigl[P(r+e+f)P(r+f)^{-1}\bigr]\bigl[P(r+f)P(r)^{-1}\bigr]
 =P(r+e+f)P(r)^{-1}.
 \end{aligned}
\]
Both orders give the same product of patches and hence differ at most by a scalar phase.

We collect the conditions used in this construction into the following definition.

\begin{definition}[Compatible symmetry patches]\label{def:compatible-symmetry-patches}
On a triangulation $Y$, let $0\in\Omega\subseteq C^{q-1}(Y,G)$ and let $\{P(r)\}_{r\in\Omega}$ be a family of unitary patches with $P(0)=\id$. The family is \textbf{compatible} if it satisfies the following conditions.
\begin{enumerate}[label=(\roman*)]
\item Each $r\in\Omega$ is a sum of elementary cochains whose successive partial sums all belong to $\Omega$. Explicitly, there exist $e_1,\ldots,e_n$, each nonzero on only one $(q-1)$-simplex, such that
\[
r=\sum_{i=1}^{n}e_i,\qquad
\sum_{i=1}^{j}e_i\in\Omega\quad (j=0,\ldots,n),
\]
where the empty sum is $0$.
 \item For every elementary move $e$ and every local input $A|_{N_e}$, there exists $r$ such that $\{r,r+e\}\subseteq\Omega$ and $(-\dd r)|_{N_e}=A|_{N_e}$. The operators $P(r+e)P(r)^{-1}$ obtained from all such $r$ agree up to a scalar phase.
 \item For each such pair of patches, the support of the operator on the many-body Hilbert space satisfies $\supp\!\left(P(r+e)P(r)^{-1}\right)\subseteq N_e$.
 \item For every pair of elementary moves $e,f$ with overlapping neighborhoods and every local input $A|_{N_e\cup N_f}$, there is a common $r$ such that $(-\dd r)|_{N_e\cup N_f}=A|_{N_e\cup N_f}$ and
 \begin{equation}
 	\{r,r+e,r+f,r+e+f\}\subseteq\Omega.
 \end{equation}

\end{enumerate}
\end{definition}

A typical choice starts with a cover $\mathcal D$ of $Y$ by $d$-dimensional disks. Whenever $N_e\cap N_f\ne\varnothing$, we require their union $N_e\cup N_f$ to lie in the interior of a single disk. Set
\[
 \Omega=\bigcup_{D\in\mathcal D}
 \bigl\{r\in C^{q-1}(Y,G)\mid \supp r\subseteq\operatorname{int}D\bigr\},
\]
where $\supp r$ is the union of the closed simplices on which $r$ is nonzero. Whenever the compatibility conditions hold, we can construct a family $\{T_A(u)\}$ satisfying the projective multiplication law for all cochains $u$. Condition~(i) ensures that this construction recovers the supplied patches,
\begin{equation}
	T_0(r)\propto P(r).
\end{equation}
The definition of compatible symmetry patches does not require a global symmetry as input. Compatibility already ensures that the resulting $T_0(z)$ satisfy the projective multiplication law of a global symmetry,
\begin{equation}
	T_0(z)T_0(z')\propto T_0(z+z').
\end{equation}
If these compatible patches arise by truncating an existing symmetry $S$, then $T_0(z)$ should implement $S(z)$. We say that the patches \emph{match} $S$ if $T_0(z)\propto S(z)$ for every closed parameter $z$.

Returning to Eq.~\eqref{eq:elementary-gauge-relations}, write $B$ for the background on $Y$. For $N_e\cap N_f=\varnothing$, the interchange phase $\gamma(f,e;B)$ vanishes. When the neighborhoods overlap, the matter matrices compared in a fusion or interchange read only $B|_{N_e\cup N_f}$, so their relative phase depends only on this restriction.

For general $s,t\in C^{q-1}(Y,G)$, choose representatives of $T^Y$ by fixing an order of the cells. The phase exponent in the product is the sum of $\gamma$ over all interchanges and fusions. If the moves already applied have total parameter $v$, the next move reads the background $B-\dd v$. Since the order is fixed, the values of $\dd v$ on the cells it reads are determined by the values of $s,t$ on their faces. Thus
\begin{equation}
 \begin{aligned}
 T^Y_{B-\dd s}(t)\,T^Y_B(s)
   &=\exp\!\bigl(2\pi\ii\,c_Y(t,s;B)\bigr)\,T^Y_B(s+t),\\
 c_Y(t,s;B)
   &=\sum_\alpha c_\alpha\bigl(t|_{N_\alpha},s|_{N_\alpha};B|_{N_\alpha}\bigr)
      \pmod 1 .
 \end{aligned}
 \label{eq:patch-fine-neighborhoods}
\end{equation}
Here $N_\alpha$ may be taken to be $N_e$ for a fusion or $N_e\cup N_f$ for a nontrivial interchange, in which case $N_e\cap N_f\ne\varnothing$. These subcomplexes contain all faces of the cells being read. The two moves under comparison come from $t$ and $s$, respectively, and their comparison phase vanishes if either move is zero. Hence $c_\alpha=0$ whenever $t|_{N_\alpha}=0$ or $s|_{N_\alpha}=0$.

This form of locality is weaker than the one required for a transport algebra. A neighborhood $N_\alpha$ can meet several top-dimensional simplices, whereas Definition~\ref{def:local-function} requires each term to read the input on a single top-dimensional simplex. We therefore regard $Y$ as a subdivision of a coarse triangulation $X$ and choose a simplicial map $p:Y\to X$ that sends each $N_\alpha$ into one coarse top-dimensional simplex. After pulling back the coarse inputs along $p$, each neighborhood term reads only the data on that simplex.

Figure~\ref{fig:coarse-four-sites} illustrates the pullback for a chain with $q=1$. A vertex parameter pulls back as $(p^*s)(i)=s(p(i))$, so the four fine vertices in each group take the same value.

\begin{figure}[H]
\centering
\begin{tikzpicture}[
  x=.77cm,y=1cm,line cap=round,
  every node/.style={font=\small,text=ink},
  fine site/.style={circle,draw=ink,fill=white,inner sep=0pt,
                    minimum size=4.2mm,font=\scriptsize},
  correspondence/.style={draw=muted!75,densely dashed,line width=.55pt},
  chain/.style={draw=ink,line width=.85pt},
  field edge/.style={draw=accent,line width=1.2pt}
]
  \foreach \first/\last/\center in {-3/0/-1.5,1/4/2.5,5/8/6.5,9/12/10.5}
    \foreach \i in {\first,...,\last}
      \draw[correspondence] (\i,1.25)--(\center,0);

  \draw[chain] (-3.68,1.25)--(12.68,1.25);
  \draw[chain] (-3.68,0)--(12.68,0);
  \foreach \y in {0,1.25}{
    \node[anchor=east,inner sep=1pt] at (-3.68,\y) {$\cdots$};
    \node[anchor=west,inner sep=1pt] at (12.68,\y) {$\cdots$};
  }
  \foreach \i/\j in {0/{j-1},4/j,8/{j+1}}{
    \draw[field edge] (\i,1.25)--(\i+1,1.25)
      node[midway,above=4pt,text=accent] {$A_{\j}$};
    \draw[field edge] (\i-1.5,0)--(\i+2.5,0)
      node[midway,below=3pt,text=accent] {$A_{\j}$};
  }
  \foreach \i in {-3,-2,-1,1,2,3,5,6,7,9,10,11}
    \node[above=4pt] at (\i+.5,1.25) {$0$};
  \foreach \i in {-3,...,12}
    \node[fine site] at (\i,1.25) {$\i$};
  \foreach \center/\j in {-1.5/{j-1},2.5/j,6.5/{j+1},10.5/{j+2}}{
    \fill[ink] (\center,0) circle[radius=2.4pt];
    \node[below=3pt] at (\center,0) {$B_{\j}$};
  }
  \node[anchor=east] at (-4.38,1.25) {fine $Y$, $p^*A$};
  \node[anchor=east] at (-4.38,0) {coarse $X$, $A$};
\end{tikzpicture}
\caption{Pullback on a one-dimensional chain. Dashed lines connect fine vertices to their images under $p:Y\to X$. All bonds point from left to right. The three fine bonds within each group map to a single coarse vertex, so their values are $0$. The fine bonds $[0,1]$, $[4,5]$, and $[8,9]$ between groups inherit the values $A_{j-1},A_j,A_{j+1}$ of their respective coarse bonds.}
\label{fig:coarse-four-sites}
\end{figure}

\begin{lemma}[Coarse-graining local functions]\label{lem:local-function-coarse}
Let $X$ be a pure $d$-dimensional simplicial complex and fix a neighborhood radius $D\geq1$. There exist a subdivision $Y$ and a simplicial map $p:Y\to X$ such that each fine vertex maps to a vertex of the smallest coarse simplex containing it, and $p$ sends the union of any two overlapping $D$-step neighborhoods $N_e,N_f$ into a single coarse top-dimensional simplex. Taking $f=e$ includes the case of a single neighborhood.
\end{lemma}

See Appendix~\ref{app:local-function-coarse-proof} for the proof. After pullback, each term in Eq.~\eqref{eq:patch-fine-neighborhoods} reads the input on just one coarse top-dimensional simplex $\Delta$. Summing the terms assigned to the same $\Delta$ gives
\begin{equation}
 F_Y(p^*t,p^*s;p^*A)=\sum_{\Delta\in X_d}
 F_\Delta(t|_\Delta,s|_\Delta;A|_\Delta)\pmod 1 .
 \label{eq:patch-coarse-function}
\end{equation}
If $s|_\Delta=0$, every term assigned to $\Delta$ reads a zero parameter, so $F_\Delta$ remains normalized in $s$. The same argument applies to $t$ or to a function of a single move parameter.

\begin{theorem}[Transport algebra from compatible patches]\label{thm:compatible-patch-transport}
Suppose a family of symmetry patches on $Y$ satisfies Definition~\ref{def:compatible-symmetry-patches}. If $Y$ and $p:Y\to X$ have the neighborhood property in Lemma~\ref{lem:local-function-coarse}, the gauge transformations assembled from these patches satisfy, after coarse-graining,
\[
 T^X_{A-\dd s}(t)\,T^X_A(s)
   =\exp(2\pi\ii c_X(t,s;A))\,T^X_A(s+t),
\]
where $c_X$ is a local function in the sense of Definition~\ref{def:local-function}, normalized separately in $t$ and $s$. This gives a transport algebra on the coarse lattice, represented on the matter and background space by $U(s)=T^X(-s)$. If the patches match $S$, the transformation for every closed parameter $z$ on the coarse lattice is proportional to $S(p^*z)$.
\end{theorem}

\begin{proof}
Choose $p$ as in Lemma~\ref{lem:local-function-coarse} and set $T^X_A(s)=T^Y_{p^*A}(p^*s)$. Since $\dd p^*=p^*\dd$, applying $s$ and then $t$ on the coarse lattice gives precisely $p^*(A-\dd s)$ as the fine input background for the second step. Substituting the coarse inputs into Eq.~\eqref{eq:patch-fine-neighborhoods} gives the stated multiplication law with
\[
 c_X(t,s;A)=c_Y(p^*t,p^*s;p^*A).
\]
The lemma ensures that the right-hand side decomposes into terms on coarse top-dimensional simplices, each normalized separately in $t$ and $s$. If $\dd z=0$, then $p^*z$ is also closed, so matching with $S$ gives $T^X_0(z)=T^Y_0(p^*z)\propto S(p^*z)$.
\end{proof}

For a fixed family of compatible symmetry patches $P$, we may rechoose the phase of each elementary move $T_A(e)$ on the fine lattice $Y$ as a function of $(e,A|_{N_e})$. After coarse-graining, Lemma~\ref{lem:local-function-coarse} turns this freedom into the local basis change of Definition~\ref{def:local-basis-change}, or a perturbation of the second kind in the terminology of the introduction.

Our framework divides the problem of gauging a microscopic symmetry into three steps.
\begin{center}
\begin{tikzpicture}[>=Stealth,every node/.style={font=\small,text=ink},patchArrow/.style={->,line width=.75pt,draw=accent},op/.style={above=2pt,font=\footnotesize,text=muted}]
 \node[align=center] (s) at (0,0) {global symmetry\\$S$};
 \node[align=center] (p) at (4.4,0) {compatible\\symmetry patches};
 \node[align=center] (t) at (8.9,0) {transport\\algebra};
 \node (g) at (12.5,0) {gauging};
 \draw[patchArrow] (s) -- node[op,align=center]{choose\\truncations} (p);
 \draw[patchArrow] (p) -- node[op,align=center]{assemble and\\coarse-grain} (t);
 \draw[patchArrow] (t) -- node[op,align=center]{statistical\\process}
    node[below=2pt,font=\footnotesize,text=muted] {if $[\omega]=0$} (g);
\end{tikzpicture}
\end{center}

This decomposition applies to any global symmetry that can be gauged, since its Gauss law operators automatically supply compatible symmetry patches. Our results address the last two steps. Given compatible symmetry patches, Theorem~\ref{thm:compatible-patch-transport} produces the transport algebra of Definition~\ref{def:hopping-algebra} by assembly and coarse-graining. Under the hypotheses of the classification theorem (Theorem~\ref{thm:hopping-statistics}), if every statistical process has a trivial phase, a local basis change makes the Gauss operators obey the group law exactly.

Two questions remain.
\begin{enumerate}
 \item \emph{Existence.} Does a given global symmetry admit compatible symmetry patches?
 \item \emph{Uniqueness.} If the same symmetry admits two families of compatible symmetry patches, do statistical processes in their transport algebras give the same $[\omega]\in H^{d+2}(K(G,q),\RZ)$? Uniqueness is needed for $[\omega]$ to characterize $S$ itself, or equivalently, for its anomaly to be well defined as a cohomology class.
\end{enumerate}

In higher dimensions, the existence of compatible truncations may encounter obstructions beyond $H^{d+2}(K(G,q),\RZ)$. Shirley et al.\ gave a two-dimensional ordinary $\ZZ_2$ symmetry for which $H^4(\ZZ_2,\RZ)=0$, yet the boundary action of a disk truncation obstructs compatible assembly~\cite{ShirleyZhangJiLevin2026}. Appendix~\ref{app:gnvw-toolkit} explains how to detect the obstruction at the disk boundary. For uniqueness, we compare two preparations of the same background. They differ by a change of basis on the matter space, leaving only the relative phases of elementary moves to affect the statistics. The next lemma expresses this comparison as the difference of two multipliers.

\begin{lemma}[Reduction to the identity symmetry]\label{lem:relative-transport}
Let $T,T'$ satisfy Eq.~\eqref{eq:patch-required-product} on the same matter lattice. Assume that each elementary move is supported in a neighborhood of its cell, that its matrix depends only on the background in that neighborhood, and that $T_0(z)\propto T'_0(z)\propto S(z)$ for every closed parameter $z$. On the backgrounds reachable from zero there exists a family of gauge transformations $R$, with elementary moves still local, satisfying
\[
 R_0(z)\propto\id,\qquad c_R(t,s;A)=c_{T'}(t,s;A)-c_T(t,s;A)\pmod1.
\]
Consequently, the multiplication laws of $T$ and $T'$ are equivalent under a local basis change if and only if a local basis change can set the phase exponent in the multiplication law of $R$ to zero.
\end{lemma}

The proof is given in Appendix~\ref{app:patch-uniqueness}.

We have no general existence or uniqueness criterion in arbitrary dimension. Analyzing compatible symmetry patches on a many-body Hilbert space requires control of quantum entanglement and QCAs. Our framework isolates the cohomological obstruction to gauging and characterizes it through the classification of transport algebras, which admits a formal proof in Lean. For ordinary symmetries in two dimensions and $(d-1)$-form symmetries in $d$ dimensions, the structure of many-body entanglement is simple enough to give explicit criteria. We assume a tensor product of finite-dimensional bosonic site Hilbert spaces on a lattice of uniformly bounded coordination.

We work on a space without boundary. The cells of the parameter triangulation $Y$ are much larger than the range of the microscopic circuits. Closed parameters and the matching condition are defined on $Y$. If $S$ was originally labeled by cochains on a finer triangulation, $S(z)$ here denotes the action after pulling $z$ back to that lattice. To read the statistics, we then coarse-grain from $Y$ to $X$ as in Theorem~\ref{thm:compatible-patch-transport}. When the next two theorems use $[\omega]$ to decide whether a symmetry can be gauged, we assume the hypotheses of Theorem~\ref{thm:hopping-statistics}, including condition \textup{(ZZ)} on the vacuum orbit of the triangulated sphere $X$. For a matching family this means that the $S(p^*z)$ for closed coarse parameters commute pairwise exactly.

\begin{theorem}[Gauging criterion for an ordinary symmetry in two dimensions]\label{thm:gauging-2d-patches}
Let a finite Abelian group $G$ act on a two-dimensional lattice by FDQCs implementing an ordinary unitary internal symmetry $S$. Compatible symmetry patches matching $S$ exist if and only if the following disk-boundary condition holds. On each connected component, choose any sufficiently large disk $D$. It must be possible to choose truncations $U_D(a)$ simultaneously for all $a\in G$ so that every fusion error $U_D(a)U_D(b)U_D(a+b)^{-1}$ is implemented by an FDQC within a strip along the boundary. Appendix~\ref{app:gnvw-toolkit} explains why one representative disk suffices and gives the equivalent cohomological criterion. When the condition holds, all matching families of compatible truncations yield the same $[\omega]\in H^4(G,\RZ)$, and $S$ can be gauged if and only if $[\omega]=0$.
\end{theorem}

\begin{theorem}[Gauging criterion for point defects]\label{thm:gauging-point-patches}
Let $d\geq1$, and let a finite Abelian group $G$ act on a $d$-dimensional lattice by FDQCs implementing a $(d-1)$-form symmetry $S$. Suppose that for each closed parameter $z$, the action of $S(z)$ is supported in a microscopic neighborhood of the dual closed lines labeled by $z$. Then there exist compatible symmetry patches matching $S$. All such truncations give the same $[\omega]\in H^{d+2}(K(G,d),\RZ)$, and $S$ can be gauged if and only if $[\omega]=0$. On a connected one-dimensional chain, a nonzero closed $0$-cochain is constant on the whole chain, so an ordinary internal symmetry automatically satisfies the support assumption.
\end{theorem}

Appendix~\ref{app:patch-existence} constructs compatible truncations from local operators on segments, disks, and junctions. Appendix~\ref{app:patch-uniqueness} first proves Lemma~\ref{lem:relative-transport}, reducing the comparison of two truncations to the identity symmetry, and then shows that every compatible truncation of the identity symmetry gives trivial statistics. The facts about truncation to a segment used in both arguments are collected in Appendix~\ref{app:gnvw-toolkit}.

\begin{remark}
 In some cases symmetry patches suffice to read the anomaly before constructing the full family of gauge transformations.
 The T-junction in Fig.~\ref{fig:patch-tjunction-exchange} involves only a few local backgrounds. Its exchange phase can be read directly by comparing two histories. Choose preparations $P_i$ with a common far end, and let $F_i=P_iP_0^{-1}$ move the central defect to the end of arm $i$. The three moves through the center are $W_1=F_2F_3^{-1}$, $W_2=F_3F_1^{-1}$, and $W_3=F_1F_2^{-1}$. Start with a single defect and move it along $3\to2\to1\to3$. Each preparation cancels against its adjacent inverse, giving $W_2W_3W_1=\id$. Now place identical defects at $1$ and $3$, and apply $W_1,W_2,W_3$ in order. At each step the defect on the third arm lies outside the neighborhood read by that move, so the same $W_i$ applies. The three moves exchange the defects, and $W_3W_2W_1$ gives the exchange phase. Each $W_i$ occurs once in each history with the same local input, so the phases from a local basis change cancel pairwise. For a $\ZZ_N$ one-form symmetry in $(2+1)$ dimensions this gives an anomaly readout~\cite[Eqs.~(75)--(81) and (87)]{FengKobayashiChenRyu2026}.
	\begin{figure}[htbp]
\centering
\begin{tikzpicture}[
  x=1cm,y=1cm,>=Stealth,line cap=round,
  every node/.style={font=\small,text=ink,inner sep=1.5pt},
  arm/.style={draw=muted!50,line width=.65pt},
  move/.style={->,line width=1.05pt,shorten <=4pt,shorten >=4pt},
  defect/.style={circle,inner sep=0pt,minimum size=5.5pt,line width=.8pt},
  red defect/.style={defect,draw=speciesA,fill=speciesA},
  blue defect/.style={defect,draw=speciesB,fill=white}
]
  \foreach \offset/\word/\source/\target in {0/1/3/2,4.3/2/1/3,8.6/3/2/1}{
    \begin{scope}[xshift=\offset cm]
      \coordinate (v0) at (0,0);
      \coordinate (v1) at (-1.15,0);
      \coordinate (v2) at (1.15,0);
      \coordinate (v3) at (0,1.3);
      \foreach \v in {1,2,3}{\draw[arm] (v0)--(v\v);}
      \foreach \v in {0,1,2,3}{\fill[ink] (v\v) circle (1.1pt);}
      \node[below=3pt] at (v0) {$0$};
      \node[left=5pt] at (v1) {$1$};
      \node[right=5pt] at (v2) {$2$};
      \node[above=5pt] at (v3) {$3$};
      \node at (0,2.18) {$W_{\word}$};
      \node at (0,-.85) {$\source\to 0\to\target$};
      \ifnum\word=2
        \draw[move,speciesB] (v\source)--(v0);
        \draw[move,speciesB] (v0)--(v\target);
        \node[blue defect] at (v1) {};
        \node[red defect] at (v2) {};
      \else
        \draw[move,speciesA] (v\source)--(v0);
        \draw[move,speciesA] (v0)--(v\target);
        \node[red defect] at (v\source) {};
        \ifnum\word=1
          \node[blue defect] at (v1) {};
        \else
          \node[blue defect] at (v3) {};
        \fi
      \fi
    \end{scope}
  }
\end{tikzpicture}
\caption{Six-step exchange at a T-junction. The filled and open dots track two identical defects and show their positions at the start of each pair of moves. Each pair takes one defect through the center $0$ to the empty endpoint while the other remains at the far end of the gray arm. After all three pairs the defects have exchanged positions. On a closed space, any additional defects needed to place the configuration in the vacuum orbit can be placed outside the neighborhoods read by all six steps.}
\label{fig:patch-tjunction-exchange}
\end{figure}
\end{remark}

\section{A generalization of braiding}
\label{sec:geometry}

A family of commutator formulas generalizes anyon braiding and allows certain anomalies to be measured directly with symmetry patches.

Consider the two crossing arcs $L_1,L_2$ on $S^2$ in Fig.~\ref{fig:boundary-avoidance}(a). Represent them by $\ZZ_3$-valued $1$-cochains $s_i$, oriented so that $\int_{S^2}s_1\smile s_2=1\bmod 3$. The patch $P(s_i)$ along $L_i$ leaves a background $-\dd s_i$ at its two endpoints, both of which lie away from the other arc. Applying moves $1$ and $2$ in that order and returning along the other path therefore gives the gauge-transformation phase $[P(s_2),P(s_1)]$.

Let $\zeta=e^{2\pi\ii/3}$. For $\omega=A_1\smile A_2/3$, choosing $\Theta^1(\phi;A)=(\phi_1\smile D_{A_2}\phi_2+A_1\smile\phi_2)/3$ in Eq.~\eqref{eq:R-prism-definition} gives the following patches at zero background:

\begin{equation}
 \begin{aligned}
 P(s_1)\ket{\phi_1,\phi_2}&=\zeta^{\int_{S^2}s_1\smile\phi_2}\ket{\phi_1+s_1,\phi_2},\\
 P(s_2)\ket{\phi_1,\phi_2}&=\ket{\phi_1,\phi_2+s_2}.
 \end{aligned}
 \label{eq:parallel-R-kernels}
\end{equation}
Applying $P(s_2)$ first increases the exponent in $P(s_1)$ by $\int_{S^2}s_1\smile s_2$. Thus the specified order gives $[P(s_2),P(s_1)]=\zeta^{-1}\id$, the anomaly phase measured by the crossing arcs.

Now consider hopping along the same arcs, with the ordering in Eq.~\eqref{eq:intro-commutator}. Substituting $\Theta(\phi)=\phi_1\smile\dd\phi_2/3$ into the zero-background hopping formula~\eqref{eq:raw-hopping} gives
\begin{equation}
 \begin{aligned}
 U_1\ket{\phi_1,\phi_2}&=\ket{\phi_1+s_1,\phi_2},\\
 U_2\ket{\phi_1,\phi_2}&=\zeta^{\int_{S^2}\phi_1\smile s_2}\ket{\phi_1,\phi_2+s_2}.
 \end{aligned}
 \label{eq:parallel-U-kernels}
\end{equation}
Applying $U_1$ first increases the exponent in $U_2$ by the same intersection number, so $[U_2,U_1]=\zeta\,\id$. On these arcs the antilinear duality of Theorem~\ref{thm:dual-operator-algebras} pairs the two complex-conjugate phases $\zeta$ and $\zeta^{-1}$.

\subsection{Boundary avoidance and the general braiding phase}
Consider two species of excitations on a closed oriented $d$-dimensional space $X$, with dimensions $p_1,p_2$ and fusion groups $G_1,G_2$, respectively. Write their codimensions as $q=d-p_1$ and $q'=d-p_2$. A hop sweeps out a $(p_i+1)$-chain $L_i$ and creates excitations on $\partial L_i$. We describe these supports by cochains on a triangulation of $X$.

Choose the coarse-graining scale to exceed both the range of the microscopic operators and the distance over which they read the configuration. On the coarse lattice, $s_1\in C^{q-1}(X;G_1)$ and $s_2\in C^{q'-1}(X;G_2)$ describe $L_1$ and $L_2$, respectively, and $\dd s_i$ records the excitations on their boundaries.

The boundary-avoidance condition in Eq.~\eqref{eq:mutual-avoidance} has the following simplexwise formulation.
\begin{definition}[Boundary avoidance]
\label{def:boundary-avoidance}
Two cochains $s_1,s_2$ satisfy boundary avoidance if at least one of the following holds on every simplex $\Delta$ of $X$:
\begin{equation}
 \begin{aligned}
 \text{(1)}\quad &\dd(s_1|_\Delta)=\dd(s_2|_\Delta)=0
   &&\text{(both parameters are closed)};\\
 \text{(2)}\quad &s_2|_\Delta=0
   &&\text{($s_1|_\Delta$ is arbitrary)};\\
 \text{(3)}\quad &s_1|_\Delta=0
   &&\text{($s_2|_\Delta$ is arbitrary)}.
 \end{aligned}
 \label{eq:geometry-simplex-avoidance}
\end{equation}
\end{definition}
Figure~\ref{fig:cochain-braiding} shows a lattice realization of two crossing $\ZZ_2$ arcs.
\begin{figure}[htbp]
\centering
\begin{tikzpicture}[x=1.05cm,y=1.05cm,font=\small,line cap=round,line join=round]
  \fill[speciesA!17] (0,2)--(0,3)--(1,3)--cycle;
  \fill[speciesA!17] (4,2)--(5,2)--(5,3)--cycle;
  \fill[speciesB!17] (2,0)--(3,0)--(3,1)--cycle;
  \fill[speciesB!17] (2,4)--(2,5)--(3,5)--cycle;
  \fill[gray!8] (2,2) rectangle (3,3);

  \draw[step=1,gray!35,line width=.35pt] (0,0) grid (5,5);
  \foreach \xx in {0,...,4}
    \foreach \yy in {0,...,4}
      \draw[gray!35,line width=.35pt] (\xx,\yy)--({\xx+1},{\yy+1});
  \foreach \xx in {0,...,5}
    \foreach \yy in {0,...,5}
      \fill[gray!50] (\xx,\yy) circle (.8pt);

  \foreach \xx in {0,1,3,4}
    \draw[speciesA,line width=1.65pt] (\xx,2)--({\xx+1},3);
  \foreach \xx in {1,...,4}
    \draw[speciesA,line width=1.65pt] (\xx,2)--(\xx,3);
  \foreach \yy in {0,1,3,4}
    \draw[speciesB,line width=1.65pt] (2,\yy)--(3,{\yy+1});
  \foreach \yy in {1,...,4}
    \draw[speciesB,line width=1.65pt] (2,\yy)--(3,\yy);
  \draw[speciesA,line width=1.55pt] (1.96,2.04)--(2.96,3.04);
  \draw[speciesB,line width=1.55pt] (2.04,1.96)--(3.04,2.96);

  \node at (2.5,5.45) {$\ZZ_2$-valued $1$-cochains};
  \begin{scope}[shift={(6.05,0)}]
    \draw[speciesA,line width=1.65pt] (0,4.65)--(.65,4.65);
    \node[anchor=west] at (.85,4.65) {$s_1=1$};
    \draw[speciesB,line width=1.65pt] (2.1,4.65)--(2.75,4.65);
    \node[anchor=west] at (2.95,4.65) {$s_2=1$};

    \filldraw[fill=speciesA!17,draw=speciesA!55]
      (0,3.35)--(0,3.85)--(.55,3.85)--cycle;
    \node[anchor=west,align=left] at (.85,3.6)
      {$\dd s_1=1$ and $s_2|_\Delta=0$};
    \filldraw[fill=speciesB!17,draw=speciesB!55]
      (0,2.4)--(.55,2.4)--(.55,2.9)--cycle;
    \node[anchor=west,align=left] at (.85,2.65)
      {$\dd s_2=1$ and $s_1|_\Delta=0$};
    \filldraw[fill=gray!8,draw=gray!40] (0,1.45) rectangle (.55,2);
    \node[anchor=west,align=left] at (.85,1.72)
      {Elsewhere, $\dd s_1=\dd s_2=0$};
  \end{scope}
\end{tikzpicture}
\caption{Cochain model of two crossing dual strings. The colored segments show cochain values on edges. Red and blue shading identify the endpoint triangles where $\dd s_1$ and $\dd s_2$, respectively, are nonzero. The other parameter vanishes on every edge of each such triangle. All other triangles have $\dd s_1=\dd s_2=0$, including the two at the crossing, each of which has two red edges and two blue edges.}
\label{fig:cochain-braiding}
\end{figure}

For any degree $k$, let $L(G,k)_n=C^{k-1}(\Delta^n;G)$. The closed cochains form the simplicial subset $K(G,k-1)$, and the zero cochain gives the basepoint. Restricting $s_1,s_2$ to each simplex defines simplicial maps from $X$ to $L(G_1,q)$ and $L(G_2,q')$, respectively. The three cases of Definition~\ref{def:boundary-avoidance} correspond, in order, to the three simplicial subsets in

\begin{equation}
 \begin{aligned}
 \mathcal P={}&\bigl(K(G_1,q-1)\times K(G_2,q'-1)\bigr)\\
 &\cup\bigl(L(G_1,q)\times\{0\}\bigr)
 \cup\bigl(\{0\}\times L(G_2,q')\bigr).
 \end{aligned}
 \label{eq:geometry-cochain-target}
\end{equation}
Consequently, pairs of cochains satisfying boundary avoidance correspond precisely to simplicial maps into $\mathcal P$:

\begin{equation}
	\Hom(X,\mathcal P)
	\cong\bigl\{(s_1,s_2)\ \big|\ \text{Eq.~\eqref{eq:geometry-simplex-avoidance} holds}\bigr\}.
	\label{eq:geometry-classification}
\end{equation}

Each pair in Eq.~\eqref{eq:geometry-classification} determines a homotopy class of maps between geometric realizations. Write $[X,\mathcal P]$ for the set of all such homotopy classes. A fixed triangulation gives the classes that have simplicial representatives on that triangulation. Allowing subdivisions of $X$ realizes every class. To compute $[X,\mathcal P]$, observe that
\[
 \bigl(L(G_1,q)\times\{0\}\bigr)
 \cup\bigl(\{0\}\times L(G_2,q')\bigr)
\]
is a contractible simplicial subset of $\mathcal{P}$. Collapsing this subset identifies the two coordinate axes of $K(G_1,q-1)\times K(G_2,q'-1)$ to a point. The resulting quotient is the smash product, and the quotient map is a homotopy equivalence:
\begin{equation}
 \mathcal P
 \xrightarrow{\;\simeq\;} K(G_1,q-1)\wedge K(G_2,q'-1).
 \label{eq:geometry-smash-quotient}
\end{equation}
Thus braiding geometries up to homotopy are classified by
\begin{equation}
	[X,K(G_1,q-1)\wedge K(G_2,q'-1)].
\end{equation}
When $S^d$ is the one-point compactification of $\RR^d$, moves with bounded support satisfy $s_1=s_2=0$ near the point at infinity $x_\infty$. Homotopies then fix the image of $x_\infty$ at the basepoint, and the classification is $\pi_d(K(G_1,q-1)\wedge K(G_2,q'-1))$.

For the two particle species moving along the arcs in Fig.~\ref{fig:boundary-avoidance}(a), $d=q=q'=2$. The space $BG_1\wedge BG_2$ is simply connected, so the Hurewicz theorem and the reduced K\"unneth formula give
\[
 \pi_2(BG_1\wedge BG_2)\cong H_2(BG_1\wedge BG_2;\ZZ)\cong G_1\otimes G_2.
\]

We now compute the relative phase between the two orders of hopping along the same supports. At zero background, let $U_1(s_1)$ and $U_2(s_2)$ move only the first and second species, respectively. The two moves are interpolated independently, and the cochains $s_1,s_2$ can have different degrees. The statistics cocycle $\omega$ is therefore defined on $K(G_1,q)\times K(G_2,q')$. Its two arguments are the spacetime configurations of the two species, represented by cocycles of degrees $q$ and $q'$.

The two histories form the sides of the square in Fig.~\ref{fig:parameter-square}. The horizontal move is $s_1$, with $\eta_1$ equal to $0$ on the left and $1$ on the right. The vertical move is $s_2$, with $\eta_2$ equal to $0$ at the bottom and $1$ at the top. Interpolating each move as in Eq.~\eqref{eq:raw-hopping} gives the matter fields $\tilde\phi_1=\eta_1\smile s_1$ and $\tilde\phi_2=\eta_2\smile s_2$ on the square. Along the right edge, for example, the first component stays at $s_1$ while the second changes from $0$ to $s_2$.

\begin{figure}[htbp]
\centering
\begin{tikzpicture}[>=Latex,x=2.6cm,y=1.4cm]
\fill[gray!5] (0,0) rectangle (3,2);
\node[below left] at (0,0){$(0,0)$};
\node[above left] at (0,2){$(0,s_2)$};
\node[above right] at (3,2){$(s_1,s_2)$};
\node[below right] at (3,0){$(s_1,0)$};
\node[above=3pt,font=\small,text=muted] at (1.5,0) {$\eta_2=0$};
\node[below=3pt,font=\small,text=muted] at (1.5,2) {$\eta_2=1$};
\node[right=3pt,font=\small,text=muted] at (0,1) {$\eta_1=0$};
\node[left=3pt,font=\small,text=muted] at (3,1) {$\eta_1=1$};
\draw[->,speciesA,thick] (0,0)--node[below]{$U_1(s_1)$}(3,0);
\draw[->,speciesB,thick] (3,0)--node[right]{$U_2(s_2)$}(3,2);
\draw[->,speciesA,thick] (3,2)--node[above]{$U_1(s_1)^{-1}$}(0,2);
\draw[->,speciesB,thick] (0,2)--node[left]{$U_2(s_2)^{-1}$}(0,0);
\end{tikzpicture}
\caption{Two histories from the vacuum to $(s_1,s_2)$. Applying $s_1$ and then $s_2$ follows the bottom and right edges. Applying $s_2$ and then $s_1$ follows the left and top edges. The return path uses the inverses of the second history's operators. The four arrows specify the order of the braiding moves.}
\label{fig:parameter-square}
\end{figure}

Following the bottom, right, top, and left edges in order gives the commutator $U_2^{-1}U_1^{-1}U_2U_1$. By Eq.~\eqref{eq:raw-hopping}, its phase on the zero initial configuration is $\exp[-2\pi\ii\int_{\partial I^2\times X}\Theta(\tilde\phi_1,\tilde\phi_2)]$. Since $X$ is closed, Stokes' theorem and the descent relation~\eqref{eq:first-descent} give
\[
 \int_{\partial I^2\times X}\Theta(\tilde\phi_1,\tilde\phi_2)
 =\int_{I^2\times X}\dd\Theta(\tilde\phi_1,\tilde\phi_2)
 =\int_{I^2\times X}\omega(\dd\tilde\phi_1,\dd\tilde\phi_2).
\]
Integrating over the square alone gives a $d$-cochain on $X$,
\begin{equation}
 \boxed{\Phi_\omega(s_1,s_2):=\int_{I^2}
 \omega\bigl(\dd(\eta_1\smile s_1),\,\dd(\eta_2\smile s_2)\bigr).}
 \label{eq:prism-braiding-cocycle}
\end{equation}
On each $d$-simplex $\Delta$ of $X$, this integral is the oriented sum of the values of $\omega$ on the top-dimensional simplices of $I^2\times\Delta$. For any $(d+1)$-simplex $\Delta$ of $\mathcal P$, the cocycle condition $\dd\omega=0$ and Stokes' theorem give
\[
 \langle\dd\Phi_\omega,\Delta\rangle
 =\int_{I^2\times\partial\Delta}\omega(\dd\tilde\phi_1,\dd\tilde\phi_2)
 =-\int_{\partial I^2\times\Delta}\omega(\dd\tilde\phi_1,\dd\tilde\phi_2).
\]
Boundary avoidance makes the integral over the boundary of the square vanish, so $\dd\Phi_\omega=0$.

\begin{theorem}
\label{thm:braiding-phase}
Let $X$ be a closed oriented $d$-dimensional space, and let $s_1,s_2$ move two species of excitations and satisfy boundary avoidance. At zero reference background, the hopping operators constructed in Eq.~\eqref{eq:raw-hopping} satisfy
\begin{equation}
 [U_2(s_2),U_1(s_1)]
 =\exp\!\left[-2\pi\ii\int_X\Phi_\omega(s_1,s_2)\right]\id.
 \label{eq:braiding-phase}
\end{equation}
\end{theorem}
\begin{proof}
The forward and reverse steps of move $i$ in Eq.~\eqref{eq:hopping-local-choice} read configurations that differ by $\dd s_j$ ($j\ne i$). On every simplex where $s_i|_\Delta\ne0$, boundary avoidance~\eqref{eq:geometry-simplex-avoidance} gives $(\dd s_j)|_\Delta=0$. Thus the phases contributed by a local basis change cancel between the two steps. The commutator is therefore a statistical process, and Theorem~\ref{thm:process-initial-state} makes its phase constant on each configuration orbit. Evaluating at $\phi=0$ in the vacuum orbit gives Eq.~\eqref{eq:braiding-phase} by the square integral above.
\end{proof}

When $\omega$ is replaced by $\omega+\dd\lambda$, boundary avoidance again cancels the boundary contributions from the square in Stokes' theorem. Hence on $\mathcal P$,
\[
 \Phi_{\omega+\dd\lambda}-\Phi_\omega
 =\dd\!\int_{I^2}\lambda(\dd\tilde\phi_1,\dd\tilde\phi_2).
\]
If $f:X\to\mathcal P$ represents $(s_1,s_2)$, then $[\omega]\mapsto[\Phi_\omega]\mapsto\langle[\Phi_\omega],f_*[X]\rangle$ gives first the response class and then the integral in Eq.~\eqref{eq:braiding-phase}. A statistics class in the kernel of $\Phi$ therefore gives a trivial phase on every allowed geometry. Likewise, a geometry with $f_*[X]=0$ gives a trivial phase for every statistics class.

Since $\RZ$ is divisible, every nonzero $[\Phi_\omega]\in H^d(\mathcal P,\RZ)$ pairs nontrivially with some integral homology class of $\mathcal P$. For a fixed physical space $X$, the measurable pairings are those with classes of the form $f_*[X]$ for some $f:X\to\mathcal P$.

Only simplices on which both $s_1$ and $s_2$ are present contribute to Eq.~\eqref{eq:prism-braiding-cocycle}. On every other simplex, the fields are independent of one parameter direction, and the square integral vanishes. Where both cochains are present, boundary avoidance gives $\dd s_1=\dd s_2=0$. Hence $\dd(\eta_1\smile s_1)=\dd\eta_1\smile s_1$ and $\dd(\eta_2\smile s_2)=\dd\eta_2\smile s_2$. The configuration fields now agree on opposite edges of the square, so we can identify the ends of each interval. Then $\dd\eta_i$ becomes a degree-one cocycle $e_i$ on the $i$th circle, with $\int_{S^1}e_i=1$.

It therefore suffices to substitute $e_1s_1,e_2s_2$ into $\omega$ and move $e_1,e_2$ to the far left in that order. Integration over the two circles extracts the coefficient of $e_1e_2$. Suppressing the cup products, we can write this rule for a representative as
\begin{equation}
 \Phi_\omega(s_1,s_2)\ \sim\
 \operatorname{coeff}_{e_1e_2}\!\bigl(\omega(e_1s_1,e_2s_2)\bigr).
 \label{eq:braiding-coefficient-rule}
\end{equation}
Juxtaposition denotes the cup product, and the coefficient is taken using multiplication in cohomology. The symbol $\sim$ means that the right-hand side gives a cochain representative of the same response. Evaluate it on the original $s_1,s_2$ where both moves are present, set it to zero on the remaining simplices, and integrate over $X$.

Return to the geometry of two disks meeting transversely at one point. For $\omega=kA_1\smile A_2/N$ and $d=q+q'-2$, substitute $e_1s_1,e_2s_2$ into Eq.~\eqref{eq:braiding-coefficient-rule}. Moving $e_2$ past $s_1$ gives $(-1)^{q-1}$. Combining this with the minus sign in Eq.~\eqref{eq:braiding-phase} gives the phase in terms of the oriented intersection number,
\begin{equation}
 [U_2(s_2),U_1(s_1)]=\exp\!\left[
 \frac{2\pi\ii k(-1)^{q}}{N}\int_X s_1\smile s_2\right]\id.
 \label{eq:ordinary-braiding-readout}
\end{equation}
Two disks of dimensions $(q'-1)$ and $(q-1)$ with oriented intersection number $+1$ therefore give the phase $e^{2\pi\ii k(-1)^{q}/N}$. For $q=q'=2$, taking $N=3$ and $k=1$ recovers the hopping commutator phase $\zeta$ obtained from Eq.~\eqref{eq:parallel-U-kernels}.

The cochains $s_1,s_2$ describe open supports and need only be closed in the region where both act. Their intersection number can therefore be nonzero even when the lower-degree cohomology of the sphere vanishes.

\subsection{Constraints on braiding phases from boundary avoidance}
\label{subsec:braiding-consequences}

\begin{proposition}
\label{prop:braiding-paired-readouts}
At zero reference background, consider two moves satisfying boundary avoidance. The commutator of the paired symmetry patches is the inverse of the hopping commutator with the same labels. Simultaneously negating both hopping labels leaves the phase unchanged.
\end{proposition}

Starting at zero background, apply $u_1$ and $u_2$ in order, then undo $u_1$ and $u_2$ in the same order. At every step, the background left by the other move lies outside the neighborhood read by the current move. With the conventions of Sec.~\ref{sec:patch-compatibility}, we may therefore use the zero-background patch for each forward step and its inverse for the return step:
\begin{equation}
 T_0(u_2)^{-1}\,T_{-\dd u_2}(u_1)^{-1}\,
 T_{-\dd u_1}(u_2)\,T_0(u_1)
 =[P(u_2),P(u_1)] .
 \label{eq:background-R-word}
\end{equation}

Equation~\eqref{eq:dual-operator-expansion} pairs each $T(u_i)$ with $U(-u_i)$ and complex conjugates scalar phases. If $\theta_U,\theta_P$ denote the two commutator phases in units of $2\pi$, then
\begin{equation}
 \theta_P(u_1,u_2)=-\theta_U(-u_1,-u_2).
 \label{eq:paired-antiparticle-braiding}
\end{equation}
In Eq.~\eqref{eq:braiding-coefficient-rule}, negating $u_1,u_2$ simultaneously is equivalent to negating the generators $e_1,e_2$ of the two parameter circles. Reversing both circles implements this substitution and preserves the orientation of $S^1\times S^1$, so the class obtained by integration is unchanged. This establishes the equality on the product space of the two universal closed cochains. The reduced classes on either coordinate axis extend by projection, so the quotient map from the product to the smash product induces an injection on reduced cohomology. By Eq.~\eqref{eq:geometry-smash-quotient}, the equality thus holds on all of $\mathcal P$. Evaluation on the given geometry using Eq.~\eqref{eq:braiding-phase} yields $\theta_U(-u_1,-u_2)=\theta_U(u_1,u_2)$. Substituting into Eq.~\eqref{eq:paired-antiparticle-braiding} gives
\begin{equation}
 [P(u_2),P(u_1)]=[U_0(u_2),U_0(u_1)]^{-1}.
 \label{eq:inverse-operator}
\end{equation}
Equation~\eqref{eq:background-R-word} expresses the patch commutator as four gauge transformations. Composing them using Eq.~\eqref{eq:normalized-R-multiplication} gives exactly the phase associated with $\omega$ in Eq.~\eqref{eq:closed-prism-comparison} for this braiding process.

\begin{corollary}
\label{cor:braiding-charge-conjugation}
Let $\iota$ negate the group labels of both species. The statistics classes obey
\[
 [\Phi_{\iota^*\omega}]=[\Phi_\omega],\qquad
 [\Phi_{\omega-\iota^*\omega}]=0.
\]
In particular, if $[\omega]$ has odd order and $\iota^*[\omega]=-[\omega]$, then $[\Phi_\omega]=0$.
\end{corollary}

Inserting $\iota^*\omega$ into the square integral negates both input labels, so the first identity follows from orientation invariance as above. Linearity of integration gives the second. In the odd-order case, both $2$ and the order of $[\omega]$ annihilate $[\Phi_\omega]$, so it vanishes.

\begin{corollary}
\label{cor:braiding-double-role-phase}
Suppose that a family of hopping operators with statistics class $[\omega]$ also realizes compatible patches of a symmetry with anomaly $[\nu]$ in the interiors of the supports and at their fusion junctions. Then on every geometry satisfying boundary avoidance,
\[
 \int_X\Phi_\omega(s_1,s_2)=-\int_X\Phi_\nu(s_1,s_2)\pmod1.
\]
\end{corollary}
A change of truncation at the boundary of either support is disjoint from the entire other support, so the patch commutator agrees with the commutator of these compatible representatives. Evaluating this commutator using the statistics phase~\eqref{eq:braiding-phase} and the paired patch phase~\eqref{eq:inverse-operator} gives the claim.

\begin{corollary}
\label{cor:symmetric-patch-braiding}
Suppose the hopping operators for both species are assembled from local moves and local configuration phases that commute with the same symmetries. If these hopping operators also serve as patches of those symmetries, their generalized braiding is $\id$ whenever boundary avoidance holds.
\end{corollary}
Write the second hopping operator as a product of local symmetric hopping operators and local configuration phases. Boundary avoidance ensures that the parameter of the first patch is closed throughout the neighborhood read by each step. We take this neighborhood to be contractible, so the locally closed parameter extends to a globally closed one. Within this neighborhood the first patch acts by conjugation in the same way as the corresponding global symmetry and therefore commutes with the step. Multiplying over all steps gives the identity commutator. This also applies to constructions that require every compatible patch to commute with the same global symmetries. By Eq.~\eqref{eq:patch-reconstruct}, the product $P(r+e)P(r)^{-1}$ of adjacent truncations defines a local increment. If the two truncations are symmetric, so is this increment. Assembling the increments along the support gives the local commutativity argument of the corollary.

Returning to the crossing arcs in Fig.~\ref{fig:boundary-avoidance}(a), the hopping operators along these arcs can also serve as patches of another symmetry. Let $P_i^{-\omega}(s_i)$ be the patch obtained by replacing $\omega$ with $-\omega$ in Eq.~\eqref{eq:parallel-R-kernels}, and let $V$ be the local circuit that applies the diagonal phase below simplex by simplex. Comparing Eqs.~\eqref{eq:parallel-U-kernels} and~\eqref{eq:parallel-R-kernels} term by term gives
\[
 V\ket{\phi}=e^{\frac{2\pi\ii}{3}\int_X\phi_1\smile\phi_2}\ket{\phi},
 \qquad
 U_i=V P_i^{-\omega}(s_i)V^{-1}\quad(i=1,2).
\]
These hopping operators commute with the original symmetry. In this example the diagonal circuit $V$ maps the transformations of a symmetry with anomaly $-[\omega]$ to the operators $U_i$ with closed parameters.

\subsection{Two M\"obius bands and the Bockstein response}
\label{subsec:mobius-detector}
\label{subsec:bockstein-detectors}
\begin{figure}[htbp]
\centering
\includegraphics{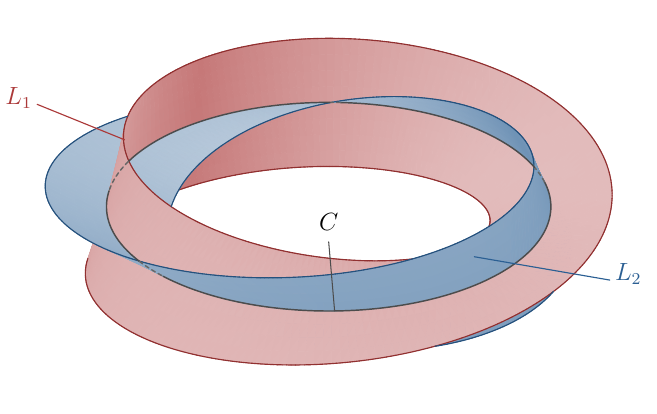}
\caption{A pair of M\"obius bands in three dimensions. The red band $L_1$ and blue band $L_2$ meet only along the core circle $C$. The outer edge of each band avoids the other band. The dark gray curve marks the intersection circle, with hidden segments dashed. The red and blue outlines mark the visible physical boundaries.}
\label{fig:intro-mobius}
\end{figure}

The two M\"obius bands in Fig.~\ref{fig:intro-mobius} meet along their common core circle $C$ in three-dimensional space. The boundary of each avoids the other band. Acting on the vacuum, $U_i(L_i)$ is supported on the entire band and creates a loop on its boundary $\partial L_i$. Both species of loops have fusion group $\ZZ_2$, so orientation reversal within a band leaves the label unchanged.

Take $q=q'=2$. Imagine a point moving around the core circle $C$ as in Fig.~\ref{fig:mobius-cross-transport}, carrying two orthogonal unit vectors normal to its direction of motion. When the point returns, the frame has rotated through $180^\circ$ in the normal plane, so both vectors point opposite to their initial directions. A short line segment centered on the moving point and aligned with either vector sweeps out a M\"obius band.
\begin{figure}[htbp]
\centering
\begin{tikzpicture}[>=Latex, font=\small,
  frame/.style={-{Latex[length=2.2mm,width=1.6mm]},line width=1.3pt},
  startframe/.style={frame,densely dashed}]
  \begin{scope}[shift={(-3.4,-3.15)}]
    \def\coreR{2.15}
    \def\arm{.72}
    \def\cx{3.4}
    \def\cy{3.15}

    \node at (3.4,5.78) {One circuit around the core $C$};
    \path[fill=gray!3]
      plot[domain=0:360,samples=121,variable=\ang]
        ({\cx+\coreR*(cos(\ang)+.42*sin(\ang))},
         {\cy+.5*\coreR*sin(\ang)}) -- cycle;
    \draw[gray!65,densely dashed]
      plot[domain=0:180,samples=61,variable=\ang]
        ({\cx+\coreR*(cos(\ang)+.42*sin(\ang))},
         {\cy+.5*\coreR*sin(\ang)});
    \draw[gray!65]
      plot[domain=180:360,samples=61,variable=\ang]
        ({\cx+\coreR*(cos(\ang)+.42*sin(\ang))},
         {\cy+.5*\coreR*sin(\ang)});
    \draw[->,gray!75]
      plot[domain=305:331,samples=18,variable=\ang]
        ({\cx+\coreR*(cos(\ang)+.42*sin(\ang))},
         {\cy+.5*\coreR*sin(\ang)});
    \node[gray!80,below] at (4.4,2.3) {$C$};

    \draw[speciesA!75,densely dashed,thin]
      plot[domain=0:360,samples=161,variable=\ang]
        ({\cx+(\coreR+\arm*cos(\ang/2))*(cos(\ang)+.42*sin(\ang))},
         {\cy+.5*(\coreR+\arm*cos(\ang/2))*sin(\ang)+.86*\arm*sin(\ang/2)});
    \draw[speciesB!75,densely dashed,thin]
      plot[domain=0:360,samples=161,variable=\ang]
        ({\cx+(\coreR-\arm*sin(\ang/2))*(cos(\ang)+.42*sin(\ang))},
         {\cy+.5*(\coreR-\arm*sin(\ang/2))*sin(\ang)+.86*\arm*cos(\ang/2)});
    \draw[->,speciesA!90,thin]
      plot[domain=110:133,samples=15,variable=\ang]
        ({\cx+(\coreR+\arm*cos(\ang/2))*(cos(\ang)+.42*sin(\ang))},
         {\cy+.5*(\coreR+\arm*cos(\ang/2))*sin(\ang)+.86*\arm*sin(\ang/2)});
    \draw[->,speciesB!90,thin]
      plot[domain=210:235,samples=15,variable=\ang]
        ({\cx+(\coreR-\arm*sin(\ang/2))*(cos(\ang)+.42*sin(\ang))},
         {\cy+.5*(\coreR-\arm*sin(\ang/2))*sin(\ang)+.86*\arm*cos(\ang/2)});

    \foreach \ang in {0,90,180,270} {
      \pgfmathsetmacro\px{\cx+\coreR*(cos(\ang)+.42*sin(\ang))}
      \pgfmathsetmacro\py{\cy+.5*\coreR*sin(\ang)}
      \pgfmathsetmacro\rx{\arm*cos(\ang/2)*(cos(\ang)+.42*sin(\ang))}
      \pgfmathsetmacro\ry{\arm*(.5*cos(\ang/2)*sin(\ang)+.86*sin(\ang/2))}
      \pgfmathsetmacro\bx{-\arm*sin(\ang/2)*(cos(\ang)+.42*sin(\ang))}
      \pgfmathsetmacro\by{\arm*(-.5*sin(\ang/2)*sin(\ang)+.86*cos(\ang/2))}
      \draw[speciesA,line width=1.6pt] ({\px-\rx},{\py-\ry})--({\px+\rx},{\py+\ry});
      \draw[speciesB,line width=1.6pt] ({\px-\bx},{\py-\by})--({\px+\bx},{\py+\by});
      \fill[speciesA] ({\px+\rx},{\py+\ry}) circle (2.6pt);
      \fill[speciesB] ({\px+\bx},{\py+\by}) circle (2.6pt);
      \fill[gray!70] (\px,\py) circle (1.2pt);
    }
    \draw[speciesA,fill=white,line width=1pt]
      ({\cx+\coreR+\arm},\cy) circle (2.8pt);
    \draw[speciesB,fill=white,line width=1pt]
      ({\cx+\coreR},{\cy+.86*\arm}) circle (2.8pt);
    \fill[speciesA] ({\cx+\coreR-\arm},\cy) circle (2.6pt);
    \fill[speciesB] ({\cx+\coreR},{\cy-.86*\arm}) circle (2.6pt);
    \node at (4.3,5.2) {$\theta=\pi/2$};
    \node[anchor=east] at (.8,3.8) {$\pi$};
    \node at (2.5,1.05) {$3\pi/2$};
    \node[anchor=west,align=left] at (6.4,3.15) {$0$\\[-2pt]$2\pi$};
  \end{scope}
  \begin{scope}[shift={(6.65,0)}]
    \node at (0,2.63) {Tangent view at the starting point};
    \draw[startframe,speciesA] (0,0)--(1.05,0);
    \draw[startframe,speciesB] (0,0)--(0,1.05);
    \draw[frame,speciesA] (0,0)--(-1.05,0);
    \draw[frame,speciesB] (0,0)--(0,-1.05);
    \draw[black!60,-{Latex[length=2mm]}] (1.28,0)
      arc[start angle=0,end angle=180,radius=1.28];
    \node[fill=white,inner sep=1.5pt] at (0,1.3) {$\pi$};
    \fill (0,0) circle (1.7pt);
  \end{scope}
\end{tikzpicture}
\caption{The center of the cross on the left moves along the core circle $C$. The red and blue dashed curves trace the tips of the two frame vectors. Open dots mark their initial positions. On the right, the initial frame (dashed) and the returning frame (solid) are superimposed in the same normal plane at the starting point. Both vectors have rotated by $180^\circ$ after one circuit.}
\label{fig:mobius-cross-transport}
\end{figure}

The generalized statistics of these two species of loops are classified by $H^5(K(\ZZ_2,2)\times K(\ZZ_2,2),\RZ)$~\cite{Xue2026Statistics}. Consider the statistics class
\begin{equation}
 \omega=\frac12A_1\Sq^1A_2,
 \label{eq:mobius-statistics-class}
\end{equation}
and extract the coefficient as in Eq.~\eqref{eq:braiding-coefficient-rule}. Since $\Sq^1 e_2=e_2^2=0$ on the circle, the Cartan formula gives $\Sq^1(e_2s_2)=e_2\Sq^1s_2$, so the phase exponent is $\frac12\int_{S^3}s_1\Sq^1s_2$. We can compute this integral in the normal directions to the core circle. Its rank-two normal bundle splits into two M\"obius line bundles $E\oplus F$, and $s_1,s_2$ are relative Thom cochains in the $F,E$ directions, respectively. The Thom identity gives
\begin{equation}
 \int_{S^3}s_1\Sq^1s_2
 =\langle w_1(E),[C]\rangle=1\pmod2,
 \qquad [U_2(s_2),U_1(s_1)]=-\id.
 \label{eq:mobius-readout}
\end{equation}
By Corollary~\ref{cor:symmetric-patch-braiding}, these hopping operators cannot simultaneously serve as patches of the symmetry they preserve. The measured phase has order two, so complex conjugation leaves $-1$ fixed. Geometrically, a small push-off of one M\"obius band is specified along its core by a section of a twisted normal line. A generic section has an odd number of zeros and can be chosen to have exactly one. This zero gives the intersection counted by the phase. The linking number of the two outer boundaries is $\pm2$, in agreement with Eq.~\eqref{eq:book-linking-calibration} below for $N=2$.

When $\dim\partial L_1>0$, one can also consider a third kind of perturbation: multiplying by a hopping operator for a lower-dimensional excitation, such as a point excitation, inside $L_1$. This decorates $\partial L_1$ with lower-dimensional excitations. In ordinary braiding, where two disks intersect transversely at one point, the braiding of an attached point excitation with the other excitation is always topologically trivial. For the two M\"obius bands in Fig.~\ref{fig:intro-mobius}, however, the inserted hopping operator can transport along $L_1$ a point excitation transforming in the nontrivial one-dimensional representation of $\ZZ_2$. Its ordinary braiding with $L_2$ can then contribute a nontrivial phase. Both notions of robustness considered here hold these lower-dimensional decorations fixed.

To obtain an arbitrary cyclic order, let the moving point carry two sets of $N$ equally spaced rays, each rotating by $2\pi/N$ in one trip around the core. Figure~\ref{fig:bockstein-book-sections} shows the initial and final positions of a ray for $N=3$. If $\theta$ is the angular coordinate along the core, the first set of rays in the normal plane is given by
\begin{equation}
 N\alpha\equiv\theta\pmod{2\pi},\qquad z=r e^{\ii\alpha},
 \qquad 0\le r\le\varepsilon,
 \label{eq:bockstein-book-parameterization}
\end{equation}
and the second set is rotated as a whole by $\pi/N$. Each set sweeps out a support consisting locally of $N$ sheets joined along the core. One trip around the core cyclically permutes the sheets. The two supports meet only on the core circle. The $N$ contributions to the integral boundary there vanish modulo $N$, expressing fusion, while the physical excitation lies on the outer boundary. For $N=2$ the support is a smooth M\"obius band. For $N>2$ it is a surface branched along its core circle.

\begin{figure}[htbp]
\centering
\begin{tikzpicture}[x=1cm,y=1cm,>=Latex,font=\small]
  \fill[gray!4] (0,0) circle (1.12);
  \foreach \j in {0,1,2}{
    \draw[speciesA,line width=1.7pt] (0,0)--({120*\j}:1.12);
    \draw[speciesB,line width=1.7pt] (0,0)--({60+120*\j}:1.12);
    \fill[speciesA] ({120*\j}:1.12) circle (2.5pt);
    \fill[speciesB] ({60+120*\j}:1.12) circle (2.5pt);
  }
  \fill[ink] (0,0) circle (1.8pt);
  \foreach \a in {0,120}{
    \draw[muted,densely dotted] (\a:1.27)--(\a:1.48);
    \draw[ink,line width=.65pt] (\a:1.12) circle (4.2pt);
  }
  \draw[ink,->,line width=.85pt] (0:1.48) arc[start angle=0,end angle=120,radius=1.48];
  \node[anchor=south west,text=ink] at (55:1.52) {$2\pi/3$};
  \node[anchor=west,text=muted] at (1.59,0) {$\theta=0$};
  \node[anchor=east,text=muted] at (-.86,1.31) {$\theta=2\pi$};
\end{tikzpicture}
\caption{Normal sections of the two branched surfaces for $N=3$. The three red rays alternate with the three blue rays, offset by $\pi/3$.
In one circuit around the core ($\theta=0$ to $2\pi$), the point on the red outer edge tracked by the curved arrow rotates through $2\pi/3$
and reaches the next red ray. The blue outer-edge point rotates through the same angle, cyclically permuting the sheets of each surface.}
\label{fig:bockstein-book-sections}
\end{figure}

Take $\ZZ_N$ coefficients and the Bockstein statistics class
\begin{equation}
 \omega_\beta=\frac{k}{N}A_1\beta_N A_2
 \ \in H^{q+q'+1}(K(\ZZ_N,q)\times K(\ZZ_N,q'),\RZ).
 \label{eq:bockstein-detector-class}
\end{equation}
Where both moves are present, $\beta_N s_2=\dd\bar s_2/N$, where the bar denotes the standard integer lift. Since $e_2$ has a closed integral representative, $e_2\bar s_2$ lifts $e_2s_2$. Taking its coboundary and dividing by $N$ gives $\beta_N(e_2s_2)=-e_2\beta_N s_2$. Equation~\eqref{eq:braiding-coefficient-rule} therefore gives
\begin{equation}
 \Phi_{\omega_\beta}(s_1,s_2)\ \sim\ (-1)^{q}\frac{k}{N^2}\,\bar s_1\smile\dd\bar s_2
 \pmod 1 .
 \label{eq:bock-family-readout}
\end{equation}
On a simplex where $\dd s_2\ne0\bmod N$, boundary avoidance forces $s_1=0$, so this expression vanishes there as well. In the three-dimensional geometry above, $q=q'=2$, and hence
\begin{equation}
 [U_2(s_2),U_1(s_1)]=\exp\!\left[-\frac{2\pi\ii k}{N^2}
 \int_X\bar s_1\smile\dd\bar s_2\right]\id.
 \label{eq:bockstein-detector-phase}
\end{equation}
The outer boundaries of the two supports are parallel $(N,1)$ curves with linking number $\pm N$. To compare this with Eq.~\eqref{eq:bockstein-detector-phase}, choose integral Poincar\'e dual cocycles $a,b$ for the outer boundaries that lift $\dd s_1,\dd s_2$, respectively, and preserve boundary avoidance. On $S^3$ we can choose integral primitives $\dd\alpha=a,\dd\beta=b$ such that $\alpha\equiv s_1,\beta\equiv s_2\pmod N$. Write $\bar s_1=\alpha+Nx$ and $\bar s_2=\beta+Ny$. Boundary avoidance gives $\bar s_1\smile b=a\smile\bar s_2=0$. Expanding the integral and applying Stokes' theorem gives
\begin{equation}
 \frac{1}{N^2}\int_X\bar s_1\smile\dd\bar s_2
 \equiv -\frac{1}{N^2}\int_X a\smile\beta
 =-\frac{\operatorname{Lk}(\partial L_1,\partial L_2)}{N^2}\pmod 1.
 \label{eq:book-linking-calibration}
\end{equation}
Here the orientations are chosen so that $\operatorname{Lk}(\partial L_1,\partial L_2)=\int_X a\smile\beta$. For $k=1$, Eq.~\eqref{eq:bockstein-detector-phase} gives a primitive $N$th root of unity. If the same hopping operators also realize compatible patches with Bockstein coefficient $\ell$, then Corollary~\ref{cor:braiding-double-role-phase} requires $\ell=-k\bmod N$.

Adding $q'-2$ and $q-2$ ordinary disk directions to the respective supports preserves the phase by the integral Thom isomorphism. This gives a probe of the same order on the sphere $S^{q+q'-1}$:
\begin{equation}
 \omega_\beta=\frac{k}{N}A_1\beta_N A_2,
 \qquad d=q+q'-1,\qquad q,q'\ge2.
 \label{eq:bockstein-all-degrees}
\end{equation}

\begin{figure}[htbp]
\centering
\includegraphics[width=.76\linewidth]{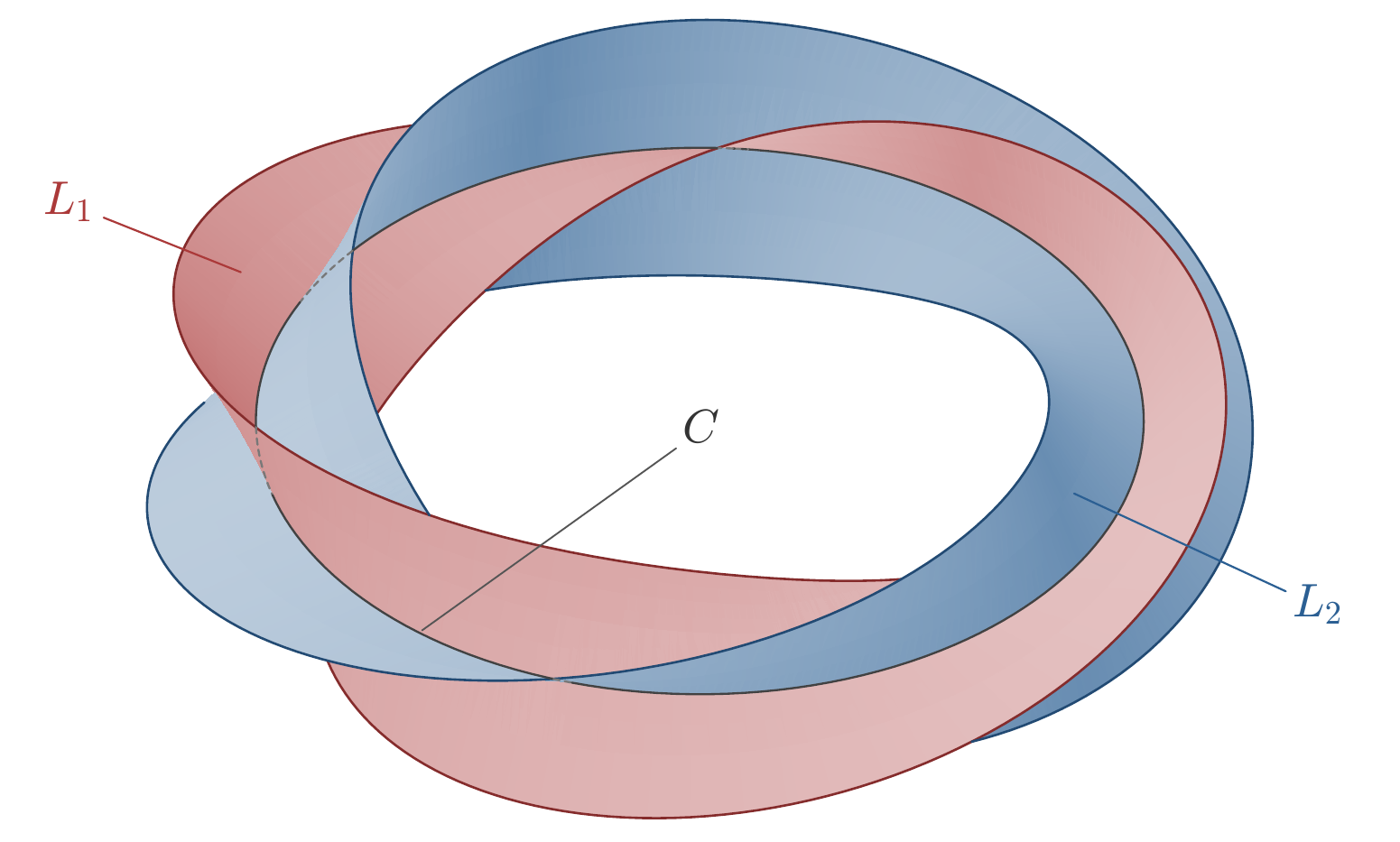}
\caption{Two ribbons whose normal frame turns through $360^\circ$. The red ribbon $L_1$ and blue ribbon $L_2$ are annuli.
They meet only along the core circle $C$, and the boundary of each ribbon avoids the other ribbon. The two shades on opposite faces reveal the twist.
The dark gray curve marks the intersection circle, with hidden segments dashed.}
\label{fig:full-twist-ribbons}
\end{figure}

\paragraph{A geometry with no Abelian braiding phase.}
Let the moving point carry the orthogonal frame of Fig.~\ref{fig:mobius-cross-transport}, this time rotating by $360^\circ$ in one circuit. Each vector returns to its original direction, and the line segment along it sweeps out one of the orientable annuli in Fig.~\ref{fig:full-twist-ribbons}. These annuli are the preimages of two short crossing arcs under the Hopf fibration. The crossing lifts to the core circle and the four endpoints lift to the four boundary circles. Each pair of boundary circles of different colors forms a Hopf link, and their linking numbers sum to zero with the induced orientations.

Give each arc in the base the unit integer label. Their parameter space is homotopy equivalent to $S^1\wedge S^1\simeq S^2$. Pulling the two arcs back along the Hopf fibration and then sending the unit labels to arbitrary Abelian labels $g_i\in G_i$ factors the geometry map as
\begin{equation}
 S^3\xrightarrow{\mathrm{Hopf}}S^2
 \longrightarrow K(G_1,1)\wedge K(G_2,1)\simeq\mathcal P .
 \label{eq:hopf-blind-factorization}
\end{equation}
Since $H^3(S^2,\RZ)=0$, every $[\Phi_\omega]$ integrates to zero on this pair of annuli. Thus even a nontrivial geometry can give a trivial phase for every Abelian statistics class.

\begin{remark}
\label{subsec:hopf-disk-detector}
Further generalized braiding geometries arise by realizing $L_1,L_2$ in the two orthogonal summands of the normal bundle of $C=L_1\cap L_2$ in $X$. For a six-dimensional example, replace the core of the two M\"obius bands by $C=S^2$ and the short line segments in each normal slice by two disks. Choose complex line bundles $E,F$ over $C$ with Chern numbers $1,-1$, respectively. Their direct sum is trivial over $S^2$ and can serve as the normal bundle of $C\subset S^6$. The two disk bundles $L_1=D(E)$ and $L_2=D(F)$ meet only along their common zero section $C$, and the boundary $S^3$ of each avoids the other support.

Take two species of $\ZZ_2$ excitations with $q=q'=3$, and let $s_i$ denote the mod-two Thom cochain of $L_i$. For $\omega=\tfrac12 A_1\Sq^2 A_2$, Eq.~\eqref{eq:braiding-coefficient-rule} reduces the phase calculation to $\int_{S^6}s_1\Sq^2s_2$. Using $s_1$ to restrict the integral to $L_1$, we find that $s_2$ there represents the core sphere $C$. Its cup square counts the self-intersection of $C$, whose normal bundle in $L_1$ is $E$. Thus $w_1(E)$ in Eq.~\eqref{eq:mobius-readout} is replaced here by $w_2(E)=c_1(E)\bmod2$, giving
\[
 [U_2(s_2),U_1(s_1)]
 =(-1)^{\langle w_2(E),[C]\rangle}\id=-\id .
\]
\end{remark}

The construction in Eq.~\eqref{eq:bockstein-all-degrees} requires both species to have codimension at least $2$. If one codimension drops to $1$, the Bockstein class can still be nontrivial, but its commutator is always $+\id$ under boundary avoidance.

\section{Bockstein statistics beyond generalized braiding}
\label{sec:bockstein_word}

When one species of excitation has codimension $1$, its open hopping operator sweeps out a full-dimensional region. Passing another open operator through this region violates boundary avoidance. Equation~\eqref{eq:bock-family-readout} gives the same restriction. If $q'=1$, then wherever both moves act, $s_2$ is a locally constant degree-zero cocycle. Its standard integer lift is locally constant as well, so $\dd\bar s_2=0$. If $q=1$, then $\bar s_1$ is locally constant and the readout is represented by $-\dd(k\,\bar s_1\smile\bar s_2/N^2)$. This primitive vanishes wherever either move is zero and thus extends by zero along the two axes of $\mathcal P$. In both cases the Bockstein response lies in $\ker\Phi$.

For $q+q'=d+1$, another closed sequence $W_N$ measures the Bockstein class~\cite{Xue2026BocksteinLoops,HsinChen2026Bockstein}. We perform two hopping operations in each of the two possible orders for $N$ rounds and compare the phases of the resulting histories. We first fix the background to zero.
Let the two species of excitation have dimensions $p_1,p_2\ge0$, each with fusion group $\ZZ_N$,
and let the spatial dimension be $d=p_1+p_2+1$.
Choose two oriented disks $L_1^{p_1+1},L_2^{p_2+1}$ whose intersection is a line segment
with one endpoint in $L_1\cap\partial L_2$ and the other in $\partial L_1\cap L_2$,
with $\partial L_1\cap\partial L_2=\varnothing$.
For example, in $\RR\times\RR^{p_1}\times\RR^{p_2}$ we may take
$L_1=[0,2]\times D^{p_1}\times\{0\}$ and
$L_2=[1,3]\times\{0\}\times D^{p_2}$. This pair of disks fits inside a coordinate neighborhood of a closed oriented space $X$.
For $p_1=p_2=0$, the disks are two intervals with interleaved endpoints on a circle.
For $p_1=p_2=1$, their boundaries form a Hopf link.
At each endpoint of the intersection, a local basis change for one hopping operator can depend on the excitation on the boundary of the other disk. The term-by-term cancellation in the four-step braiding process of the introduction therefore fails.

Let the cochains $s_1,s_2$ of Sec.~\ref{sec:spacetime} describe the moves on the two disks, and denote the corresponding hopping operators by $\mathsf X,\mathsf Y$. Define
\begin{equation}
 W_N(\mathsf X,\mathsf Y)=(\mathsf X\mathsf Y)^{-N}(\mathsf Y\mathsf X)^N,
 \label{eq:bockstein-word}
\end{equation}
where $(\mathsf Y\mathsf X)^N$ alternates $\mathsf X$ and $\mathsf Y$, and $(\mathsf X\mathsf Y)^N$ reverses their order in each round.
A negative power denotes the inverse of the original operation. Both histories use the same hopping operators.
The two histories end in the same configuration. Following one forward and the other backward gives $W_N$. In this closed spacetime history, the $p_1$-dimensional fusion locus of the first species links the $(p_2+1)$-dimensional worldvolume of the second, and the $p_2$-dimensional fusion locus of the second links the $(p_1+1)$-dimensional worldvolume of the first.

\begin{figure}[htbp]
\centering
\begin{tikzpicture}[x=1cm,y=1cm,>=Latex,font=\small]
  \begin{scope}[x=1.35cm,y=1.35cm]
    \draw[gray!25,step=1] (0,0) grid (3,3);
    \foreach \i in {0,1,2,3}{
      \node[below] at (\i,-.08) {$\i$};
      \node[left] at (-.08,\i) {$\i$};
    }
    \foreach \i in {0,1,2}{
      \draw[->,speciesA,line width=1.4pt] (\i,\i)--({\i+1},\i);
      \draw[->,speciesB,line width=1.4pt] ({\i+1},\i)--({\i+1},{\i+1});
      \draw[->,speciesB,line width=1.1pt,densely dashed] (\i,\i)--(\i,{\i+1});
      \draw[->,speciesA,line width=1.1pt,densely dashed] (\i,{\i+1})--({\i+1},{\i+1});
    }
    \foreach \i in {0,1,2,3}{\fill[black!80] (\i,\i) circle (1.7pt);}
    \node[above right,align=left] at (3,3) {$(3,3)$\\$\equiv(0,0)$};
    \node[below] at (1.5,-.48) {Moves $i$ of species 1};
    \node[rotate=90,above] at (-.6,1.5) {Moves $j$ of species 2};
  \end{scope}
  \begin{scope}[shift={(7.1,0)}]
    \draw[speciesA,->,line width=1.4pt] (0,3.8)--(.9,3.8);
    \node[right] at (1.05,3.8) {$\mathsf X\colon i\mapsto i+1$};
    \draw[speciesB,->,line width=1.4pt] (0,3.1)--(.9,3.1);
    \node[right] at (1.05,3.1) {$\mathsf Y\colon j\mapsto j+1$};
    \draw[black!75,line width=1.2pt] (0,2.25)--(.9,2.25);
    \node[right] at (1.05,2.25) {Solid $(\mathsf Y\mathsf X)^3$};
    \draw[black!75,line width=1.2pt,densely dashed] (0,1.55)--(.9,1.55);
    \node[right] at (1.05,1.55) {Dashed $(\mathsf X\mathsf Y)^3$};
    \node[align=left,text width=4.5cm,anchor=west] at (0,.35)
      {Solid path out, dashed path back.\\[5pt]
       $W_3=(\mathsf X\mathsf Y)^{-3}(\mathsf Y\mathsf X)^3$};
  \end{scope}
\end{tikzpicture}
\caption{Two alternating histories for $N=3$. The lattice point $(i,j)$ gives the configuration reached from the same initial state
after $i$ moves of the first species and $j$ moves of the second. The labels are taken modulo three.
The arrows show the forward direction of each history.
The process $W_3$ compares their phases by traversing the second history in reverse.
At its $(j+1)$st move, the second species reads label $i=j+1$ of the first species on the solid path and $i=j$ on the dashed path, for $j=0,1,2$.}
\label{fig:bockstein-staircases}
\end{figure}

\subsection{Cancellation of local basis changes and the spacetime integral}
\label{subsec:wn-general}

Apply the local basis change~\eqref{eq:hopping-local-choice} to an elementary move near $x\in L_1$ in $\mathsf X$. On input configuration $a$, this step acquires an extra phase $e^{2\pi\ii f_x(a)}$, where $f_x$ reads only a neighborhood of $x$. Write the initial configuration as $a=(a_1,a_2)$. Before the $\mathsf X$ step of round $i+1$, both staircases have applied $\mathsf X$ exactly $i$ times, whereas they have applied $\mathsf Y$ respectively $i$ and $i+1$ times. The resulting change in the phase exponent of $W_N$ is
\begin{equation}
\begin{aligned}
 \delta\vartheta_W(a)=\sum_{i=0}^{N-1}\big[&
 f_x(a_1+i\,\dd s_1,\ a_2+i\,\dd s_2)\\[-2pt]
 &{}-f_x(a_1+i\,\dd s_1,\ a_2+(i+1)\dd s_2)\big].
\end{aligned}
 \label{eq:wn-local-cancellation-main}
\end{equation}
Earlier elementary moves within $\mathsf X$ shift the inputs of both terms by the same configuration, which we absorb into $a$.

Section~\ref{sec:geometry} required the boundary of each disk to avoid the other disk. Here the two boundaries need only be separated, so that on each simplex $\Delta$ read by the move, either $\dd s_1|_\Delta=0$ or $\dd s_2|_\Delta=0$. If $\dd s_2|_\Delta=0$, the two terms in Eq.~\eqref{eq:wn-local-cancellation-main} cancel in each round. If $\dd s_1|_\Delta=0$, the first term equals the second term of the previous round, so the sum over $i\bmod N$ again vanishes. This second case occurs at $L_1\cap\partial L_2$ in Fig.~\ref{fig:bockstein-loop-disks}. Order the four boundary points along the extension of the intersection line as $x_1<y_1<x_2<y_2$, with $x_1,x_2$ on the boundary of the first disk. Starting from the vacuum, the two moves near $y_1$ both read the second-species label $i$, so
\begin{equation}
 f_x\!\left(\OverlapWord{i}{i}\right)
 =f_x\!\left(\OverlapWord{(i-1)}{i}\right).
 \label{eq:wn-interior-cancellation}
\end{equation}
Exchanging the disks gives the same cancellation for every elementary move in $\mathsf Y$. Thus $W_N$ is a statistical process.

\begin{figure}[htbp]
\centering
\resizebox{5.5cm}{!}{%
\begin{tikzpicture}[
  x=1cm,y=1cm,font=\small,line cap=round,line join=round,
  disk support/.style={line width=.95pt},
  disk hidden/.style={line width=.72pt,dash pattern=on 2.3pt off 1.8pt}
]
  \begin{scope}
    \clip (-.65,0) circle[radius=1];
    \fill[speciesA,opacity=.4,even odd rule]
      (-.65,0) circle[radius=1]
      (.66,0) ellipse[x radius=1.12,y radius=.58];
  \end{scope}
  \begin{scope}
    \clip (.66,0) ellipse[x radius=1.12,y radius=.58];
    \fill[speciesB,opacity=.4,even odd rule]
      (-.65,0) circle[radius=1]
      (.66,0) ellipse[x radius=1.12,y radius=.58];
  \end{scope}
  \begin{scope}
    \clip (-.65,0) circle[radius=1];
    \clip (.66,0) ellipse[x radius=1.12,y radius=.58];
    \fill[speciesA!62.5!speciesB,opacity=.64]
      (-1.8,0) rectangle (1.9,1.08);
    \fill[speciesB!62.5!speciesA,opacity=.64]
      (-1.8,-1.08) rectangle (1.9,0);
  \end{scope}
  \draw[speciesB!65!speciesA,disk hidden]
    (.198890,.528561)
    arc[start angle=114.31183,end angle=180,x radius=1.12,y radius=.58];
  \draw[speciesA!65!speciesB,disk hidden]
    (.198890,-.528561)
    arc[start angle=328.09148,end angle=360,radius=1];
  \draw[black!75,line width=.9pt] (-.46,0)--(.35,0);
  \draw[speciesA,disk support]
    (.35,0) arc[start angle=0,end angle=328.09148,radius=1];
  \draw[speciesB,disk support]
    (-.46,0) arc[start angle=180,end angle=474.31183,x radius=1.12,y radius=.58];
  \node[speciesA!85!black] at (-.94,.22) {$L_1$};
  \node[speciesB!85!black] at (1.02,.1) {$L_2$};
\end{tikzpicture}%
}
\caption{Disk supports in the three-dimensional Bockstein process. The red disk $L_1$ and the blue disk $L_2$ support
$\mathsf X$ and $\mathsf Y$, respectively, and intersect along a line segment. Their boundaries form a Hopf link and carry the two species of loop excitations.}
\label{fig:bockstein-loop-disks}
\end{figure}

After $N$ moves of each species the net move vanishes, so $(\mathsf X\mathsf Y)^N=F(\hat a_1,\hat a_2)$ is a unitary function of the configuration. Using $(\mathsf Y\mathsf X)^N=\mathsf X^{-1}(\mathsf X\mathsf Y)^N\mathsf X$ and the shift relation in Definition~\ref{def:hopping-algebra}, we obtain
\begin{equation}
 W_N=\frac{F(\hat a_1+\dd s_1,\hat a_2)}{F(\hat a_1,\hat a_2)}.
 \label{eq:wn-quantization-main}
\end{equation}
By Theorem~\ref{thm:process-initial-state}, this ratio has a constant value $w$ on the vacuum orbit. Shift $a_1$ successively by $\dd s_1$ and multiply the $N$ ratios. The configuration returns to its starting point, giving $w^N=1$.

\paragraph{Bockstein statistics from the spacetime integral.}
Here $q=d-p_1$ and $q'=d-p_2$, so $q+q'=d+1$. For the Bockstein statistics class, Eq.~\eqref{eq:background-coupling-construction} at zero background gives
\begin{equation}
 \omega_k(A)=\frac{k}{N}\bar A_1\smile\beta_N A_2,\qquad
 \Theta^1_k(\phi;0)=\frac{k}{N}\bar\phi_1\smile\beta_N(\dd\phi_2),
 \qquad k\in\ZZ_N .
 \label{eq:wn-input}
\end{equation}
The bar denotes the standard integer lift on each simplex, and $\beta_N f=\dd\bar f/N$. Compare the two staircases in Fig.~\ref{fig:bockstein-staircases} using Eq.~\eqref{eq:raw-hopping}. On an $\mathsf X$ step, $\phi_2$ is fixed and the integral vanishes. The $(j+1)$st $\mathsf Y$ step on each staircase takes the second field from $\phi_2+j s_2$ to $\phi_2+(j+1)s_2$. At this step $(\mathsf Y\mathsf X)^N$ has already performed one additional translation $s_1$ of the first field.

Glue the time intervals for these $N$ consecutive $\mathsf Y$ steps together. Since $Ns_2=0\bmod N$, the second field returns to its starting value, giving a matter history $\widetilde\phi_2$ on the time circle $S^1_{\rm t}$. Here $\dd\widetilde\phi_2$ is the spacetime coboundary. In the standard prism for each interval, only the simplices whose first edge is vertical contribute to $\int_{I_{\rm t}}\overline{\dd\widetilde\phi_2}$. Each interval therefore gives $\bar s_2$. The time circle has no boundary, so Stokes' theorem gives
\begin{equation}
 \int_{S^1_{\rm t}}\beta_N(\dd\widetilde\phi_2)
 =-\frac1N\,\dd\!\int_{S^1_{\rm t}}\overline{\dd\widetilde\phi_2}
 =-\frac1N\,\dd(N\bar s_2)=-\dd\bar s_2 .
 \label{eq:wn-time-layers}
\end{equation}

In the difference of the two exponents, the integer lifts of the first field differ by $\bar s_1\bmod N$. Moving the time direction past $\bar s_1$ gives $(-1)^{q-1}$. The minus signs in Eqs.~\eqref{eq:raw-hopping} and~\eqref{eq:wn-time-layers} cancel. Thus for any initial matter state,
\begin{equation}
 W_N(\mathsf X,\mathsf Y)
 =\exp\!\left[\frac{2\pi\ii k}{N}(-1)^{q-1}
      \int_X\bar s_1\smile\dd\bar s_2\right]\id .
 \label{eq:wn-value}
\end{equation}
Modulo $N$, the spatial integral is the oriented intersection number of $L_1$ with $\partial L_2$. Since $X$ is closed, Stokes' theorem also expresses the phase in terms of the intersections of $\partial L_1$ with $L_2$,
\[
 \int_X\bar s_1\smile\dd\bar s_2
 =(-1)^{q}\int_X\dd\bar s_1\smile\bar s_2 .
\]
Choose compatible orientations so that the signed intersection number in Eq.~\eqref{eq:wn-value} is $+1$. The measured phases are then $e^{2\pi\ii k/N}$ for the respective values of $k$. Along the gauge direction, update the background at each step and negate both move labels. Equation~\eqref{eq:dual-operator-expansion} gives the phase $e^{-2\pi\ii k/N}$ for the paired symmetry. Appendix~\ref{app:wn-homology} constructs the integral homology cycle that pairs with this phase. For two species of $\ZZ_2$ loops in three dimensions, $W_2$ and the two M\"obius bands of Sec.~\ref{subsec:mobius-detector} detect the same mixed component.

\subsection{\texorpdfstring{$W_2$}{W2} on interleaved intervals in one dimension}
\label{subsec:wn-one-dimensional}

Choose four points $x_1<y_1<x_2<y_2$ along an arc of a one-dimensional circle. The operator $\mathsf X$ creates a particle and an antiparticle of the first species at $x_1,x_2$, and $\mathsf Y$ creates a pair of the second species at $y_1,y_2$. Figure~\ref{fig:bockstein-fusion-history} depicts the eight steps of $W_2$ as worldlines. Starting from the vacuum, two $\ZZ_2$ particles of the same species can fuse to the vacuum when they meet again. Each circled fusion event links the worldline of the other species.

\paragraph{Domain wall statistics.}
Let $\chi_I,\chi_J$ be the indicator functions of $I=[x_1,x_2]$ and $J=[y_1,y_2]$. Place two $\ZZ_2$ fields $\phi_1,\phi_2$ at each site of the circle, and write $U_1,U_2$ for their respective hopping operators at zero background. For the statistics class $\omega=\tfrac12 A_1\smile A_2\smile A_2$, we may take $\Theta^1(\phi;0)=\tfrac12\phi_1\smile\dd\phi_2\smile\dd\phi_2$. Dividing the time layer over each bond into two triangles and substituting into Eq.~\eqref{eq:raw-hopping} gives
\begin{equation}
\begin{aligned}
 U_1(\chi_I)\ket{\phi_1,\phi_2}
 &=\ket{\phi_1+\chi_I,\phi_2},\\
 U_2(\chi_J)\ket{\phi_1,\phi_2}
 &=\left(\prod_{\substack{[ij]\\\chi_J(i)\ne\chi_J(j)}}
       (-1)^{\phi_1(i)[\phi_2(j)-\phi_2(i)+\chi_J(i)]}\right)
       \ket{\phi_1,\phi_2+\chi_J}.
\end{aligned}
\label{eq:bock-z2-hopping}
\end{equation}
The interval $I$ contains exactly one endpoint $y_1$ of $J$. Taking $k=1,N=2$ in Eq.~\eqref{eq:wn-value} gives
\[
 W_2\bigl(U_1(\chi_I),U_2(\chi_J)\bigr)=-\id .
\]
Equation~\eqref{eq:bock-z2-hopping} also identifies the source of this minus sign. The phase of $U_2$ reads only the bonds crossing the two ends of $J$, and $U_1$ flips $\phi_1$ only inside $I$. Hence only the bond crossing $y_1$ contributes to the relative phase.

\paragraph{The same history as a sequence of gauge transformations.}
Starting at zero background, Eq.~\eqref{eq:dual-operator-expansion} maps these eight hopping steps to the gauge direction and complex conjugates the relative phase of the two histories. Since each label in $\ZZ_2$ equals its own inverse, following the same path in Fig.~\ref{fig:bockstein-fusion-history} should give $-1$. We now compute the sign step by step.
\begin{figure}[htbp]
\centering
\begin{minipage}[c]{.43\linewidth}
\centering
\resizebox{5.3cm}{!}{
\begin{tikzpicture}[
    x=1cm,y=1cm,line cap=round,line join=round,
    font=\small,
    support/.style={line width=.92pt},
    mark forward/.style={
      postaction={decorate},decoration={markings,
        mark=at position #1 with {\arrow{Stealth[length=3.5pt,width=3.0pt]}}}
    }
  ]
  \draw[black!35,line width=.5pt] (-2.18,0) -- (2.18,0);
  \draw[speciesA,support,mark forward=.66]
    (-1.99,0) .. controls (-1.33,.99) and (.25,.99) .. (.93,0);
  \draw[speciesB,support,mark forward=.66]
    (1.99,0) .. controls (1.33,-.99) and (-.25,-.99) .. (-.93,0);
  \foreach \p in {-1.99,.93}{\fill[speciesA] (\p,0) circle[radius=1.5pt];}
  \foreach \p in {-.93,1.99}{\fill[speciesB] (\p,0) circle[radius=1.5pt];}
  \node[text=speciesA,above=1.5pt] at (-.53,.75) {$\mathsf X$};
  \node[text=speciesB,below=1.5pt] at (.53,-.75) {$\mathsf Y$};
  \foreach \p/\s in {-1.99/x_1,-.93/y_1,.93/x_2,1.99/y_2}{%
    \node[anchor=base,text height=5pt,text depth=2pt] at (\p,-.34) {$\s$};
  }
  \node[font=\small] at (0,-1.55) {$y_1$ lies inside the support of $\mathsf X$};
\end{tikzpicture}}
\end{minipage}\hfill
\begin{minipage}[c]{.53\linewidth}
\centering
\resizebox{6.1cm}{!}{
\begin{tikzpicture}[
  x=1cm,y=.66cm,font=\small,line cap=round,line join=round,
  world/.style={line width=1.45pt},
  moving/.style={world,postaction={decorate},decoration={markings,
    mark=at position #1 with {\arrow{Stealth[length=3.8pt,width=3.8pt]}}}},
  moving/.default=.57,
  bluefront/.style={preaction={draw=white,line width=3.8pt}}
]
  \foreach \xx in {0,1.1,2.2,3.3}
    \draw[gray!35,densely dashed,line width=.35pt] (\xx,-.15)--(\xx,8.3);
  \foreach \yy in {0,...,8}
    \draw[gray!16,line width=.35pt] (-.12,\yy)--(3.48,\yy);

  \draw[-{Stealth[length=4.5pt,width=4pt]},gray!75,line width=.7pt]
    (-1.28,.05)--(-1.28,8.35);
  \node[anchor=south,text=gray!75] at (-1.28,8.4) {$t$};
  \foreach \yy/\lab in {.5/\mathsf X,2.5/\mathsf X,4.5/\mathsf X^{-1},6.5/\mathsf X^{-1}}
    \node[anchor=east,text=speciesA] at (-.27,\yy) {$\lab$};
  \foreach \yy/\lab in {1.5/\mathsf Y,3.5/\mathsf Y,5.5/\mathsf Y^{-1},7.5/\mathsf Y^{-1}}
    \node[anchor=east,text=speciesB] at (-.27,\yy) {$\lab$};

  \draw[world,speciesA] (0,0)--(0,2);
  \draw[world,speciesA] (2.2,1)--(2.2,3);
  \draw[moving,speciesA] (0,0)--(2.2,1);
  \draw[moving,speciesA] (0,2)--(2.2,3);
  \draw[world,speciesA] (2.2,4)--(2.2,6);
  \draw[world,speciesA] (0,5)--(0,7);
  \draw[moving,speciesA] (2.2,4)--(0,5);
  \draw[moving,speciesA] (2.2,6)--(0,7);

  \draw[world,bluefront,speciesB]
    (3.3,1)--(3.3,3)--(1.1,4)--(1.1,2)--cycle;
  \draw[world,bluefront,speciesB]
    (1.1,5)--(1.1,7)--(3.3,8)--(3.3,6)--cycle;
  \draw[moving=.76,speciesB] (3.3,1)--(1.1,2);
  \draw[moving,speciesB] (3.3,3)--(1.1,4);
  \draw[moving=.76,speciesB] (1.1,5)--(3.3,6);
  \draw[moving,speciesB] (1.1,7)--(3.3,8);

  \foreach \xx/\yy in {0/0,2.2/3,2.2/4,0/7} {
    \draw[speciesA,line width=1.75pt]
      (\xx,\yy)+(-.075,-.115)--+(.075,.115)
      (\xx,\yy)+(-.075,.115)--+(.075,-.115);
  }
  \foreach \xx/\yy in {3.3/1,1.1/4,1.1/5,3.3/8} {
    \draw[speciesB,line width=1.75pt]
      (\xx,\yy)+(-.075,-.115)--+(.075,.115)
      (\xx,\yy)+(-.075,.115)--+(.075,-.115);
  }
  \foreach \xx/\yy in {2.2/3,1.1/5}
    \draw[gray!65,densely dashed,line width=.5pt]
      (\xx,\yy) ellipse[x radius=.21,y radius=.32];

  \foreach \yy in {1.5,5.5}
    \draw[black!85,line width=.7pt,preaction={draw=white,line width=1.8pt}]
      (2.2,\yy) +(0,.23)--+(.15,0)--+(0,-.23)--+(-.15,0)--cycle;
  \node[anchor=west,inner sep=1pt] at (2.42,1.79) {$-1$};
  \node[anchor=west,inner sep=1pt] at (2.42,5.21) {$+1$};

  \foreach \xx/\lab in {0/x_1,1.1/y_1,2.2/x_2,3.3/y_2}
    \node[anchor=north] at (\xx,-.28) {$\lab$};
\end{tikzpicture}}
\end{minipage}
\caption{Left, the intervals $I=[x_1,x_2]$ and $J=[y_1,y_2]$ have interleaved endpoints. Right, the eight steps of $W_2$ run from bottom to top. The red and blue worldlines follow the two excitation species. Sloping lines indicate motion and vertical lines indicate rest. Crosses mark pair creation or fusion. The dashed oval encloses an event surrounded by a worldline of the other species. For gauge transformations along the same history, $\mathsf Y$ at step 2 and $\mathsf Y^{-1}$ at step 6 read the red background defect at $x_2\in\partial I\cap J$, as marked by diamonds. Starting at zero background and $\phi_2=0$, the factors $-1,+1$ correct these two steps relative to the zero-background patch.}
\label{fig:bockstein-fusion-history}
\end{figure}

For the above $\omega$, Eq.~\eqref{eq:background-coupling-construction} gives
\[
 \Theta^1(\phi;A)=\tfrac12\bigl[
 \phi_1\smile D_{A_2}\phi_2\smile D_{A_2}\phi_2
 +A_1\smile\phi_2\smile D_{A_2}\phi_2+A_1\smile A_2\smile\phi_2\bigr].
\]
First take each step to be a symmetry patch at zero background. From Eq.~\eqref{eq:R-prism-definition}, the two patches act on the matter space as
\[
\begin{aligned}
 P_1(\chi_I)\ket{\phi_1,\phi_2}
 &=(-1)^{\int_X\chi_I\smile\phi_2\smile\dd\phi_2}\ket{\phi_1+\chi_I,\phi_2},\\
 P_2(\chi_J)&=\prod_{j\in J}X_j^{(2)} .
\end{aligned}
\]
Along each staircase, the first patch acts once on $\phi_2$ and once on $\phi_2+\chi_J$. The two staircases acquire the same pair of phases, so $W_2\bigl(P_1(\chi_I),P_2(\chi_J)\bigr)=+\id$.

Along the gauge direction, each step inherits the background left by its predecessor, as in Eq.~\eqref{eq:patch-reconstruct}. The second gauge transformation preserves $D_{A_2}\phi_2$. On an edge in the gauge direction, only the last term $A_1\smile A_2\smile\phi_2$ contributes. Here $A_2$ takes the value $\chi_J$ on that edge, and the matter field at its endpoint is $\phi_2+\chi_J$. By Eq.~\eqref{eq:R-prism-definition}, the matter matrix starting at background $A$ is
\begin{equation}
 T_{2,A}(\chi_J)\ket{\phi_1,\phi_2}
 =(-1)^{\int_X A_1\smile[\chi_J(\phi_2+\chi_J)]}
   \ket{\phi_1,\phi_2+\chi_J}.
 \label{eq:bockstein-T2-defect}
\end{equation}
The square brackets denote a pointwise product of lattice functions, and the output background is $(A_1,A_2-\dd\chi_J)$. The phase of the first gauge transformation depends only on $\phi_2,A_2$. Both staircases read the same two pairs of values, so their contributions cancel in the relative phase.

The two diamonds in Fig.~\ref{fig:bockstein-fusion-history} mark the steps at which the second gauge transformation encounters $A_1=\dd\chi_I$. Step 2 acts forward from the initial $\phi_2$. Step 6 undoes the forward move on the other staircase that started at $\phi_2+\chi_J$. By Eq.~\eqref{eq:bockstein-T2-defect}, the corrections to the zero-background patches at these steps are
\[
 \underbrace{(-1)^{\int_X\dd\chi_I\smile[\chi_J(\phi_2+\chi_J)]}}_{\text{step 2}}
 \ \underbrace{\left((-1)^{\int_X\dd\chi_I\smile[\chi_J\phi_2]}\right)^{-1}}_{\text{step 6}}
 =(-1)^{\int_X\dd\chi_I\smile\chi_J}=-1 .
\]
The last integral counts $x_2$, the unique endpoint of $I$ inside $J$. For $\phi_2=0$, the corrections at the two diamonds are $-1,+1$, respectively. The hopping operator crosses the endpoint $y_1$ of $J$ inside $I$, while the gauge transformation reads the endpoint $x_2$ of $I$ inside $J$. Both give $W_2=-\id$.

\paragraph{Symmetric patches and closed hopping operators.}
We can now compare the two interleaved intervals directly with the one-dimensional model of Hsin and Chen~\cite[Eqs.~(10), (11) and Appendix~E]{HsinChen2026Bockstein}. Extend the first patch to the entire circle. Each bond contributes $\phi_2(j)[\phi_2(j+1)-\phi_2(j)]=\phi_2(j)+\phi_2(j)\phi_2(j+1)\bmod 2$ to its phase exponent. The global generators of the paired symmetry are therefore precisely those in their Eq.~(10),
\[
 P_1(\mathbf{1})=\left(\prod_j X_j^{(1)}\right)
         \left(\prod_j Z_j^{(2)}\,\mathrm{CZ}^{(2)}_{j,j+1}\right),
 \qquad P_2(\mathbf{1})=\prod_j X_j^{(2)} .
\]
In this model Hsin and Chen found a way to measure the anomaly directly with symmetry restrictions. Their $X_I$ already commutes with both global generators. If it is paired with the bare restriction $\prod_{j\in J}X_j^{(2)}$ of the second generator, $X_I$ acts once on each of the same two $\phi_2$ configurations along both staircases, yielding $W_2=+\id$. They add local gates at the endpoints of $J$ to obtain $Y_J$, which also commutes with both global generators. With these symmetric restrictions, $W_2(X_I,Y_J)=-\id$, reproducing the mixed anomaly of this symmetry~\cite[Eqs.~(124)--(134), (147)]{HsinChen2026Bockstein}.

Applying this method to a general symmetry requires an entire compatible family of symmetric restrictions, which may fail to exist. Consider the nontrivial Bockstein phase of two species of $\ZZ_2$ loops in three-dimensional space. The two M\"obius bands satisfy boundary avoidance, and Eqs.~\eqref{eq:inverse-operator} and~\eqref{eq:mobius-readout} give $-\id$ for the patch commutator. If every compatible restriction commuted with the global symmetry, the local increments between neighboring restrictions would also be symmetric, and Corollary~\ref{cor:symmetric-patch-braiding} would force this commutator to equal $\id$. Performing gauge transformations in their actual backgrounds requires only compatibility of the restrictions. Equation~\eqref{eq:bockstein-T2-defect} gives the corresponding background correction in the one-dimensional example.

Now extend the two \emph{hopping operators} in Eq.~\eqref{eq:bock-z2-hopping} to the full circle. Both indicator functions become $\mathbf{1}$, so the product over endpoints in the second line equals $1$, giving
\[
 U_1(\mathbf{1})=\prod_j X_j^{(1)},\qquad U_2(\mathbf{1})=\prod_j X_j^{(2)} .
\]
This onsite symmetry has zero anomaly, although the two hopping operators on interleaved intervals have statistics $W_2=-\id$. The two quantities therefore differ.

Table~\ref{tab:statistics-anomaly-methods} lists ways to apply a statistical process to measure statistics or an anomaly. Table~\ref{tab:statistics-anomaly-pitfalls} explains why other prescriptions can give a different result.

\begin{table}[!htbp]
\centering
\caption{Using a statistical process to measure statistics or an anomaly.}
\label{tab:statistics-anomaly-methods}
\small
\begin{tabularx}{\linewidth}{@{}>{\raggedright\arraybackslash}p{.29\linewidth}@{\hspace{1em}}>{\raggedright\arraybackslash}X@{\hspace{1em}}>{\raggedright\arraybackslash}p{.19\linewidth}@{}}
\toprule
\textbf{Operator used at each step} & \textbf{Quantity measured} & \textbf{Reference}\\
\midrule
Hopping operator $U(s)$ for a topological excitation or a domain wall &
Measures statistics. By gauge--time duality, the paired anomaly phase is the complex conjugate of this result. &
Theorems~\ref{thm:hopping-statistics} and~\ref{thm:dual-operator-algebras}\\
\addlinespace[.5em]
Gauge transformation $T_A(u)$ moving a symmetry defect &
Measures the anomaly. Each step starts in the background $A$ left by its predecessor and updates it to $A-\dd u$. &
Theorem~\ref{thm:dual-operator-algebras}\newline Sec.~\ref{subsec:wn-one-dimensional}\\
\addlinespace[.5em]
Symmetry patch at zero background &
Measures the anomaly directly when the process is invariant under perturbations of the first kind, as in generalized braiding with boundary avoidance. &
Proposition~\ref{prop:braiding-paired-readouts}\\
\bottomrule
\end{tabularx}
\end{table}

\begin{table}[!htbp]
\centering
\caption{Uses of a statistical process that can fail to give the intended result.}
\label{tab:statistics-anomaly-pitfalls}
\small
\begin{tabularx}{\linewidth}{@{}>{\raggedright\arraybackslash}p{.29\linewidth}@{\hspace{1em}}>{\raggedright\arraybackslash}X@{\hspace{1em}}>{\raggedright\arraybackslash}p{.19\linewidth}@{}}
\toprule
\textbf{Prescription} & \textbf{Limitation} & \textbf{Reference}\\
\midrule
Substitute symmetry patches directly~\cite[Sec.~V, Eqs.~(106)--(108)]{KobayashiEtAl2026Generalized} &
For a process sensitive to perturbations of the first kind, including the one-dimensional example, the result depends on the choice of restriction. &
Sec.~\ref{subsec:wn-one-dimensional}\\
\addlinespace[.5em]
Require all patches used to commute with the global symmetry~\cite[Appendix~E]{HsinChen2026Bockstein} &
Such compatible restrictions need not exist. A nontrivial type-III anomaly in one dimension, or a nontrivial anomaly detectable by generalized braiding in higher dimensions, rules out this choice. &
Corollary~\ref{cor:symmetric-patch-braiding}\newline Sec.~\ref{subsec:wn-one-dimensional}\\
\addlinespace[.5em]
Treat hopping operators extended to the entire space as a symmetry~\cite[last paragraph of the introduction]{HsinChen2026Bockstein} &
In one dimension the type-III anomaly of this symmetry is $-2$ times the type-III statistics of the original hopping operators~\cite[Theorem~4.1]{Xue2026Comparison}. &
Sec.~\ref{sec:typeiii-zero}\newline Eq.~\eqref{eq:typeiii-factor-minus-two}\\
\bottomrule
\end{tabularx}
\end{table}

\section{Type-III statistics on a circle, closed hopping operators, and twisted boundary conditions}
\label{sec:typeiii-zero}

Consider the type-III class for $G=\ZZ_N^3$ on $X=S^1$. In this example the closed hopping operators $U_0(z)$ obey the group law, so we can take them as a group action in their own right and compare its anomaly with that of the original symmetry $S(z)$. Let $\mathbf{1}_i$ be the constant parameter with $i$th component $1$ and all other components $0$ at every vertex, and let $\zeta=e^{2\pi\ii/N}$. The three matter fields $\phi_i$ live on vertices. We suppress the cup product symbol and retain the order of its factors. One solution to Eq.~\eqref{eq:first-descent} is
\begin{equation}
 \Theta(\phi)=\frac{1}{N}\phi_1\,\dd\phi_2\,\dd\phi_3,\qquad
 \omega(a)=\frac{1}{N}a_1a_2a_3 .
 \label{eq:typeiii-stat-cocycle}
\end{equation}
Here the product is taken in $\ZZ_N$, and division by $N$ gives an additive phase in $\RZ$.

\paragraph{Symmetry transformations and hopping.}
A global shift by $z$ changes only the undifferentiated $\phi_1$ in Eq.~\eqref{eq:typeiii-stat-cocycle}, increasing $\Theta$ by $\dd(z_1\phi_2\,\dd\phi_3)/N$. Substituting the endpoint term $-z_1\phi_2\,\dd\phi_3/N$ into Eq.~\eqref{eq:symmetry-descent-primitive} and setting $\bra{z}S(z)\ket{0}=1$ gives
\begin{equation}
 S(z)\ket{\phi}=\zeta^{-z_1\int_X\phi_2\,\dd\phi_3}\ket{\phi+z}.
 \label{eq:typeiii-original-symmetry}
\end{equation}
In this example we choose the phases of the zero-background gauge transformations so that $T_0(z)=S(z)$.
For the moves in the time direction, use Eq.~\eqref{eq:raw-hopping}. For any vertex field $t$, write $U_i(t)=U(t\mathbf{1}_i)$. Divide the square swept out by each spatial edge into two triangles. When only $\phi_2$ moves, only the triangle in which the temporal edge precedes the spatial edge contributes. When only $\phi_3$ moves, the spatial edge comes first and the orientation changes sign. With the order of cup factors in Eq.~\eqref{eq:typeiii-stat-cocycle}, we obtain
\begin{equation}
 \begin{aligned}
 U_1(t)\ket{\phi}&=\ket{\phi+t\mathbf{1}_1},\\
 U_2(t)\ket{\phi}&=\zeta^{-\int_X\phi_1\,t\,\dd\phi_3}\ket{\phi+t\mathbf{1}_2},\\
 U_3(t)\ket{\phi}&=\zeta^{ \int_X\phi_1\,\dd\phi_2\,t}\ket{\phi+t\mathbf{1}_3}.
 \end{aligned}
 \label{eq:typeiii-zero-hops}
\end{equation}
When only $\phi_1$ moves, both coboundaries are pulled back from the spatial circle, so the phase in the first line is one. In each line the exponent is linear in $t$ and depends only on the other two components. Thus $U_i(t)$ can also be implemented vertex by vertex. The phases in Eqs.~\eqref{eq:typeiii-original-symmetry} and~\eqref{eq:typeiii-zero-hops} ensure that symmetry transformations and hopping operators commute. For instance, applying $U_2(t)$ first shifts the exponent in $S(\mathbf{1}_1)$ by $-\int_X t\,\dd\phi_3$. Applying $S(\mathbf{1}_1)$ first shifts the exponent of $\zeta$ in $U_2(t)$ by the same amount.

\paragraph{A statistical process on intervals with alternating endpoints.}
Take the intervals $I,J$ with alternating endpoints in Fig.~\ref{fig:typeiii-four-endpoints-demo}, and write $U_i(L)=U_i(\chi_L)$ for an interval $L$. By Eq.~\eqref{eq:typeiii-zero-hops}, $U_1(I)$ shifts by $\chi_I$ the field $\phi_1$ that enters the hopping operator $U_2(\mathbf{1})$ on the whole circle. With the commutator convention of Eq.~\eqref{eq:intro-commutator}, we have
\begin{equation}
 \begin{aligned}
 [U_1(I),U_2(\mathbf{1})]\ket{\phi}&=\zeta^{\int_X\chi_I\,\dd\phi_3}\ket{\phi},\\
 \bigl[U_3(J),[U_1(I),U_2(\mathbf{1})]\bigr]
 &=\zeta^{-\int_X\chi_I\,\dd\chi_J}\,\id=\zeta\,\id .
 \end{aligned}
 \label{eq:typeiii-zero-process}
\end{equation}
The first line depends only on the third component at the two endpoints of $I$. The operator $U_3(J)$ adds one to $\phi_3$ at one of these endpoints, and its own phase cancels when it conjugates this diagonal operator. With the orientation in the figure, the right endpoint of $J$ lies inside $I$ and the left endpoint outside, so $\int_X\chi_I\,\dd\chi_J=-1$. Setting $N=2$ gives $-\id$~\cite{Xue2026Comparison}.

\begin{figure}[htbp]
\centering
\begin{tikzpicture}[x=1cm,y=1cm,>=Stealth,line cap=round,
    every node/.style={font=\small,text=ink,inner sep=2pt}]
  \foreach \x/\j in {0/0,2.2/1,4.4/2,6.6/3}{
    \draw[muted!45,densely dashed,thin] (\x,1.64)--(\x,-.12);
    \node[above] at (\x,1.66) {$x_{\j}$};
  }
  \draw[accent,line width=1.15pt] (-.65,1.38)--(7.25,1.38);
  \draw[accent,-{Stealth[length=2.3mm]}] (3.02,1.38)--(3.47,1.38);
  \foreach \x in {-.65,7.25}
    \draw[accent] (\x-.04,1.29)--(\x+.04,1.47);
  \node[anchor=east,text=accent] at (-.90,1.38) {$U_2(\mathbf{1})$};

  \draw[speciesB,line width=1.6pt] (2.2,.75)--(6.6,.75);
  \fill[speciesB] (2.2,.75) circle (2.3pt);
  \draw[speciesB,fill=white,line width=1pt] (6.6,.75) circle (2.3pt);
  \node[anchor=east,text=speciesB] at (-.90,.75) {$U_1(I)$};

  \draw[speciesA,line width=1.6pt] (0,.12)--(4.4,.12);
  \fill[speciesA] (0,.12) circle (2.3pt);
  \draw[speciesA,fill=white,line width=1pt] (4.4,.12) circle (2.3pt);
  \node[anchor=east,text=speciesA] at (-.90,.12) {$U_3(J)$};

  \node[anchor=east] at (-.90,-.47) {After $U_3(J)$};
  \node at (2.2,-.47) {$\phi_3(x_1)+1$};
  \node at (6.6,-.47) {$\phi_3(x_3)$};
\end{tikzpicture}
\caption{The circle is cut outside $I\cup J$, with its left and right ends identified. The inner commutator reads the third spin species at both endpoints of $I$. Only the left endpoint $x_1$ lies in $J$.}
\label{fig:typeiii-four-endpoints-demo}
\end{figure}

We now check how a local basis change in Eq.~\eqref{eq:zero-hopping-rephasing} affects the measured phase. Since $U_2(\mathbf{1})$ preserves the configuration $a=\dd\phi$, the extra phase of $U_1(I)$ cancels in the inner commutator. This commutator is diagonal, so the extra phase of $U_3(J)$ also cancels in the outer commutator. If one step implementing $U_2(\mathbf{1})$ acquires a phase $e^{2\pi\ii f_x(a)}$ near $x$, Eq.~\eqref{eq:intro-commutator} gives the additive correction
\[
 \begin{aligned}
 &f_x(a)-f_x(a+(\dd\chi_I,0,0))\\
 &\quad{}-f_x(a+(0,0,\dd\chi_J))+f_x(a+(\dd\chi_I,0,\dd\chi_J)).
 \end{aligned}
\]
Separate the two sets of endpoints by more than the diameter of the neighborhood on which $f_x$ depends. At least one of $\dd\chi_I$ and $\dd\chi_J$ then vanishes near $x$, and the four terms cancel in pairs. Any local moves performed before this step shift all four inputs by the same configuration. Thus the phase in Eq.~\eqref{eq:typeiii-zero-process} remains $\zeta$ under a local basis change. By the antilinear isomorphism $\alpha$ of Eq.~\eqref{eq:dual-operator-expansion}, replacing $U(s)$ by $T(-s)$ and updating the background at each step gives a paired process with phase $\overline\zeta$. The two phases are distinct for $N>2$.

\paragraph{Taking the closed hopping operators as a symmetry.}
The operator $U_2(\mathbf{1})$ above shifts the second component on the whole circle. Now take all three parameters in Eq.~\eqref{eq:typeiii-zero-hops} to be constant and shift the third, second, and first components in that order. Global shifts preserve spatial differences, giving
\begin{equation}
 U_0(z)\ket{\phi}=\zeta^{\int_X \phi_1(z_3\,\dd\phi_2-z_2\,\dd\phi_3)}
 \ket{\phi+z}.
 \label{eq:typeiii-strip-actual-U}
\end{equation}
Since $\int_X\dd\phi_i=0$, the exponent is unchanged under any constant shift, and hence $U_0(z')U_0(z)=U_0(z+z')$. Comparing with Eq.~\eqref{eq:typeiii-original-symmetry}, we see that $S(\mathbf{1}_1)$ and $U_0(\mathbf{1}_1)$ both shift $\phi_1$ by one everywhere, while $S(\mathbf{1}_1)$ also contributes a phase depending on $\phi_2,\phi_3$.

Take $\widetilde S(z)=U_0(z)$ as the symmetry and consider the diagonal circuit acting edge by edge,
\begin{equation}
 V\ket{\phi}=\zeta^{\int_X\phi_1\,\dd(\phi_2\phi_3)}\ket{\phi},
 \label{eq:typeiii-local-basis}
\end{equation}
where $\phi_2\phi_3$ is the pointwise product of vertex fields. Shifting $\phi\mapsto\phi+z$ increases the exponent by $\int_X\phi_1(z_2\,\dd\phi_3+z_3\,\dd\phi_2)$. The terms containing $z_1$ are total coboundaries and integrate to zero on the circle. On the basis vectors $V\ket{\phi}$, Eq.~\eqref{eq:typeiii-strip-actual-U} therefore gives
\[
 \widetilde S(z)\,V\ket{\phi}
 =\zeta^{-2z_2\int_X\phi_1\,\dd\phi_3}\,V\ket{\phi+z}.
\]
The right-hand side is precisely the action in Eq.~\eqref{eq:typeiii-original-symmetry} with the first two components interchanged and the coefficient multiplied by $2$. Graded commutativity of the degree-one classes then gives the type-III anomaly of the symmetry defined by closed hopping operators~\cite[Theorem~4.1]{Xue2026Comparison},
\begin{equation}
 [\widetilde\omega]=2\left[\frac{1}{N}a_2a_1a_3\right]=-2[\omega].
 \label{eq:typeiii-factor-minus-two}
\end{equation}
For $N=2$, $\widetilde S(z)$ acts as an on-site shift in this basis.

\paragraph{Two commutators with a twist in the first component.}
\label{subsec:typeiii-unit-flux}

For an exact background $A=-\dd v$, Eq.~\eqref{eq:normalized-R-multiplication} gives $T_0(v)S(z)T_0(v)^{-1}\propto S_A(z)$, so the commutation phases of $S_A(z)$ agree with those at zero background. The classification in Theorem~\ref{thm:hopping-statistics} imposes condition (ZZ) on configuration state lines in the vacuum orbit. Now cut the spatial circle $X$, glue its ends using the generator $S(\mathbf{1}_1)$ of the first component, and compare the remaining two generators under this twisted boundary condition. The background satisfies
\[
 A_2=A_3=0,\qquad \int_X A_1=\int_X a_1=1,\qquad a=A+\dd\phi .
\]
Use the oriented triangle $\Delta^2=[012]$ of the multiplication laws~\eqref{eq:dual-triangle-products} and~\eqref{eq:normalized-R-multiplication}. Applying $\mathbf{1}_2$ before $\mathbf{1}_3$, or in the reverse order, gives respectively
\[
 (\dd\eta_1\smile\dd\eta_2)[012]=1,\qquad
 (\dd\eta_2\smile\dd\eta_1)[012]=0 .
\]
Substituting $\omega(a)=a_1a_2a_3/N$ from Eq.~\eqref{eq:typeiii-stat-cocycle} leaves only the integral of the first component around the spatial circle. Both parameters of the gauge triangle carry a minus sign, so their product has a plus sign. With the convention of Eq.~\eqref{eq:intro-commutator}, the two multiplication laws give
\begin{align}
 [S_A(\mathbf{1}_3),S_A(\mathbf{1}_2)]
 &=\exp\!\left(\frac{2\pi\ii}{N}\int_X A_1\right)\id
   =e^{2\pi\ii/N}\id ,
 \label{eq:typeiii-unit-flux-symmetry}\\
 [U_A(\mathbf{1}_3),U_A(\mathbf{1}_2)]
 &=\exp\!\left(-\frac{2\pi\ii}{N}\int_X a_1\right)\id
   =e^{-2\pi\ii/N}\id .
 \label{eq:typeiii-unit-flux-hopping}
\end{align}
Take $N=2$. For a unit twist in the first component, the hopping commutator equals $-\id$ on every $\mathcal H_a$ in this sector. Thus the two closed hopping operators preserve no common state line in this eight-dimensional space. The operator $S_A(\mathbf{1}_3)$ reverses the eigenvalue of $S_A(\mathbf{1}_2)$, so the energy levels of any Hamiltonian preserving both original symmetry transformations occur in pairs~\cite{WangSantosWen2015}. With periodic boundary conditions the same hopping commutator equals $+\id$.

Equation~\eqref{eq:typeiii-unit-flux-hopping} is the statistical process for closed parameters in Sec.~\ref{subsec:hopping-statistics}. On the source configuration $a$, the product gives $e^{-2\pi\ii/N}\Pi^a_a$. Summing over $a$ in this orbit gives the identity operator with the coefficient shown in the equation. Decomposing $\mathbf{1}_i$ into local steps can only change the configuration-dependent phase of the closed operator, and this phase cancels in the commutator.

Now twist the spatial boundary condition by the closed hopping symmetry $\widetilde S(\mathbf{1}_1)=U_0(\mathbf{1}_1)$. Substituting $[\widetilde\omega]=-2[\omega]$ from Eq.~\eqref{eq:typeiii-factor-minus-two} into Eq.~\eqref{eq:normalized-R-multiplication} gives the commutation phase $e^{-4\pi\ii/N}$ for the two generators, which equals $1$ at $N=2$. For $N=4$ and a circle twisted by $S(\mathbf{1}_1)$, the commutation phases of $S_A$ and $U_A$ are $\ii$ and $-\ii$, respectively. Twisting the circle by $\widetilde S(\mathbf{1}_1)$ instead gives $-1$ for the two generators of $\widetilde S$.

\section*{Acknowledgments}
\addcontentsline{toc}{section}{Acknowledgments}

We thank Yitao Feng, Shenghan Jiang, Shuheng Shao, Bowen Shi, Xiao-Gang Wen, Yunlong Zang, and Carolyn Zhang for helpful discussions. We also acknowledge assistance from GPT 5.6-Sol, GPT 6-Astra, and GPT 6.2-Astra in preparing this work.

\appendix
\section{An integral homology interpretation of \texorpdfstring{$W_N$}{WN}}
\label{app:wn-homology}

In Fig.~\ref{fig:bockstein-fusion-history}, $N$ excitations of the same type fuse to the vacuum. To record this filling by integral chains, take $K_1=K(\ZZ_N,q)$ and $K_2=K(\ZZ_N,q')$. Their first positive-degree homology groups $H_q(K_1;\ZZ)$ and $H_{q'}(K_2;\ZZ)$ are both $\ZZ_N$. Choose integral cycles $u_1,u_2$ representing their respective generators. A chain $b_i$ satisfying $\partial b_i=Nu_i$ fills the fusion process. There are two fillings of $Nu_1\times u_2=u_1\times Nu_2$, using $b_i$ in the first or second factor. Subtracting them with the product orientation gives the cycle
\begin{equation}
 c_{\rm Tor}=b_1\times u_2-(-1)^{q}u_1\times b_2,\qquad
 \partial(b_1\times b_2)=(-1)^{q+1}N c_{\rm Tor}.
 \label{eq:wn-tor-cycle}
\end{equation}
Its $N$th multiple is a boundary. This is precisely the degree of the mixed torsion summand in the reduced K\"unneth formula. Since $H_{q+1}(K_1;\ZZ)=H_{q'+1}(K_2;\ZZ)=0$, in degree $d+2=q+q'+1$ that formula gives
\[
 \widetilde H_{d+2}(K_1\wedge K_2;\ZZ)
 \simeq\operatorname{Tor}(\ZZ_N,\ZZ_N)\simeq\ZZ_N .
\]

The phase measured in Sec.~\ref{sec:bockstein_word} distinguishes these $N$ classes. Take universal mod-$N$ cocycles $A_i$ satisfying $\langle A_i,u_i\rangle=1\bmod N$. On the second filling the Bockstein gives
\[
 \langle\beta_N A_2,b_2\rangle
 =\frac1N\langle\bar A_2,\partial b_2\rangle
 =\langle\bar A_2,u_2\rangle\equiv1\pmod N .
\]
The bidegree of $\omega_k=(k/N)\bar A_1\smile\beta_N A_2$ is $(q,q'+1)$, so only $u_1\times b_2$ contributes, with the sign of its coefficient. Thus
\begin{equation}
 \langle\omega_k,c_{\rm Tor}\rangle
 =(-1)^{q+1}\frac{k}{N}\pmod1 .
 \label{eq:wn-tor-pairing}
\end{equation}
The phase measured for $k=1$ has order $N$, so $[c_{\rm Tor}]$ generates the entire mixed summand. For $\int_X\bar s_1\smile\dd\bar s_2=1$, this is precisely the phase of $W_N$ in Eq.~\eqref{eq:wn-value}. Projecting to the first factor reduces both staircases to $\mathsf X^N$, and projecting to the second reduces both to $\mathsf Y^N$, so both pure components vanish. This identifies the image of $W_N$ in $H_{d+2}(K_1\times K_2;\ZZ)$ as $[c_{\rm Tor}]$. For general $X$, the measurement determines the lattice statistics homology class up to the kernel of this map.

\section{Existence of compatible patches and uniqueness of the statistics class}
\label{app:patch-transport}

\subsection{Mapping a local neighborhood into one coarse simplex}
\label{app:coarse-hopping-locality}
\label{app:local-function-coarse-proof}

Figure~\ref{fig:coarse-four-sites} assigns the same coarse vertex to a group of fine vertices. In general, a simplex of $K=\operatorname{sd}X$ is a chain of nested coarse faces. The chains in which every face contains $v$ form the closed dual cell $D_v$ of the coarse vertex $v$. These cells cover $K$. Subdivide $k$ more times to obtain $Y$. For each fine vertex $y$, choose a cell $D_v$ containing it and set $p(y)=v$. This label lies in the smallest coarse simplex containing $y$, so $p$ is a simplicial map.

Any path from $D_v$ to a coarse simplex that does not contain $v$ must traverse at least $2^k$ fine edges. To see this, map $K$ to an edge $[0,1]$ by sending the vertices corresponding to coarse faces containing $v$ to $0$ and all others to $1$. Subdivide both complexes $k$ times. The image of an edge path from $D_v$ to that coarse simplex must then cover all $2^k$ steps from $0$ to $1$.

Choose $2^k>2D+1$ and a vertex $x$ in $N_e\cap N_f$. From any vertex $y$ of the union, one can reach $x$ in at most $D+1+D$ steps, with the middle step inside the supporting simplex. Starting at $y\in D_{p(y)}$, one therefore reaches $x$ in a coarse simplex whose smallest face containing $x$ still contains $p(y)$. Thus all the labels lie in a single coarse simplex, proving Lemma~\ref{lem:local-function-coarse}.

\subsection{The GNVW index and restriction to a segment}\label{app:gnvw-toolkit}

\paragraph{Restriction to a segment.}
Group each cross-section of a strip into a single site. The GNVW index of a one-dimensional QCA is a positive rational number measuring net transport across a cut. It is multiplicative under composition. The index is $1$ exactly when a finite-depth quantum circuit (FDQC) in the strip implements the action. In that case, the action can be restricted to the interior of any segment. Every symmetry transformation of finite projective order on a closed line admits such restrictions~\cite{GrossNesmeVogtsWerner2012}.

\paragraph{The disk-boundary index.}
Suppose a finite Abelian group $G$ acts as an ordinary symmetry on a two-dimensional bosonic lattice by FDQCs. Choose a large disk $D$ and a disk restriction $U_D(a)$ for each $a\in G$, with $U_D(0)=\id$. The fusion error $U_D(a)U_D(b)U_D(a+b)^{-1}$ is supported near the disk boundary. The GNVW indices of these errors, evaluated in the induced orientation of the boundary, define a cohomology class $[\nu]\in H^2(G,\mathbb Q_{>0}^{\times})$, the disk-boundary index. It is independent of the restrictions and can be tested on one large disk in each connected component~\cite{KawagoeShirley2025,ShirleyZhangJiLevin2026}.

The disk-boundary index $[\nu]$ is trivial if and only if one can choose disk restrictions for all group elements on the original matter Hilbert space so that every fusion error is an FDQC within the boundary strip.

For any two such choices $U_D,U'_D$ and each group element $a$, the relative action $U'_D(a)U_D(a)^{-1}$ is also an FDQC within the boundary strip and can be restricted to any boundary segment.

\paragraph{The common edge of two disks.}
Suppose $[\nu]=0$ and choose restrictions on two neighboring disks as above. If both disks carry the label $a$, their restrictions implement $S(a)$ in the two interiors. In the gap between them one can insert a finite-depth circuit that implements the same $S(a)$ across the middle of the common edge. The circuit can stop short of the junction at either end.

\subsection{Constructing compatible patches from local restrictions}\label{app:patch-existence}

\emph{The domain of a patch.}
Definition~\ref{def:compatible-symmetry-patches} requires $P(r)$ only on $\Omega$. For instance, $r$ may be supported in a disk while the support of $r+z$ extends over the entire space. Suppose a preparation $Q(r)$ is available for all cochains and its elementary moves are local. Then compatibility and matching to $S$ are equivalent to the following condition for every closed parameter $z$,
\begin{equation}
 Q(r+z)\ \propto\ Q(r)S(z),\qquad \dd z=0,\quad Q(0)=\id .
 \label{eq:patch-preparation-shift}
\end{equation}
For an ordinary symmetry, $z$ is constant on each connected component. This is precisely the condition of Shirley et al.\ for shifting all labels together~\cite{ShirleyZhangJiLevin2026}. Given compatible patches that match $S$, the assembled preparation $Q(r)=T_0(r)$ is defined for every $r$ and satisfies $Q(r)\propto P(r)$ on $\Omega$. Conversely, the two constructions below produce $Q$ directly, and we take $P=Q|_\Omega$.

\paragraph{Point defects ($q=d\geq1$).}
Choose coarse dual cells so that each symmetry transformation acts along the dual edges marked by its closed parameter. On a given edge, closed parameters with the same label $a$ have the same action in its interior. The finite-order result of Appendix~\ref{app:gnvw-toolkit} allows us to restrict this action to a segment. We thus choose an edge circuit $F_\rho(a)$ once for each dual edge and label. This gives a consistent definition for every cochain $r$,
\[
 F(r)=\prod_\rho F_\rho\bigl(r(\rho)\bigr).
\]
where $\rho$ is the $(d-1)$-face dual to the edge, and the zero label gives the identity. The circuits on the segments in Fig.~\ref{fig:patch-existence-local}(a) are now fixed. It remains to specify the action where they meet.

\begin{figure}[htbp]
\centering
\begin{tikzpicture}[
  x=1cm,y=1cm,line cap=round,line join=round,
  every node/.style={font=\small,text=ink,inner sep=1.5pt},
  heading/.style={font=\small\bfseries,anchor=base},
  skeleton/.style={draw=muted!68,dash pattern=on 2pt off 2pt,line width=.6pt},
  segment/.style={draw=speciesB!80,fill=speciesB!16,line width=.8pt},
  disk/.style={draw=speciesB!65,fill=speciesB!9,line width=.65pt},
  junction/.style={draw=speciesA!85,fill=speciesA!17,line width=.85pt},
  leader/.style={draw=ink!72,line width=.55pt}
]
  \begin{scope}
    \node[heading] at (0,1.98) {(a) Point defects};
    \foreach \side in {-1,1}{
      \draw[skeleton] (\side*2.65,.88)--(\side*1.68,0)--(\side*2.65,-.88);
    }
    \draw[skeleton] (-1.68,0)--(1.68,0);
    \draw[segment,rounded corners=3pt] (-1.38,-.15) rectangle (1.38,.15);
    \foreach \p/\q in {-1.68/x,1.68/y}{
      \draw[junction] (\p,0) circle[radius=.35];
      \fill[ink] (\p,0) circle[radius=1.4pt];
      \node[below] at (\p,-.4) {$\q$};
    }
    \node at (0,.5) {$F_\rho(a)$};
    \draw[leader] (0,.35)--(0,.17);
    \node at (-1.58,.9) {$K_x$};
    \draw[leader] (-1.58,.75)--(-1.65,.37);
    \node at (1.58,.9) {$K_y$};
    \draw[leader] (1.58,.75)--(1.65,.37);
  \end{scope}
  \begin{scope}[xshift=7.1cm]
    \node[heading] at (0,1.98) {(b) Line defects in two dimensions};
    \begin{scope}
      \clip[rounded corners=6pt] (-1.83,-1.27) rectangle (1.83,1.47);
      \begin{scope}[shift={(150:.16)}]
        \filldraw[disk] (0,0)--(90:3)--(150:3)--(210:3)--cycle;
      \end{scope}
      \begin{scope}[shift={(30:.16)}]
        \filldraw[disk] (0,0)--(-30:3)--(30:3)--(90:3)--cycle;
      \end{scope}
      \begin{scope}[shift={(270:.16)}]
        \filldraw[disk] (0,0)--(210:3)--(270:3)--(330:3)--cycle;
      \end{scope}
      \foreach \angle in {90,210,330}{
        \draw[skeleton] (0,0)--(\angle:3);
        \draw[draw=accent!85,line width=4.4pt] (\angle:.26)--(\angle:2.4);
      }
      \draw[junction] (0,0) circle[radius=.28];
      \fill[ink] (0,0) circle[radius=1.4pt];
    \end{scope}
    \node at (-.94,.48) {$U_i(a)$};
    \node at ( .94,.48) {$U_j(b)$};
    \node at (0,-.89) {$U_k(c)$};
    \draw[leader] (.06,1.04)--(.35,1.26)--(.48,1.26);
    \node[right] at (.5,1.26) {$K_{ij}$};
    \draw[leader] (.28,.02)--(.56,-.14)--(.78,-.14);
    \node[right] at (.8,-.14) {$K_{ijk}$};
  \end{scope}
\end{tikzpicture}
\caption{Local construction of compatible patches. (a) The circuit $F_\rho(a)$ implements the action in the middle of a dual edge, and $K_x,K_y$ correct its endpoints. (b) The three disk restrictions $U_i,U_j,U_k$ leave strips between them. Apply $K_{ij}$ in the middle of a common edge, then $K_{ijk}$ at the junction of the three disks.}
\label{fig:patch-existence-local}
\end{figure}

At a dual vertex $x$, the background $A(x)$ records the net label contributed by the incident edges. For example, if a line with label $a$ enters $x$ along one edge and leaves along another, the two contributions $a$ and $-a$ cancel. Both a line passing through $x$ and an empty set of incident edges therefore give $A(x)=0$.

To compare preparations with the same $A(x)$, fix one incident edge at $x$ and list it first. Denote the signed contributions of the edges to $A(x)=-(\dd r)(x)$ by $t=(t_1,\ldots,t_{d+1})$, so $A(x)=\sum_i t_i$. As reference, let the first edge contribute all of $A(x)$ and leave the others empty, giving $t^{\mathrm{ref}}=(A(x),0,\ldots,0)$. This choice serves only to compare the actions near $x$.

Subtract the reference labels edge by edge. The signed sum of the differences is zero, precisely the condition for a parameter to be closed at $x$. We may therefore choose a global closed parameter $z$ that realizes these differences on the incident edges.\footnote{On a disk, a closed cochain of positive degree can be written as $\dd v$. Extend $v$ and take its coboundary $\dd v$ to obtain the extension. A closed cochain of degree zero extends as a constant on each connected component.} Let $F_x(t)$ be the product of the circuits on the incident edges, with the input to each $F_\rho$ expressed in the original orientation of that edge. Applying the closed action $S(z)$ followed by the reference preparation $F_x(t^{\mathrm{ref}})$ agrees with $F_x(t)$ in each edge interior. A vertex unitary $K_x(t)$ supplies the difference at $x$. For every $r$ we can thus take
\begin{equation}
 Q(r)=\Bigl(\prod_x K_x(t_x)\Bigr) F(r),
 \qquad t_x=\text{labels of the edges incident on $x$}.
 \label{eq:patch-existence-reference}
\end{equation}
Adding a closed parameter preserves the reference at every vertex, so the actions in the edge interiors and at the vertices satisfy Eq.~\eqref{eq:patch-preparation-shift}.

\paragraph{Line defects in two dimensions ($d=2,\ q=1$).}
Here $r$ assigns a group element to each dual disk. Suppose compatible patches matching $S$ exist. The resulting $T_0(r)$ obeys the simultaneous-shift relation above, and a local change of labels changes only the action nearby. Shirley et al.\ proved that for such a globally defined preparation, every fusion error of the disk restrictions $U_D(a)=T_0(a\mathbf1_D)$ is a boundary circuit~\cite{ShirleyZhangJiLevin2026}. Hence the disk-boundary index $[\nu]$ is trivial.

Conversely, suppose $[\nu]=0$. Choose restrictions as in Appendix~\ref{app:gnvw-toolkit} and apply the corresponding restriction inside each disk in Fig.~\ref{fig:patch-existence-local}(b). Consider a left disk $i$ and a right disk $j$. If both carry $a$, completing the common edge gives $S(a)$ on both sides and throughout the middle of the edge.

If the left disk carries $a$ and the right disk $b$, first apply $S(a)$ on both sides, then apply $U_j(b-a)$ on the right alone. The left still implements $S(a)$, the right implements $S(b)$, and this prescription also fixes the action in the middle of their common edge. On the right, applying $a$ followed by $b-a$ differs from applying $b$ directly by a disk-boundary circuit. Restrict this circuit to the common edge and combine it with the correction used to join the two disks labeled $a$. The result is the edge circuit $K_{ij}(a,b)$ shown in the figure.

Finally, consider a junction where three disks carry labels $a,b,c$. Choose the action near the junction for $(0,b-a,c-a)$ using the disk and edge circuits just constructed. For $(a,b,c)$, prescribe $S(a)$ near the junction followed by this reference preparation. The three disk interiors and the middles of the three edges already implement this prescription. The difference at the junction can be supplied by a local unitary $K_{ijk}$. This defines $Q(r)$ for arbitrary disk labels. Adding $h$ to every label leaves the two differences unchanged and multiplies the entire preparation on the right by $S(h)$.

For either construction, take $\Omega$ to consist of the disk-supported labels in Definition~\ref{def:compatible-symmetry-patches}. An elementary move $e$ changes only one edge and its endpoints, or one disk and its edges and junctions. Two assignments that give the same local background differ there by a closed parameter $z$. The corresponding $S(z)$ cancels in the ratio $Q(r+e)Q(r)^{-1}$, so the projective matrix of this move depends only on the nearby background. Starting from zero, the labels in a disk can be filled in one at a time. If two moves need a common preparation, copy the labels on $N_e\cup N_f$ into a single disk containing the union. Composing the elementary moves gives $T_0(z)\propto Q(z)\propto S(z)$, as required for matching.

\subsection{Independence of the statistics class from the choice of restriction}\label{app:patch-uniqueness}

\begin{proof}[Proof of Lemma~\ref{lem:relative-transport}]
Use the same parameter $r$ to prepare $A=-\dd r$ from the zero background in both constructions. Their matter matrices differ by a change of basis
\[
 C_A\propto T'_0(r)T_0(r)^{-1},\qquad C_0=\id .
\]
Replacing $r$ by $r+z$ multiplies both preparations on the right by the same closed action $S(z)$. Hence the projective class of $C_A$ depends only on $A$. Now take a step $s$ from $A$ to $B=A-\dd s$. The projective group law gives $T'_A(s)C_A\propto C_B T_A(s)$. Choose a representative of each $C_A$ and define the matter matrix of the relative gauge transformation by
\[
 R_A(s)=T'_A(s)C_A T_A(s)^{-1}C_A^{-1}
       =\exp(2\pi\ii h(s;A))\,C_B C_A^{-1}.
\]
For closed parameters we have $R_0(z)\propto\id$. Comparing consecutive steps $s,t$ with the single step $s+t$, the intermediate changes of basis cancel and give
\[
 c_R(t,s;A)=h(s;A)+h(t;B)-h(s+t;A)
           =c_{T'}(t,s;A)-c_T(t,s;A)\pmod1 .
\]
It remains to check locality of the elementary moves. Arrange the preparation into a bounded number of layers of elementary moves. Then $C_A O C_A^{-1}$ depends only on the values of $r$ near the support of the local operator $O$. If two backgrounds agree on a disk around this support, the difference of their preparation parameters is closed on the disk. Extend that difference to a global closed parameter as in Appendix~\ref{app:patch-existence}. One can then use the same local preparation to compute the conjugated operator for both backgrounds. Taking $O=T_A(e)^{-1}$ shows that the matrix $R_A(e)$ itself is local.
\end{proof}

By the lemma, it suffices to show that every compatible restriction of the trivial symmetry gives trivial statistics. In this case $R_0(z)\propto\id$ for all closed parameters, and all $r$ preparing the same $A=-\dd r$ give the same projective matrix $C_A\propto R_0(r)$. We need only choose representatives for the $C_A$ so that the phase $h(e;A)$ of an elementary move depends only on the background near $e$.

\paragraph{Point defects ($q=d\geq1$).}
The dual vertices are well separated. Away from them one can use the zero-background preparation, so $C_A$ acts there as the identity. Near a dual vertex $x$, its action depends only on $A(x)$. Choose a unitary $K_x(A(x))$ for each value, taking the identity at zero. Then $C_A=\prod_x K_x(A(x))$. Changing one dual edge gives a background $B=A-\dd e$ that differs from $A$ only at the two endpoints. All other factors in $C_BC_A^{-1}$ cancel point by point, leaving only the endpoint factors.

\begin{figure}[htbp]
\centering
\begin{tikzpicture}[x=1cm,y=1cm,line cap=round,line join=round,
  every node/.style={font=\small,text=ink},
  quiet/.style={draw=muted!50,line width=.6pt},
  changed/.style={draw=accent,line width=1.8pt},
  junction/.style={circle,draw=accent,fill=white,line width=1.1pt,
    minimum size=5pt,inner sep=0pt}]
  \coordinate (a) at (0,1.02);
  \coordinate (b) at (-1.23,-.99);
  \coordinate (c) at (1.23,-.99);
  \fill[pale] (a)--(b)--(c)--cycle;
  \draw[quiet] (a)--(0,1.34);
  \foreach \sgn in {-1,1}{
    \draw[quiet] (\sgn*1.23,-.99)--(\sgn*1.91,-1.25);
    \draw[quiet] (\sgn*1.91,-1.25)--(\sgn*2.20,-.88)
                   (\sgn*1.91,-1.25)--(\sgn*2.16,-1.49);
    \fill[muted!55] (\sgn*1.91,-1.25) circle (1.6pt);
  }
  \draw[quiet] (a)--(b)--(c)--cycle;
  \foreach \start/\stop in {a/b,b/c,c/a}
    \draw[changed] ($(\start)!.085!(\stop)$)--($(\start)!.915!(\stop)$);
  \foreach \v in {a,b,c} \node[junction] at (\v) {};
  \node[text=accent] at (0,-.28) {$i$};
  \node[anchor=west,font=\scriptsize] at (.28,1.03) {junction $K$};
  \node[anchor=west,font=\scriptsize] at (.88,.12) {edge segment $L$};
\end{tikzpicture}
\caption{An elementary move for line defects in two dimensions. Changing the label of dual disk $i$ from $r(i)$ to $r(i)+g$ can affect only the edge circuits $L$ and junction unitaries $K$ on its boundary. The affected segments and junctions are highlighted for a triangular disk.}
\label{fig:patch-uniqueness-cancellation}
\end{figure}

\paragraph{Line defects in two dimensions ($d=2,\ q=1$).}
First assign a label $u$ to a single disk $D$. The preparation $V_u=R_0(u\mathbf1_D)$ acts as the identity inside and outside the disk, leaving support only near its boundary. These restrictions come from compatible preparations, so Appendix~\ref{app:patch-existence} implies that each of their fusion errors is a boundary circuit. The trivial symmetry also admits the restriction $\id$ for every label. Comparing these two choices as in Appendix~\ref{app:gnvw-toolkit}, we see that $V_u$ can be restricted to a boundary segment.

If neighboring disks carry labels $a,b$, use the reference $(0,b-a)$ to compute $C_A$ on their common edge. For each edge we therefore choose a segment of the circuit $V_u$ depending only on $u=b-a$. Write $L_A$ for the product over all segments. Two restrictions of the same action to a segment differ only by a unitary at each endpoint, which can be absorbed into the junctions~\cite[Appendix~D.1, Eq.~(98)]{KawagoeShirley2025}. After removing $L_A$, choose one unitary at each three-disk junction depending only on the two independent local label differences. Let $K_A$ denote their product and take $C_A=K_A L_A$. As Fig.~\ref{fig:patch-uniqueness-cancellation} shows, changing one disk makes $C_BC_A^{-1}=K_B(L_B L_A^{-1})K_A^{-1}$ depend only on its edge segments and junctions.

In both cases the matrices $R_A(e)$ and $C_BC_A^{-1}$ depend only on the background near $e$, so their relative phase $h(e;A)$ is local. Compose elementary moves in the fixed order of Sec.~\ref{subsec:local-patch-transport}. By the coarse-graining lemma (Lemma~\ref{lem:local-function-coarse}), removing $h(e;A)$ from each step is an allowed local basis change on the coarse lattice. After this change, $\widetilde R_A(s)=C_{A-\dd s}C_A^{-1}$ for every $s$. The intermediate factors cancel exactly in two consecutive moves, so the multiplier is identically $1$. Lemma~\ref{lem:relative-transport} gives the same statistics class for both restrictions of the symmetry, and the gauging criterion follows from Theorem~\ref{thm:hopping-statistics}.

\end{document}